\documentclass[aps,pra,showpacs,twoside,twocolumn,longbibliography,10pt]{revtex4-2}
\usepackage[colorlinks=true, citecolor=blue, urlcolor=blue, linkcolor = blue ]{hyperref}
\usepackage{epsfig,newlfont,amssymb,amsfonts,amsmath,bm,subfigure,palatino,mathtools,amsthm,braket,times,soul,enumitem,color}
\usepackage[normalem]{ulem}
\newcommand{\stkout}[1]{\ifmmode\text{\sout{\ensuremath{#1}}}\else\sout{#1}\fi}
\usepackage[english]{babel}
\usepackage[utf8]{inputenc}
\usepackage{array}
\usepackage{xcolor}
\usepackage{graphics}

\def\Tr{\text{Tr}}

\newcommand{\AD}[1]{{\color{black} #1 }}

\usepackage{amsthm}
\usepackage{verbatim}
\usepackage{bbm}
\usepackage{wrapfig}

\usepackage{hyphenat}

\usepackage{float}

\usepackage{orcidlink}

\usepackage[normalem]{ulem}

\usepackage{multirow}
\usepackage{makecell}

\newlength\figureheight 
\newlength\figurewidth 

\begin{document}


\title{Optimal work extraction in measurement-based quantum Otto engines: Non-adiabaticity and generalized measurements can be beneficial}



\author{Arunabha Das$^{1,2}$\, \orcidlink{0009-0000-2201-185X}}
\author{Sayan Mondal$^{1,2}$\, \orcidlink{0000-0003-2921-403X}}\thanks{These authors contributed equally to this work.}
\author{Debarupa Saha$^{1,2}$\, \orcidlink{0009-0006-3226-7799}}\thanks{These authors contributed equally to this work.}

\author{Ujjwal Sen$^{1,2}$\, \orcidlink{0000-0002-0091-5847}}


\affiliation{$^{1}$Harish-Chandra Research Institute, Chhatnag Road, Jhunsi, Prayagraj  211 019, India\\
\(^2\)Homi Bhabha National Institute, Training School Complex, Anushakti Nagar, Mumbai 400 094, India}

\begin{abstract}
Measurement-based quantum heat engines have attracted significant interest as alternatives to conventional thermal engines, as they replace the hot thermal reservoir with quantum measurements, thereby offering greater controllability and simpler implementation. Motivated by these advantages, we investigate a measurement-driven quantum Otto engine with a qubit working substance and study the optimal work extractable from such engines along with the corresponding efficiency, including whether their performance can surpass that of conventional quantum Otto cycles. We analyze the engine in both the infinite-time (adiabatic) and finite-time (non-adiabatic) regimes, considering two distinct implementations obtained through optimization over all projection-valued measurements (PVMs) and over all two-outcome positive operator-valued measurements (POVMs). We show that measurement-based engines can outperform conventional quantum Otto engines within specific parameter regimes and that POVM-based engines can yield higher optimal work extraction than PVM-based ones. Furthermore, by incorporating the thermodynamic cost associated with resetting the auxiliary system required for POVM implementation, we demonstrate that the resulting net work output can still exceed that of PVM-based engines under suitable conditions on the spectral gaps and cold bath temperature. We also identify regimes in which non-adiabatic implementations can yield higher work output and efficiency than their adiabatic counterparts. Our study provides operational guidelines for designing improved measurement-driven quantum Otto engines.
\end{abstract}

\maketitle
\section{Introduction}

Quantum thermodynamics has attracted significant attention due to both its foundational importance~\cite{Gemmer2009, Vinjanampathy01102016, ThermoBook2018, Deffner-book2019} and its prospect for developing nanoscale technologies for energy storage, transport, and conversion. In this context, a variety of quantum thermal devices have been proposed and extensively investigated, including quantum heat engines~\cite{Engine_1,Engine_2,Engine_3,Engine_4,Engine_5,Engine_6,Engine_7,Engine_8,Engine_9,Engine_10,Engine_11,CANGEMI20241}, quantum refrigerators~\cite{refrigerator_0,refrigerator_1,refrigerator_2,PhysRevLett.105.130401,CANGEMI20241}, quantum transistors~\cite{transistor_1,transistor_2,transistor_3,transistor_4,transistor_5,transistor_6}, quantum heat transformers~\cite{transformer2,transformer_1}, and quantum batteries~\cite{battery_1,battery_2,battery_3}. Among these, quantum heat engines occupy a central role as they provide a fundamental framework for understanding how thermodynamic tasks such as the conversion of heat into useful work can be realized and potentially enhanced in the quantum regime.


The quantum analogs of standard thermodynamic cycles, such as the Carnot~\cite{Carnot_1,Carnot_2,Otto_2,Carnot_3,CANGEMI20241} and Otto~\cite{Otto_1,Otto_2,Otto_3,Otto_4,Otto_5,Otto_6,Otto_7,Otto_8,Yunger-Halpern2019,Otto_10,CANGEMI20241} cycles, have therefore been widely investigated. In this work, we focus on the quantum Otto cycle. In a conventional quantum Otto engine, the working medium undergoes two unitary work strokes interleaved with thermalization strokes involving hot and cold thermal reservoirs. The work strokes are typically assumed to be adiabatic in order to suppress unwanted transitions between instantaneous energy eigenstates. However, truly adiabatic evolution requires long operation times, resulting in vanishingly small output power. Motivated by the need to realize thermal machines operating at finite power, considerable effort has therefore been devoted to the study of finite-time quantum Otto engines and the role of non-adiabatic effects~\cite{Otto_finite_time_1,Otto_finite_time_2,Otto_finite_time_3,Otto_finite_time_4,Otto_finite_time_5,Otto_finite_time_6,Otto_finite_time_7,Otto_finite_time_8,Otto_finite_time_9,Otto_finite_time_10,Otto_finite_time_11,Otto_finite_time_12,Otto_finite_time_13,Otto_finite_time_14,Otto_finite_time_15,Otto_finite_time_16}.

These theoretical developments have been accompanied by rapid experimental progress. Quantum Otto engines and related thermal cycles have been experimentally implemented using superconducting quantum circuits~\cite{Experimental_1,Experimental_8}, ultracold atomic systems~\cite{Experimental_2}, spin-based platforms~\cite{Experimental_3}, optical simulators~\cite{Experimental_4}, cavity-QED setups~\cite{Experimental_5}, and several other experimentally accessible architectures~\cite{Experimental_6,Experimental_7,Experimental_9}. Such developments have established quantum thermal machines as experimentally relevant systems rather than merely theoretical constructs.

An interesting alternative to conventional quantum Otto engines arises when one of the thermalization strokes, typically the hot-bath stroke, is replaced by a quantum measurement process~\cite{MBO_1,MBO_2,PV_1,PV_2,PV_3,PV_4_&_finite_time_2,PV_5_&_finite_time_3,PV_6_&_finite_time_4,PV_4_&_finite_time_2,PV_5_&_finite_time_3,PV_6_&_finite_time_4,Three_&_five_stroke_engines_with_POVMs,aux_asst_1_&_finite_time_5, Jakhar2026}. In such measurement-based quantum Otto engines, energy is injected into the working medium through quantum measurements rather than via thermal contact with a hot reservoir, enabling measurement-induced work extraction under generalized thermodynamic settings. These engines have attracted considerable attention in recent years because quantum measurements offer enhanced controllability, simpler implementation, and the possibility of improved thermodynamic performance. For instance, enhanced output power has been reported in many-body spin-based engines where the thermalization stroke is replaced by a quantum measurement process~\cite{aux_asst_1_&_finite_time_5}. Measurement-driven engines have been extensively investigated both in the adiabatic regime~\cite{MBO_2} and in finite-time non-adiabatic settings~\cite{PV_4_&_finite_time_2,PV_5_&_finite_time_3,PV_6_&_finite_time_4}.

The measurements employed in such engines can be either projective-valued (PV) measurements~\cite{PV_1,PV_2,PV_3,PV_4_&_finite_time_2,PV_5_&_finite_time_3,PV_6_&_finite_time_4} or more general positive operator-valued measurements (POVMs)~\cite{Three_&_five_stroke_engines_with_POVMs,Swith_based_engine,aux_asst_1_&_finite_time_5}. Existing studies have demonstrated that measurement-based engines can outperform conventional hot-bath-driven Otto engines in terms of extractable work and efficiency. However, most previous studies focus on specific measurement protocols or restricted parameter regimes, and a systematic characterization of optimal work extraction in measurement-driven engines remains largely unexplored, particularly in finite-time non-adiabatic regimes. This naturally raises several fundamental questions: What is the maximum work extractable from measurement-based quantum Otto engines? How does this optimum compare across PVM-, POVM-, and conventional two-bath Otto engines? Furthermore, how is the performance modified once the thermodynamic cost associated with implementing measurements is taken into account?



Motivated by these open questions, in the present work we investigate measurement-based quantum Otto engines in which the hot-bath thermalization stroke is replaced by a quantum measurement process. Considering a single-qubit working substance, we analyze both projective-valued measurements (PVMs) and generalized two-outcome POVMs based quantum Otto engines, and derive the corresponding conditions for optimal work extraction in both the adiabatic and non-adiabatic regimes.


\AD{We first show that, in the adiabatic limit, the efficiency of measurement-based Otto engines remains identical to that of conventional two-bath Otto engines. We then examine whether such engines can nevertheless provide enhanced work extraction. In this regard, we also investigate finite-time measurement-based Otto cycles, where the unitary work strokes are intrinsically non-adiabatic, and compare with their conventional two-bath counterparts. In particular, we demonstrate that the maximum work obtainable using PV measurements always exceeds that of conventional Otto engines operating with finite-temperature hot baths, in the adiabatic limit as well as in the non-adiabatic regimes.

Moreover, we demonstrate that POVM-based Otto engines can yield even higher optimal work than PVM-based ones, not only in the adiabatic limit but also in non-adiabatic regimes. Particularly, we identify that, in the adiabatic limit, POVM-based Otto engines can yield even higher optimal work than their PVM-based counterparts under suitable conditions based on the cold bath temperature and the spectral gaps of the setup, even when the thermodynamic cost to recycle the auxiliary system required for the POVM-implementation is considered. In terms of the efficiency at the optimal work extraction, POVM-based engines can outperform the PVM-based engines, particularly in the non-adiabatic regime.

Furthermore, in case of measurement-based engines, non-adiabatic implementations can yield higher work output and efficiency than their adiabatic counterparts, and in particular, we identify the corresponding conditions in case of PVM-fuled Otto engines. Also, measurement-fueled Otto cycles can operate as engines even in parameter regimes where conventional Otto cycles fail to operate as engines. Finally, we provide a detailed comparative analysis of conventional two-bath, PVM-based, and POVM-based Otto engines in both adiabatic and non-adiabatic regimes, thereby identifying the operational regimes where measurement-driven quantum thermal machines exhibit enhanced thermodynamic performance.

}

The remainder of the paper is organized as follows. In Sec.~\ref{Sec - Preliminaries}, we introduce the necessary preliminaries, including the framework of quantum measurements, the conventional Otto cycle, the adopted notations and sign conventions, and the description of the working substance. The optimization of work extraction, the computation of the efficiency for the maximal work extraction, for both the PVM- and POVM-based Otto engines in Sec.~\ref{sec - adia_PVM_POVM} and a comparative analysis of performance between them also including the conventional two-bath Otto engine in Sec.~\ref{sec - comparison_adia}, in the adiabatic limit,  whereas Sec.~\ref{sec - Non-adia_PVM_POVM} and Sec.~\ref{sec - comparison_non-adia}, respectively, discuss the similar analysis in the non-adiabatic regime. In Sec.~\ref{sec - key results} we summerize our results in a nutshell based on the comparative study and key findings, establishing a hierarchy between the three classes of quantum Otto engines according to their performances. Finally, we conclude in Sec.~\ref{sec - Conclusion}.

\section{Preliminaries}
\label{Sec - Preliminaries}
In this section, we discuss the notion of PVM and POVM in the context of the measurement-based Otto engine. We discuss various aspects of the engine including its work output and efficiency.
\subsection{Quantum measurements}
In quantum mechanics, any observable $\hat O$ is a Hermitian linear operator acting on an arbitrary $d$-dimensional Hilbert space $\mathcal{H}$, that can be spectral-decomposed as $\hat O \coloneqq \sum_{i}{a_i P_i}$, where $a_i$ are the eigenvalues and $P_i$ are the corresponding eigenvectors. 
It is well known that the eigenvalues of a hermitian operators are real and the corresponding eigenstates are orthonormal and complete.
The building block of the PV measurement are these projection operators $\{P_i\}$, which are:
(i) Hermitian $P_{i}^{\dagger} = P_i$, (ii) positive semi-definite $P_i \succeq 0 $, (iii) complete $\sum_{i}{P_i} = I_d$, with $I_d$ denoting the identity operator on the $d$-dimensional Hilbert space $\mathcal{H}$, and (iv) mutually orthonormal $P_i P_j = \delta_{ij} P_i$. When the observable $\hat{O}$ is measured, for an arbitrary state $\rho$, the probability of getting a particular outcome $a_i$ is given by $p_i = \Tr(P_i \rho)$. This yields the corresponding post-measurement state as  $\rho'_i = {P_i \rho P_i}/{p_i}$. 
If the measurement outcome is not recorded (i.e., one performs a non-selective projective measurement), the resulting state is given by the Lüders update rule~\cite{Luders_2006},
$\rho_{\mathrm{PM}} \coloneqq \sum_i p_i \rho'_i = \sum_i P_i \rho P_i$, where the $\mathrm{PM}$ in the subscript signifies the post-measurement state.

The PV measurements do not constitute the most general kind of measurements allowed in quantum mechanics. There exists a more general class of measurements, called POVM in the framework of quantum mechanics. It is characterized by a collection of operators $\{E_i\}$ corresponding to the $i$-th outcome of the measurement. The probability of getting the $i$-th outcome upon measurement on $\rho$ is given by  $\tilde{p}_{i} = \Tr(E_i \rho)$. Similar to the PV measurements, the POVMs also satisfy the following properties: the Hermiticity, the semi-definite positivity, and the completeness. However, being more general than PV measurements, POVMs may not satisfy the orthogonality.
The post-measurement state corresponding to the $i$-th outcome, is given by $\bm{\rho}_{i} = {M_i \rho M_i}/{\tilde{p}_{i}}$, where $M_i$ are the Kraus operators, defined as $M_i = \sqrt{E_{i}}$, according to von Neumann-L\"uder's rule~\cite{Busch_1998}. The post measurement state without the selection of any particular outcome is given by $\bm{\rho}' = \sum_{i}{\tilde{p}_{i} \bm{\rho}_{i}} = \sum_{i}{M_i \rho M_{i}^{\dagger}}$.
According to Naimark's dilation theorem~\cite{Naimark1976Non,Peres1990Neumark,Nielsen_Chuang_2010,Paris_2012,preskill2018quantum_ch3}, any POVM acting on a Hilbert space can be realized as a PV measurement on a larger Hilbert space. It can be performed by coupling the working substance $\rho_s$ with an auxiliary system $\rho_a$, leading to $\rho_{sa} = \rho_s \otimes \rho_a$, then applying a global unitary transformation, and finally performing a PV measurement on the extended Hilbert space associated with the composite system.
\begin{figure}[t]
    \centering
    \includegraphics[
        width=8.7cm,
    ]{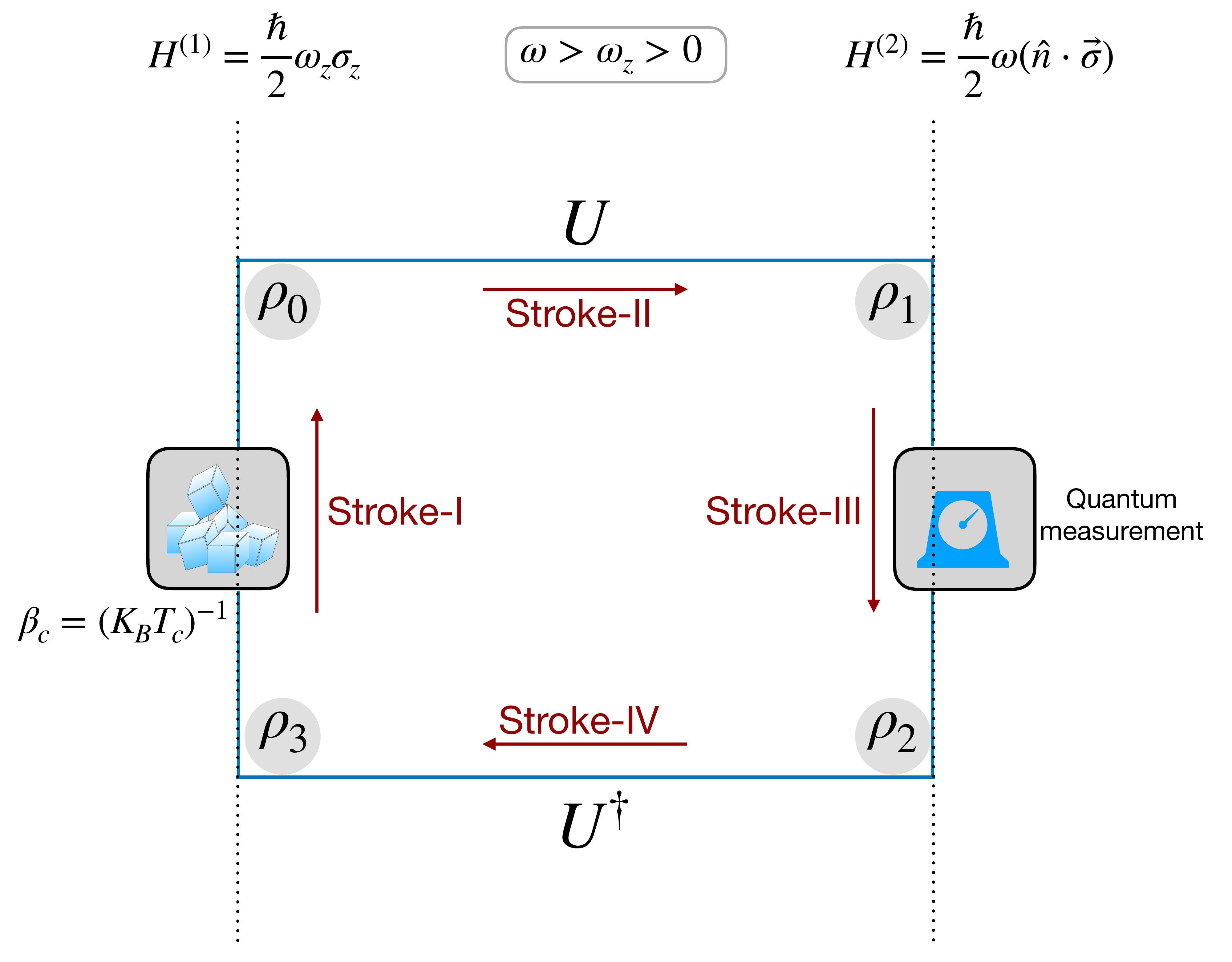}
    \caption{{\textbf{Schematic diagram of the measurement-based quantum Otto engine.} The measurement-based quantum Otto engine consists of four strokes. In Stroke-I, the working medium thermalizes with a cold bath at temperature $T_c$. Stroke-II and Stroke-IV are unitary driving strokes, generated through adiabatic or non-adiabatic driving of the Hamiltonian. Since these strokes are entropy conserving, they correspond to the work strokes of the cycle. Stroke-III is the measurement stroke, which replaces the hot-bath thermalization process present in the conventional Otto engine, where the system would otherwise interact with a bath at temperature $T_h>T_c$. The measurement can be either PVM or generalized POVM. 
    During both Stroke-I and Stroke-III, the Hamiltonian remains fixed, making them isochoric heat strokes.}}
    \label{fig: schematic}
\end{figure}
In the subsequent section we briefly discuss the functioning of a quantum Otto engine.
\subsection{Quantum Otto engines}
\subsubsection{Conventional two-bath quantum Otto engines}
\label{tradional Otto 2b adia}
The engine is a cyclic sequence of thermodynamic processes that converts heat to mechanically useful work. A quantum system, described in the arbitrary $d$-dimensional Hilbert space, serves as the working substance in the case of quantum engines. In our work, a two-level system is considered as the working substance. In each operation of the cycle, the working substance undergoes a number of distinct thermodynamic operations, called strokes.

The conventional Otto cycle consists of four strokes. These involve two heat strokes and two work strokes. In order to extract work, the Otto cycle operates with the help of two thermal baths: a hot bath at temperature $T_h$ and a cold bath at temperature $T_c$, with $T_h> T_c$. The work is extracted by driving the system between two Hamiltonian  $H^{(1)}$ and $H^{(2)}$ with varying spectral gap. We briefly discuss the four strokes in the following.
\begin{itemize}
    \item \textit{\textbf{Stroke-I.}} This is a thermalizing stroke where the system (working substance) is put in contact with the cold bath and the system-Hamiltonian is kept fixed at $H^{(1)}$. In this stroke, the energy is exchanged between the system and the bath. The entropy changes during this process, and consequently, the change in energy of the system in this stroke contributes to heat. This stroke yields the thermal state (Gibbs state) $\rho_0 = e^{-\beta_c H^{(1)}}/\Tr(e^{-\beta_c H^{(1)}})$ corresponding to the Hamiltonian $H^{(1)}$ and the cold bath temperature $T_c = 1/(k_B \beta_c)$ after complete thermalization. The exchanged heat during this stroke is given by
    \begin{align}
        q_c = \Tr(H^{(1)} \rho_0) - \Tr(H^{(1)} \rho_3) \label{qc_gen},
    \end{align}
    where $\rho_3$ is the final state after the completion of the previous cycle.
    \item \textit{\textbf{Stroke-II.}} After the stroke-I, the engine is isolated from the cold bath. The system Hamiltonian is driven from $H^{(1)}$ to $H^{(2)}$. This leads to a unitary evolution of the system, i.e., $\rho_0 \to \rho_1 = U \rho_0 U^{\dagger}$. This stroke is considered as a work stroke, as unitary evolution preserves the entropy of the system. The work associated with this stroke is given by
    \begin{align}
        w_1 = \Tr(H^{(2)} \rho_1) - \Tr(H^{(1)} \rho_0) \label{w1_gen}.
    \end{align}    
    \item \textit{\textbf{Stroke-III.}} This is also a thermalizing stroke, where the system is put in contact with the hot bath, and consequently, the exchanged energy associated with this stroke contributes to heat. The Hamiltonian is fixed at $H^{(2)}$. The energy is exchanged between the bath and the system. Similar to Stroke-I, this is also a thermalizing stroke which leads to the thermal state (Gibbs state) $\rho_2 = e^{-\beta_c H^{(2)}}/\Tr(e^{-\beta_h H^{(2)}})$ corresponding to the Hamiltonian $H^{(2)}$ and the hot bath temperature $T_h = 1/(k_B \beta_h)$ after complete thermalization. The exchanged heat during this stroke is given by
    \begin{align}
    q_h = \Tr(H^{(2)} \rho_2) - \Tr(H^{(2)} \rho_1) \label{qh_gen}.
    \end{align}   
     \item \textit{\textbf{Stroke-IV.}} The system is isolated from the hot bath, and the Hamiltonian is driven from $H^{(2)}$ to $H^{(1)}$. Consequently, the system undergoes a unitary evolution, i.e., $\rho_2 \to \rho_3 = \tilde{U} \rho_2 \tilde{U}^{\dagger}$. This is again a work stroke, because the entropy of the system is conserved during this stroke due to unitary evolution. The work associated with this stroke is given by
     \begin{align}
         w_2 = \Tr(H^{(1)} \rho_3) - \Tr(H^{(2)} \rho_2) \label{w2_gen}.
     \end{align}
     We consider the two unitary strokes to be symmetric. For the Otto cycle, the exact path of $U$ of stroke-II is retraced by the unitary $\tilde{U}$ of the stroke-IV reversibly. Consequently, we obtain $\tilde{U} = U^{\dagger}$.
\end{itemize}
The total work corresponding to the cycle is the sum total of the work from both of these two work strokes, i.e., stroke-II and stroke-IV. The Otto engine is usually analyzed in the adiabatic regime, i.e., the time taken by each of the work strokes (stroke-II and stroke-IV) is taken to be very large. Hence, the evolutions of the states in the unitary strokes follow the adiabatic theorem, while the thermalizing strokes completely thermalize the system to the corresponding Gibbs state. Recently, engines involving work strokes with finite-time duration have also been studied, and we call such engines to be operating in the non-adiabatic regime~\cite{Otto_finite_time_1,Otto_finite_time_2,Otto_finite_time_3,Otto_finite_time_4,Otto_finite_time_5,Otto_finite_time_6,Otto_finite_time_7,Otto_finite_time_8,Otto_finite_time_9,Otto_finite_time_10,Otto_finite_time_11,Otto_finite_time_12,Otto_finite_time_13,Otto_finite_time_14,Otto_finite_time_15,Otto_finite_time_16}. In both cases, it is typically considered that the thermalizing strokes (stroke-I and stroke-III) lead to complete thermalization.
The detailed analysis of the conventional Otto cycle with two baths in the adiabatic limit in Appendix~\ref{App - 2b conventional Otto adia}, while the analysis including the non-adiabatic regime is given in Appendix~\ref{App - nonadia 2b}. While these analyses are well established, we present them here for completeness, as they provide a useful basis for our comparative analysis.

\subsubsection{Measurement-based Otto engines}
In a measurement-based Otto cycle, one of the thermalization strokes is replaced by a quantum measurement stroke. The energy exchanged during this process is interpreted as heat, since the measurement changes the entropy of the working substance while the Hamiltonian remains fixed. Depending on the protocol, either the hot bath~\cite{PV_1,PV_2,PV_3,PV_4_&_finite_time_2,PV_5_&_finite_time_3,PV_6_&_finite_time_4,ent_measurement,MBO_1,MBO_2,Three_&_five_stroke_engines_with_POVMs} or the cold bath~\cite{aux_asst_1_&_finite_time_5} can be replaced by a measurement process. However, both thermalization strokes cannot simultaneously be replaced by measurements. This is because a bare quantum measurement is intrinsically entropy-increasing and, by itself, cannot sustain cyclic engine operation, as established by the no-go theorem~\cite{No-Go-MBO}. Consequently, an additional entropy-reducing mechanism, such as feedback control or bath-assisted thermalization, is necessary for the engine to operate.

In the current work, the stroke-III of the Otto cycle, i.e., the thermalizing stroke involving the hot bath, is replaced with a measurement process. The first two strokes are the same as the conventional Otto cycle. The exchanged energy is incorporated as heat, according to the entropic definition in the context of quantum thermodynamics. Fig.~\ref{fig: schematic} depicts this type of Otto cycle. Besides the projection-valued (PV) measurements, stroke-III can also be realized using projective operator-valued measurements (POVM), and such measurements are modeled with the assistance of an auxiliary system. The goal of our work is to find the optimal extractable work and the corresponding efficiency from these two classes of measurement-based quantum Otto engines. Moreover, our work provides a comparative study between them, also including the conventional two-bath one.

Before proceeding with our analysis, we first introduce the notations and sign conventions used throughout this study. These are discussed below.

\subsection{Notations and Sign Conventions}
\label{subsec - convention_work}
\textbf{\textit{Conventions used for heat, work, and efficiency.}}~In this article, we adopt the following sign convention. The exchanged heats associated with the two isochoric strokes, stroke-I and stroke-III, are $q_c$ and $q_h$, given by Eq.~\eqref{qc_gen} and Eq.~\eqref{qh_gen}, respectively. The positivity of these quantities signifies heat flow towards the working substance. In contrast, the negativity implies heat flow from the working substance.

The works associated with the two unitary strokes, stroke-II and stroke-IV, are $w_1$ and $w_2$, given by Eq.~\eqref{w1_gen} and Eq.~\eqref{w2_gen}, respectively. Regarding these quantities, the positivity signifies work done on the system (working substance), whereas the negativity implies work done by the system. The sum of $w_1$ and $w_2$ defines the total work performed during the cycle. Under the standard thermodynamic convention, this quantity is negative when the engine delivers work. To make the engine operation explicit, we consider the total work as $w_\text{total} = -(w_1 + w_2)$, incorporating an extra negative sign, such that a positive value of $w_{\mathrm{total}}$ corresponds to net work performed by the Otto cycle. This convention allows us to directly focus on maximizing the work output. Accordingly, when the net work done is performed by the cycle $(w_{\text{total}}>0)$, the Otto cycle operates as an engine. Conversely, when $w_{\mathrm{total}} \leq 0$, the cycle ceases to operate as an engine.

The heat during stroke-III is injected to the working substance $(q_h>0)$, and consequently the efficiency of the engine is defined as $\eta = w_\text{total}/{q_h}$, which lies in the domain $\eta \in (0,1]$, where $\eta=1$ is the Carnot limit.

\textbf{\textit{Notations.}}~\AD{Here, we introduce the notations used in the 
remaining part of the paper. In the adiabatic Otto cycle, we denote the states of the working substance after each stroke by $\rho^{X}_{i}$ and their corresponding energies by $E^X_i$, with $i\in\{0,1,2,3\}$ (see Fig.~\ref{fig: schematic}), where $X \in \{\mathcal{C},\Pi,\mathcal{P}\}$ represents the type of the Otto cycle. Here, $\mathcal{C}$, $\Pi$, and $\mathcal{P}$ corresponds to the conventional (traditional) two-bath, PVM-based, and POVM-based Otto cycles, respectively. The work associated with strokes-II and IV are denoted by $W^X_1$ and $W^X_2$, respectively, while the heat exchanged during strokes-I and III are denoted by $Q^X_c$ and $Q^X_h$, respectively. $W^X$ and $\eta^X$ denote the total work and the efficiency of the engine, respectively. When we discuss about the non-adiabatic Otto cycle, $\rho^{X}_{i}$, $E^X_i$, $W^X_1$, $W^X_2$, $W^X$, and $\eta^X$ are replaced by $\tilde{\rho}^{X}_{i}$, $\mathcal{E}^X_i$, $\mathcal{W}^X_1$, $\mathcal{W}^X_2$, and $\boldsymbol\eta^X$, respectively.}

Maximization of any quantity $Z$ over any particular set $Y$ gives the maximum value denoted by $Z_{M}$. Similarly, the minimization of any quantity $\mathrm{Z}$ over any particular set $\mathrm{Y}$ gives the minimum value denoted by $\mathrm{Z}_{m}$.

\textbf{\textit{Units used in plots.}}~The frequencies considered in this paper are scaled by a characteristic frequency ${\Omega}_0$~(having dimensions of frequency). Consequently, all energies, work, and heat are expressed in the units of $\hbar{\Omega}_0$, the temperatures are expressed in the units of $\hbar \Omega_0/k_B$, and the times are expressed in the units of $\Omega_0^{-1}$, where $\hbar$ and $k_B$ denote the reduced Planck constant and the Boltzmann constant, respectively.

\subsection{Description of the working substance}
\label{Subsec - Hamiltonian description}
We consider a two-level quantum system as the working substance of the engine. The two unitary strokes, stroke-II and stroke-III, change the Hamiltonian of the system between $H^{(1)}$ and $H^{(2)}$.
Without any loss of generality, we consider that the $z$-direction is along $H^{(1)}$.
\begin{align}
    H^{(1)} &= \frac{\hbar}{2} \omega_z \sigma_z = \frac{\hbar}{2} \omega_z (|0\rangle \langle0| - |1\rangle \langle1|) \label{H1}
\end{align}
During this stroke, the spectral gap is changed to $\omega$ and the direction is also rotated to $\hat{n}$ direction.
\begin{align}
    H^{(2)} &= \frac{\hbar}{2} \omega (\hat{n} \cdot \vec{\sigma}) \label{H2}
\end{align}
Without any loss of generality, we can consider the plane of rotation as the $xz$-plane, therefore $\vec{n} \cdot \vec{\sigma} = n_z \sigma_z + n_x \sigma_x$, where $n_z^2 + n_x^2 = 1$.
It can be parameterized as $n_z = \cos{\lambda}$ and $n_x = \sin{\lambda}$, where $\lambda$ represents the angle between the $\hat{z}$ and $\hat{n}$ directions, with $\lambda \in [0,\pi/2]$. Note that, the angle of rotation $\lambda$ can also be manifested as a measure of the non-commutivity, because $[H^{(1)},H^{(2)}] = (i \hbar^2 \omega \omega_z \sin \lambda) (\sigma_y / 2)$. With no rotation of basis, i.e., $\lambda = 0$, the initial and final Hamiltonians commute $[H^{(1)},H^{(2)}]=0$.

With the aforesaid parameterization, we obtain the spectral structure of $H^{(2)}$, hence $H^{(2)}$ can be expressed as
\begin{align}
    H^{(2)} = \frac{\hbar}{2} \omega (|\chi^+\rangle \langle \chi^+ | - |\chi^-\rangle \langle \chi^- |) \label{H2 parameterized},
\end{align}
where $|\chi^+\rangle \coloneqq c_{\lambda} |0\rangle + s_{\lambda} |1\rangle$ and $|\chi^-\rangle \coloneqq -s_{\lambda} |0\rangle + c_{\lambda} |1\rangle$, with $c_{\lambda} \coloneqq \cos(\lambda/2)$ and $s_{\lambda} \coloneqq \sin(\lambda/2)$. 

We first analyze the optimal work extraction and efficiency of the PVM- and POVM-fueled engines in the adiabatic regime. The results are presented below.
\section{Performance Analysis in the Adiabatic Regime}
\label{sec - adia_PVM_POVM}
\subsection{Optimal work for PVM-fueled quantum Otto engine in adiabatic regime}
\label{Sec - Opt PV_based}
In this section, we determine the optimal work extractable from a quantum Otto engine when stroke-III is replaced by a projective measurement and identify the measurement basis that maximizes this extraction. We consider the two unitary strokes to be adiabatic.

\label{PV adia}
Here, stroke-I and stroke-II are identical to those in the conventional Otto engine operating between two thermal baths~(for complete analysis of the conventional two-bath engine, see Appendix.~\ref{App - 2b conventional Otto adia}). Consequently, the energies of the working substance after Strokes I and II, denoted by $E_0^{\Pi}$ and $E_1^{\Pi}$, are the same as in the conventional Otto cycle.
{The energies corresponding to the states after stroke-I and stroke-II are 
\begin{align}
    E_0^{\Pi} &= - \frac{\hbar}{2} \omega_z \tau_z\label{E0}~\text{and}\\
    E_1^{\Pi} & = -\frac{\hbar}{2} \omega \tau_z \label{E1},
\end{align}
respectively, where $\tau_z \coloneqq \tanh{v_z}$, with $v_z \coloneqq (\beta_c \omega_z)/2$. See Appendix~\ref{App - 2b conventional Otto adia} for details. Accordingly, the work associated with stroke-II, $W_1^{\Pi} = E_1^{\Pi} - E_0^{\Pi}$ is given by
\begin{align}
    W_1^{\Pi} = \frac{\hbar}{2} \tau_z (\omega_z - \omega) \label{W1}.
\end{align}}

{We now consider stroke-III, where a projective (PV) measurement is performed. Without loss of generality, the measurement basis can be taken as $\{ |\psi^+_{\theta\phi}\rangle, |\psi^{-}_{\theta\phi}\rangle \}$, where 
\begin{align}
    &|\psi^+_{\theta\phi}\rangle \coloneqq \cos{\frac{\theta}{2}} |\chi^+\rangle + e^{\iota \phi} \sin{\frac{\theta}{2}} |\chi^-\rangle~\text{and}\nonumber\\
    &|\psi^{-}_{\theta\phi}\rangle \coloneqq \sin{\frac{\theta}{2}} |\chi^+\rangle - e^{\iota \phi} \cos{\frac{\theta}{2}} |\chi^-\rangle.
    \label{psi perp}
\end{align}
These two states are orthogonal and each of them is normalized. Here, $\theta \in [0,\pi]$ and $\phi \in [0, 2\pi)$ denote the polar angle and the azimuthal angle, respectively, in the Bloch sphere representation of the qubit with two poles representing $|\chi^+\rangle$ and $|\chi^-\rangle$.
The post-measurement state without post selection is given as
\begin{align*}
    \rho_2^{\Pi} &= 
    p_{\psi^+_{\theta\phi}} |\psi^+_{\theta\phi}\rangle \langle \psi^+_{\theta\phi}| + p_{\psi^{-}_{\theta\phi}} |\psi^{-}_{\theta\phi}\rangle \langle \psi^{-}_{\theta\phi}|,
\end{align*}
where
\begin{align*}
    &p_{\psi^\pm_{\theta\phi}} = \langle \psi^\pm_{\theta\phi} | \rho_1^{\Pi} |\psi^\pm_{\theta\phi} \rangle = \frac{1}{2}(1 \mp \tau_z \cos{\theta}).
\end{align*}
Consequently, the energy of the qubit after stroke-III comes out to be
\begin{align}
E_2^{\Pi} &= \Tr(H^{(2)} \rho_2^{\Pi})=-\frac{\hbar}{2} \omega \tau_z \cos ^2{\theta} .
\label{E2_PV}
\end{align}
The change in energy of the working substance in the stroke-III is attributed as heat absorbed by the working qubit. This heat absorbed, $Q_h$, due to the measurement stroke is given by
\begin{align}
    Q_h^{\Pi} &= E_2^{\Pi} - E_1^{\Pi} = \frac{\hbar}{2} \omega \tau_z \sin^{2}{\theta} \label{Q_h PV adia}.
\end{align}
Finally, after the measurement stroke, we have the final work stroke, the stroke-IV, which is a unitary evolution induced by the adiabatic change of the qubit  Hamiltonian from $H^{(2)}$ to $H^{(1)}$. This evolution, being adiabatic, preserves the elements of the density matrix, but the basis changes from the eigenbasis of $H^{(2)}$ to that of $H^{(1)}$.
Hence, the state  $\rho_3^{\Pi}$ after stroke-IV  is given by
\begin{align*}
    \rho_3^{\Pi} =& p_{\chi^+ \chi^+ } |0\rangle \langle 0| +p_{\chi^+ \chi^- } |0\rangle \langle 1| \\
    &+p_{\chi^- \chi^+ } |1\rangle \langle 0| + p_{\chi^- \chi^- } |1\rangle \langle 1|,
\end{align*}
and the corresponding energy $E_3$ is given by 
\begin{align*}
    E_3^{\Pi} &= \Tr(H^{(1)}\rho_3^{\Pi}) = -\frac{\hbar}{2} \omega_z \tau_z \cos^2{\theta}.
\end{align*}
\AD{Here, $\{p_{ij}\}_{i,j = \chi^\pm}$, are the elements of $\rho_2^{\Pi}$ in $\{\ket{\chi^+},\ket{\chi^-}\}$ basis are given by $p_{\chi^+ \chi^+ } = (1-\tau_z \cos^2 \theta)/2$, $p_{\chi^+ \chi^- } = -e^{-\iota \phi} \tau_z \sin{(2 \theta)}/4$, $p_{\chi^- \chi^+ }=p_{+-}^*,$ and $p_{\chi^- \chi^- } = (1-p_{\chi^+ \chi^+ })$.
} The work associated with this stroke is evaluated to be
\begin{align}
W_2^{\Pi} = E_3^{\Pi} - E_2^{\Pi} = \frac{\hbar}{2} (\omega - \omega_z) \tau_z \cos^2{\theta}.
\end{align}
Finally, the qubit undergoes thermalization in stroke-I. Consequently, it gets reset to the thermal state, corresponding to the temperature $T_c$.  
The heat given out by the qubit in this stroke is given by $$Q_c^{\Pi} = E_0^{\Pi} - E_3^{\Pi} = - \frac{\hbar}{2} \omega_z \tau_z \sin^{2}{\theta}.$$ 

With the help of Eq.~\eqref{W1}, according to the convention discussed in Sec.~\ref{subsec - convention_work} the total work comes out to be
\begin{align}
    W^{\Pi} = -(W_1^{\Pi} + W_2^{\Pi})= \frac{\hbar}{2} \tau_z (\omega - \omega_z) \sin^{2}{\theta}\label{W_PV}.
\end{align}
For $\omega > \omega_z$, the work output by the cycle, $W^{\Pi}$ is positive, indicating that the cycle operates in the work-producing regime as an engine. 
Furthermore, we note that $Q_h^{\Pi} + Q_c^{\Pi} = W^{\Pi}$, signifying that the total energy in the cyclic process is conserved, satisfying the conservation of energy.

We find that the efficiency of the PVM-based engine is given by $\eta^{\Pi} \coloneqq W^{\Pi}/Q_h^{\Pi} = 1-({\omega_z}/{\omega})$, which further matches with the expression obtained in~Eq.~\eqref{effi-rob} for more general settings. 

From Eq.~\eqref{W_PV}, we find that the total work output is maximized for $\theta = {\pi}/{2}$ and it is independent of $\phi$, where $\theta$ and $\phi$ correspond to the measurement basis of $\{\ket{\xi_{\phi}^{+}}, \ket{\xi_{\phi}^{-}} \}$ basis, where, $|\xi_{\phi}^{\pm}\rangle \coloneqq [(c_{\lambda} \mp e^{\iota \phi} s_{\lambda}) \ket{0} + (s_{\lambda} \pm e^{\iota \phi} c_{\lambda}) \ket{1}]/\sqrt{2}$. The corresponding maximum value is given by
\begin{align}
    W_{M}^{\Pi} = \frac{\hbar}{2} \tau_z (\omega - \omega_z),
    \label{abs_W_max_PV}
\end{align}
{In Appendix~\ref{App - 2b conventional Otto adia}, we show that a conventional Otto cycle operating with a hot bath with infinite temperature ($\beta_h = 0$), produces the maximum work $W_{M}^{\mathcal{C}} = \hbar \tau_z(\omega-\omega_z)/2$. This is identical to the maximum work, $W^{\Pi}_{M}$ obtained by the PVM-based Otto engine, with same parameters. The underlying reason is that the post-measurement state $\rho_3^{\Pi}$ obtained after the optimal PVM, $\theta=\pi/2$, is the maximally mixed state $I/2$, which mimics the thermal state $e^{-\beta_h H^{(2)}}/\Tr\big(e^{-\beta_h H^{(2)}}\big)$ generated by an infinite-temperature hot bath, i.e., $\beta_h=0$. Thus, in the realistic scenario where the hot bath has a finite temperature, $\beta_h>0$, the PVM-based Otto engine produces more work than its conventional counterpart. This highlights the significance of PVM-based engines, as the enhanced work output does not rely on the idealized assumption of an infinitely hot bath, but persists under physically realistic finite-temperature conditions. Hence, PVM-based engines provide a viable route to improving the work extraction of quantum heat engines beyond that achievable with conventional thermalization protocols.


Next, we turn our attention to POVM-fueled engines and analyze their optimal work output. In particular, we investigate the conditions under which the optimal work output of a POVM-fueled engine can outperform that of a PVM-based engine.

\subsection{Optimal work for POVM-fueled quantum Otto engine in adiabatic region}
\label{POVM adia}
As the work strokes are considered adiabatic, the density matrix elements directly transfer from $\rho_0$ to $\rho_1$, but the representation of the basis changes from $\{|0\rangle, |1\rangle\}$ to $\{|\chi^+\rangle, |\chi^-\rangle\}$. So, the unitary operation $U$ associated with the work strokes can be expressed as $U = |\chi^+\rangle \langle 0| + e^{\iota \kappa} |\chi^-\rangle \langle 1|$, where $\kappa \in [0,2\pi)$.
We note that $U H^{(1)} U^{\dagger} = (\hbar \omega_z/2) (|\chi^+\rangle \langle \chi^+|-|\chi^-\rangle \langle \chi^-|) = (\omega_z/\omega) H^{(2)}$. Here, we have used the fact that under adiabatic unitary evolution in stroke-II, the populations of the two levels remain invariant but the two levels change from the eigenbasis of $H^{(1)}$ to that of $H^{(2)}$.

Now, let us calculate the work done in the first unitary stroke-II. The work done ${W}_1$ is given by the difference between the final and initial energy of the working medium, i.e.,
\begin{align}
\label{W1-adibatic}
{W}^{X}_{1} &= {E}^{X}_{1} - {E}^{X}_{0}\nonumber\\
&=\Tr(H^{(2)} \rho^{X}_{1})-\Tr(H^{(1)} \rho^{X}_{0})\nonumber\\
&=\Tr(H^{(2)} \rho^{X}_{1})-\Tr(H^{(1)} U \rho^{X}_{1} U^{\dagger})\nonumber\\
&=\left(1-\frac{\omega_{z}}{\omega} \right){E}^{X}_{1}.
\end{align}
Similarly, we have
\begin{align*}
{W}_{2}^X &= {E}_{3}^X - {E}_{2}^X=-\left(1-\frac{\omega_{z}}{\omega}\right){E}_{2}^X
\end{align*}

So, the total work ${W}$ comes out to be,
\begin{align}
    {W}^X = -({W}^X_1 + {W}^X_2) 
    = \left(1-\frac{\omega_{z}}{\omega}\right)({E}^X_{2}-{E}^X_{1}) \label{W_robust},
\end{align}
where $X \in \{\mathcal{C}, \Pi, \mathcal{P} \}$.
Note that, $W^X$, given by Eq.~\eqref{W_robust}, is valid for each of the conventional two-bath, PVM-based, and POVM-based Otto engines.

As discussed in the previous sections, in measurement-based Otto engines, strokes I and II are the same as in the conventional two-bath Otto engine. Thus, we have ${E}^{\mathcal{P}}_1 = (-\hbar/2)\omega\tau_z$. According to Eq.~\eqref{W_robust}, for the POVM-based engine, ${W}^{\mathcal{P}} = \left(1-
(\omega_{z}/\omega)\right)({E}^{\mathcal{P}}_{2}-{E}^{\mathcal{P}}_{1})$. Therefore, the total work done by the engine turns out to be
\begin{align}
    W^{\mathcal{P}} = \Big(1-\frac{\omega_{z}}{\omega}\Big)
    \left(E_2^{\mathcal{P}} + \frac{\hbar}{2}\omega\tau_z\right)
    \label{abs_W_POVM_adia}.
\end{align}
Eq.~\eqref{abs_W_POVM_adia} shows that, in order to find the maximum work extractable from the engine, we require to maximize $E_2^{\mathcal{P}}$.

For stroke III, the measurement stroke, we consider a two-outcome POVM. To realize the two-outcome POVM, we introduce a qubit auxiliary system, initially in the state $\rho_a$. 
According to Naimark's dilation theorem, to implement a general two-outcome POVM, a global unitary operation is applied on the joint system-auxiliary state, and subsequently a PV measurement is applied on the auxiliary.
In accordance to this, we act a global unitary operation $V$ on the composite state $\rho_{sa} = \rho_{1}^{\mathcal{P}} \otimes \rho_{a}$, and then perform a PV measurement on the composite system, of the form $\{ I \otimes \Pi_i \}_i$, where $I$ denotes the identity operator on the system, and $\{\Pi_i\}_{i=1,2}$ denote rank-one projectors on the auxiliary system.
Following the completion of the measurement stroke, we decouple the auxiliary system.
The bipartite state after the completion of the measurement stroke-III, without post-selection, is given by
\begin{align}
    \rho'_{sa} &= \sum_{i=1}^{2}{(I \otimes \Pi_i) (V \rho_{sa} V^{\dagger}) ( I \otimes \Pi_i)}.
    \label{rhopsa}
\end{align}
Here, we can assume that the auxiliary has a trivial Hamiltonian such that all eigenvalues are zero. Thus, any change in its state does not have an energy cost. 
The energy of the system after the measurement stroke is 
\begin{align}
    {E}^{\mathcal{P}}_2 &= \Tr\left((H^{(2)} \otimes {I}) \rho'_{sa}\right)\\&= \Tr\left[(H^{(2)} \otimes {I}) (V \rho_{sa} V^{\dagger})\right]. \label{E2_POVM}
\end{align}

The maximization of $E_2^{\mathcal{P}}$ can be performed analytically by arranging the eigenvalues of $H^{(2)} \otimes {I}$ and $\rho_{sa}$ in the ascending order, denoted by $r_i$ and $s_i$, respectively and the maximum value comes out to be $\sum_{i}{r_i s_i}$. Here, the maximization is performed over all $V$, for any given auxiliary state. Thus, we maximize the energy $E_2^{\mathcal{P}}$ over all possible two-outcome POVM measurements.

For this, let us first consider the initial state of the auxiliary system as a pure state. The pure qubit auxiliary state has the eigenvalues of $1$ and $0$. Moreover, the two eigenvalues of $\rho_1^{\mathcal{P}}$ are $\exp(-v_z)/\mathcal{Z}$ and $\exp(v_z)/\mathcal{Z}$, where $\mathcal{Z} \coloneqq 2 \cosh{v_z}$. We obtain the eigenvalues of $\rho_{sa}$ in ascending order as $\{s_i\} = \{\exp(v_z)/\mathcal{Z}, \exp(-v_z)/\mathcal{Z},0,0\}$. Similarly, the eigenvalues of $H^{(2)} \otimes {I}$ in ascending order turns out to be $\{r_i\} = \{({\hbar}/{2}) \omega, ({\hbar}/{2}) \omega, -({\hbar}/{2}) \omega, -({\hbar}/{2}) \omega\}$.
Hence, we obtain the maximum value of $E_2^{\mathcal{P}}$ as
\begin{align}
    \Big[E_{2}^{\mathcal{P}}\Big]_{M} &= \sum_i r_i s_i 
    = \frac{\hbar}{2} \omega.
\end{align}
This leads to the maximum work extractable from the engine to be
\begin{align}
    W_{M}^{\mathcal{P}} = \frac{\hbar}{2} (\omega - \omega_z) (1 + \tau_z) \label{abs_W_max_POVM},
\end{align}
Note that, the optimal work can be achieved with any choice of the pure auxiliary state, owing to the fact that any pure qubit state possesses the eigenvalues $0$ and $1$. Moreover, in Appendix~\ref{App - Mixed state aux}, it is shown that any choice of mixed auxiliary state cannot achieve the same.

After the completion of the measurement stroke, the system undergoes a unitary stroke-IV and finally followed by a thermalizing stroke to reset the working qubit to start another cycle. To analyze further we need to find the post-measurement state $\rho_2^{\mathcal{P}}$ of the system.

{\textit{\textbf{Feasibility of the optimal work:}} 
Having derived the optimal work extractable from the POVM-based Otto engine, $W^{\mathcal{P}}_{M}$ (Eq.~\eqref{abs_W_max_POVM}), a natural question is whether this optimum can be achieved, and if so, which specific POVM realizes it. Here, we consider $\rho_a = \ket{\chi^+}\bra{\chi^+}$ as the initial state of the auxiliary system.
To find the POVM that optimizes work for this particular choice of $\rho_a$, we need to find a unitary operator, denoted by $V_0$, which maximizes the energy $E_2^{\mathcal{P}}$. We note that $\text{supp}(\rho_{sa}) = \mathbb{S}_{sa} = \{ |\chi^+ \chi^+ \rangle, |\chi^- \chi^+ \rangle \}$, but the energy operator $H^{(2)}\otimes I$, has $\mathbb{E}_{h^{(2)}} = \{\ket{\chi^+ \chi^+ },\ket{\chi^+ \chi^- }\}$ as a higher energy subspace. Thus a unitary that takes the state from $\mathbb{S}_{sa} \to \mathbb{E}_{h^{(2)}}$ leads to optimal work.}

One such unitary operator $V_0$, performing a swap operation, can be
expressed as a permutation matrix is given by
\begin{align}
V_0 = 
    \begin{pmatrix}
        1 & 0 & 0 & 0\\
        0 & 0 & 1 & 0\\
        0 & 1 & 0 & 0\\
        0 & 0 & 0 & 1\\
    \end{pmatrix},
    \label{our-V0}
\end{align}
in matrix representation on the basis of $\{ \ket{\chi^+ \chi^+}, \ket{\chi^+ \chi^-}, \ket{\chi^- \chi^+}, \ket{\chi^- \chi^-}\}$.
Now, let find the effect of the POVM on the system. 
Following the joint evolution with $V_0$, given by Eq.~\eqref{our-V0}, the local PV measurement is performed on the auxiliary in the basis  $\{ \Pi_i \} = \{ |\psi^+_{\theta\phi}\rangle \langle \psi^+_{\theta\phi}|, |\psi^-_{\theta\phi}\rangle \langle \psi^-_{\theta\phi}| \}$, where $ |\psi^+_{\theta\phi}\rangle$ and $ |\psi^-_{\theta\phi}\rangle$ are given by Eq.~\eqref{psi perp}.
The post-measurement state becomes $\rho_2^{\mathcal{P}} = \sum_{i} K_i \rho_1^{\mathcal{P}} K_i^\dagger$ with $i \in \{ \chi^+, \chi^- \}$, where
\begin{align*}
    K_{\chi^+} &\coloneqq {}_{a}\langle \psi^+_{\theta\phi} | V_0 | \chi^+ \rangle_{a}\\
    &= c_{\theta} \ket{\chi^+} \bra{\chi^+} + e^{-\iota \phi} s_{\theta} \ket{\chi^+} \bra{\chi^-}
    ~\text{and}\\
    K_{\chi^-} &\coloneqq {}_{a}\langle \psi^-_{\theta\phi} | V_0 | \chi^+ \rangle_{a}\\
    &= s_{\theta} \ket{\chi^+} \bra{\chi^+} - e^{-\iota \phi} c_{\theta} \ket{\chi^+} \bra{\chi^-},
\end{align*}
with $c_{\theta} \coloneqq \cos(\theta/2)$, $s_{\theta} \coloneqq \sin(\theta/2)$.
Thus, we obtain the post-measurement state of the system to be $\rho^{\mathcal{P}}_2 = \Tr_{a}(\rho'_{sa}) = \ket{\chi^+} \bra{\chi^+}$, which yields $E_2 = \Tr(\rho_2 H^{(2)}) = \hbar \omega /2$. 
Following the measurement in stroke-III, the system is adiabatically evolved by driving the Hamiltonian from $H^{(2)}$ to $H^{(1)}$. 
Hence, we have $\rho_3^{\mathcal{P}} = |0\rangle \langle 0 |$ and $E_3^{\mathcal{P}} = {(\hbar}/{2}) \omega_z$. 
 We evaluate the total work $W^{\mathcal{P}}_{M} =  ({\hbar}/{2}) (\omega - \omega_z) (1 + \tau_z)$,  which is identical to the optimal extracted work from a POVM-based Otto engine. 
These expressions match with the ones we derived at the beginning of the section without explicitly finding out the state of the  engine at intermediate points between each strokes.
}

{Moreover, we would like to reiterate that $V_0$ is not a unique unitary operation that optimizes the work. It can be said that any unitary operation which maps $\mathbb{S}_{sa}$, the support set of $\rho_{sa}$, to the highest energy subspace $\mathbb{E}_{h^{(2)}}$, can serve as the optimal unitary in this scenario. For a more detailed analysis on this, see Appendix~\ref{App - Swapping_stage}.}

Having derived the maximum work output per cycle for both the PVM- and POVM-fueled quantum Otto engines in the adiabatic limit, we now turn to the efficiency of these engines. In the following section, we calculate and compare their efficiencies in the adiabatic regime. In particular, we demonstrate that, although the work output may differ between the two engines, their efficiencies remain identical and are exactly equal to that of the conventional two-bath quantum Otto engine. The analysis is presented as follows.
\subsection{Efficiency of the conventional two-bath, PVM-fueled, and POVM-fueled Otto engines in the adiabatic limit remains same.}
\label{Subsec - adiabatic: efficiency can not be enhanced}
{In this section, we develop a general framework to analyze the efficiency of the quantum Otto cycle. We consider a quantum Otto engine in which stroke-III may correspond either to a thermalization (to a hot bath) process or to a measurement operation. We show that, irrespective of this distinction, the efficiency remains identical in both cases provided the work strokes are adiabatic. }


In order to calculate the efficiency, we need to first find the work output and heat absorbed by the engine. In the adiabatic limit, according to Eq.~\eqref{W_robust}, we find that the net output work, ${W}^{X} = (1- {\omega_z}/{\omega})({E}^{X}_2 - {E}^{X}_1)$ is independent of whether the engine operates via a conventional thermal stroke or a measurement-based process.

The heat absorbed ${Q}_h^X$ is given by ${Q}^X_h \coloneqq {E}^X_{2}-{E}^X_{1}$.}
Thus, the efficiency turns out to be
\begin{align}
    \eta^{X} \coloneqq \frac{W^X}{Q^{X}_{h}} = \left(1-\frac{\omega_{z}}{\omega}\right) \coloneqq \eta_0.
    \label{effi-rob}
\end{align}


We would like to reiterate that the efficiency of the quantum Otto engine in the adiabatic limit remains independent of the nature of the process involved in stroke-III of the cycle, as is evident from Eq.~\eqref{effi-rob}. Moreover, the efficiency at the adiabatic limit remains the same for all work output and hence also for the optimal work output.
Thus, it is important to investigate whether the work output of the engine differs for different processes in stroke-III. 
Furthermore, while considering POVM in stroke-III, we model this by introducing an additional auxiliary qubit, which along with the system-qubit, undergoes a unitary operation, followed by a PV measurement on the auxiliary qubit. In such a case, it is necessary to reset the auxiliary qubit after the current cycle, so that it can be reused in the next cycle. This qubit-reset may require additional work which can be provided from the net work ${W}$, in which case, the net work decreases, thereby reducing the efficiency. This can be prevented either by discarding the auxiliary qubit after every cycle and use a already prepared auxiliary qubit or by thermalizing the auxiliary qubit along with the work qubit during stroke-I.

Having established that the conventional two-bath, PVM-based, and POVM-based, quantum Otto engines exhibit identical efficiencies in the adiabatic limit, the optimal work output per cycle remains the relevant figure of merit for comparing their performance. The optimal work output for each of the three engine types has already been derived in the preceding sections. Therefore, in the following section, we focus on a systematic comparative analysis of these previously obtained results. We demonstrate that, although the efficiencies of the three engines are identical, their optimal work outputs differ, with the POVM-fueled engine achieving the highest maximum work output among the three.

\section{Optimal Work Extraction: Comparative Analysis in the Adiabatic Limit}
\label{sec - comparison_adia}
\subsection{Two-bath vs. PVM-based Otto engines}
\label{subsec - PV vs trad}
{First we compare the maximum work output from a PVM-based and two-bath conventional Otto engines.}

\label{comp: PV - trad, adia}
The work output of a conventional two-bath Otto engine (see Appendix~\ref{App - 2b conventional Otto adia}) is given by $W^{\mathcal{C}} = ({\hbar}/{2})(\tau_z- \tau) (\omega - \omega_z)$, according to Eq.~\eqref{W_2b} where $\tau = \tanh[(\hbar/2)\beta_h \omega]$. 

The work output is maximized in the limit $\beta_h \to 0$, corresponding to an infinitely hot bath, yielding
$W_{M}^{\mathcal{C}} = \left[W^{\mathcal{C}}\right]_{\beta_h = 0}= \frac{\hbar}{2}\tau_z(\omega - \omega_z)$.
This coincides exactly with the maximal work output of the PVM-based Otto engine, i.e.,
\begin{align}
W_{M}^{\mathcal{C}} = W_{M}^{\Pi}.
\end{align}
However, it is important to note that achieving this maximal work output in the conventional two-bath Otto engine requires the hot bath to be at infinite temperature. Since such a limit is physically realizable, the PVM-based Otto engine offers an operational advantage by attaining the same maximal work output without requiring an infinitely hot reservoir.

\subsection{PVM-based vs. POVM-based Otto engines}
\label{subsec - POVM vs PV, adia}
{In Sec.~\ref{POVM adia}, we have presented the optimal work for the POVM-based Otto engine $W^{\mathcal{P}}_{M}$ in Eq.~\eqref{abs_W_max_POVM}. On the other hand, the optimal work for the PVM-based Otto engine $W^{\Pi}_{M}$ is derived in Sec.~\ref{PV adia} as Eq.~\eqref{abs_W_max_PV}.
The difference between the two is given by
\begin{align}
    \Delta W^{\mathcal{P}\Pi} \coloneqq & W^{\mathcal{P}}_{M} - W^{\Pi}_{M} \nonumber\\
    =& \frac{\hbar}{2} (\omega - \omega_z). \label{Delta_W}
\end{align}
Since $\omega > \omega_z$, we have $W_{M}^{\mathcal{P}} > W_{M}^{\Pi}$. Interestingly, $\Delta W^{\mathcal{P}\Pi}$ is  only dependent on $\omega$ and $\omega_z$, and independent of $\beta_c$. Thus, provided that a sufficiently large supply of initially pure auxiliary states is available for implementing the POVM, such that the auxiliary states can be used without the need for resetting, the work output of the POVM-based engine is always greater than that of the PVM-based engine.

However, we also consider the scenario in which the availability of auxiliary states is limited, necessitating the inclusion of the cost associated with resetting the auxiliary states. The analysis is presented as follows.

{\textit{\textbf{Considering the auxiliary reset cost.}}
In our analysis of POVM-based Otto engine, we have considered an auxiliary system that is introduced in a pure state of $|\chi^+\rangle$ in every cycle. The analysis up to now has considered a supply of the initially pure auxiliary enough for the number cycles for which the engine will be run. This could be a difficult condition to meet. Hence, here we consider the cost of recycling the auxiliary system after each cycle. In other words, the auxiliary is reset to a specific pure state after each cycle. 
This is similar to the case of Landauer's principle~\cite{Maruyama2009,  
Reeb2014, Chattopadhyay2025}, where a system in any state is erased to a specific pure state. 
According to the Landauer's principle, any process that erases a state $\sigma_\text{er}$ to a specific pure state $|\psi_\text{erased}\rangle$, requires at least $k_B T S(\sigma_\text{er}) \ln 2$, and this energy is released to the environment as heat. Here, $k_B$ is the Boltzmann constant, $T$ is temperature of the surrounding, $S(\sigma_\text{er})$ is the von-Neumann entropy of the state $\sigma_\text{er}$ to be erased.
Thus, the minimum cost of erasing the auxiliary qubit, which is in a state $\rho_a'$ after the POVM stroke to a pure state, is given by $ W_{a} = k_B T_c S(\rho_a')\ln2$, where $S(\sigma) = -\Tr(\sigma\log_2\sigma)$ is the von-Neumann entropy~\cite{von_Neumann_entropy_English_translated_original_paper}.

Note that Eq.~\eqref{abs_W_max_POVM} remains valid even when the auxiliary system is initialized in a pure state of a $d$-dimensional (with $d \geq 2$) Hilbert space, characterized by a single nonzero eigenvalue equal to $1$ and a $(d-1)$-fold degenerate zero eigenvalue, provided the optimization is performed over all $d$-outcome POVMs. In this more general setting, the minimum erasure cost associated with recycling the auxiliary system is given by $k_B T_c S(\rho'_a) \ln d$, which is not a smaller than the qubit case. Nevertheless, throughout the present work, we restrict our analysis to the case of a qubit auxiliary system.

{\textit{\textbf{Temperature regime where the auxiliary-reset POVM Otto engine outperforms PVM-based engines.}} 
We now determine the regime of the cold-bath temperature $T_c$ for which the POVM-based Otto engine, including the auxiliary-reset cost, yields a larger net work output than the PVM-based Otto engine. The required condition is
$\Delta W^{\mathcal{P}\Pi} > W_{a}$, which leads to the bound
\begin{align*}
T_c < \frac{\hbar(\omega-\omega_z)}{2k_B \ln 2}.
\end{align*}
In deriving this expression, we have used the inequality $0 \leq S(\sigma)\leq 1$. We note that the entropy $S(\rho_a')$ can, in general, depend on $T_c$ through the joint state $\rho_{sa}'$. Incorporating this dependence results in a transcendental equation, yielding a tighter but analytically intractable bound. Therefore, we restrict ourselves to the simpler, weaker bound presented above.
}
{As mentioned above, the preceding bound on $T_c$ was derived by considering the maximum possible cost of resetting the auxiliary (corresponding to Landauer's principle), which corresponds to $S_{\rho_a'}=1$. However, in the optimal work extraction analysis presented in Sec.~\ref{POVM adia}, the post-measurement auxiliary state $\rho_a'$ is not necessarily maximally mixed, and consequently, $S_{\rho_a'}$ need not be equal to 1. Therefore, in the following section, we explicitly calculate the resetting cost corresponding to the optimal work extraction scenario of a POVM-fueled quantum Otto engine in the adiabatic limit.}

{\textbf{\textit{Auxiliary-reset cost for POVM realized with joint unitary $V_0$ given in Eq.~\eqref{our-V0}.}} The post-measurement {state} without selection is {obtained as the following:} $\rho'_{sa} = \rho_2\otimes\rho_a'$, with $\rho_2 = |\chi^+\rangle \langle \chi^+|$ and 
\begin{align}
\rho'_a =
\begin{pmatrix}
\nu^- & \lambda^- \\
\lambda^+ & \nu^+
\end{pmatrix},\label{rhosa}
\end{align}
where $\nu^{\pm} \coloneqq (1 \pm \tau_z \cos^{2}{\theta})/2$ and $\lambda^{\pm} \coloneqq -e^{\pm i \phi}(\tau_z/4)\sin{(2\theta)}$ and $\rho_a'$ is represented in the basis ${\{|\chi^+\rangle,|\chi^-\rangle\}}$ in Eq.~\eqref{rhosa}. 
The von Neumann entropy of the post-measurement state of the auxiliary system $\rho'_a$ is evaluated as
\begin{align}
    S(\rho'_a) = - \big( p_a^+ \log_2{p_a^+} + p_a^- \log_2{p_a^-} \big)\label{VN_entropy_post_measurement_aux},
\end{align}
where $p_a^{\pm} = (1 \pm \tau_z \cos{\theta})/2$, where $p_a^{\pm}$ are the eigenvalues of $\rho'_a$. 
}

{\textit{\textbf{Minimum auxiliary-reset cost for POVM realized with joint unitary $V_0$.}}
We note from Eq.~\eqref{VN_entropy_post_measurement_aux} that the auxiliary-reset cost depends on the measurement that is applied on the auxiliary qubit following the $V_0$ unitary. But the work generated from the POVM-based Otto engine is independent of this measurement, see Eq.~\eqref{abs_W_max_POVM}. Thus, we need to choose an appropriate PV measurement that minimizes the auxiliary-reset cost. We see that the choice of the PV measurement imprints on the final state $\rho_a'$ of the auxiliary and thereby onto its entropy.   
The entropy is minimum when the value of $p_a^{\pm}$ is far from $1/2$, which can be achieved by $\theta = 0$ or $\pi$. Hence, the measurement basis is to be chosen as $\{ \Pi_i\}=\{ |\chi^\pm\rangle \langle \chi^\pm| \}$ to minimize the auxiliary cost. 
The minimum auxiliary cost for POVM realized using the joint unitary $V_0$ of Eq.~\eqref{our-V0}
is 
\begin{align}
    \left[ W_a^{V_0}\right]_{m} = - \frac{1}{2}k_B T_c (\ln{2}) &\left[ t_z^+\log_2(t_z^+/2) + t_z^-\log_2(t_z^-/2) \right], 
    \label{min aux cost}
\end{align}
where $t_z^\pm = 1 \pm \tau_z$. 
The net work output of the POVM-based Otto engine is obtained by subtracting the auxiliary cost from the optimized work given by Eq.~\eqref{abs_W_POVM_adia}. Thus, we obtain net work as
\begin{align}
    W_\text{net}^{V_0}= \frac{\hbar}{2}(\omega - \omega_z) (1 + \tau_z) - \left[W_a^{V_0}\right]_{m}.
    \label{net work suff}
\end{align}
}
{Let us reiterate here that the auxiliary-reset cost for the current case attains its maximum value, i.e., $[W_a]_{M} = k_B T_c \ln{2}$, obtained when $\theta = {\pi}/{2}$.}  
\begin{figure}[!t]
    \centering
    \includegraphics[
    ]{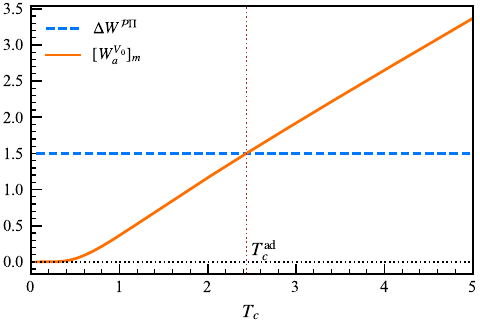}
    \caption{\textit{\textbf{Comparison between the advantage of the POVM-based engine over the PVM-based engine and the corresponding auxiliary reset cost.}} The enhancement in work extraction offered by the POVM-based engine over the PVM-based engine is quantified by the difference between their respective optimal work outputs, $\Delta W^{\mathcal{P}\Pi}$~[Eq.~\eqref{Delta_W}], shown by the dashed blue curve. The solid orange curve represents the auxiliary reset cost, $[W_{a}^{V_0}]_{m}$~[Eq.~\eqref{min aux cost}], associated with reinitializing the auxiliary system required for implementing the POVM. Both quantities are plotted as functions of the cold-bath temperature $T_c$. The results are shown for the compression ratio $\gamma=2.5$, corresponding to $(\omega,\omega_z)=(5,2)$. The vertical dashed brown line illustrates the critical temperature $T_c^{\text{ad}}$ beyond which $\Delta W^{\mathcal{P}\Pi}<[W_{a}^{V_0}]_{m}$ holds. Note that, $\Delta W^{\mathcal{P}\Pi}$ and $[W_{a}^{V_0}]_{m}$ are plotted along the vertical axis in the units of $\hbar{\Omega}_0$, and the temperature, expressed in the units of $\hbar \Omega_0/k_B$, is plotted along the horizontal axis.
    }
    \label{fig: surpass}
\end{figure}

{\textit{\textbf{Graphical visualization.}}
In Fig.~\ref{fig: surpass}, we plot the difference between the optimal work extracted from the POVM-based and PVM-based Otto engines, $\Delta W^{\mathcal{P}\Pi}$,
shown by the blue dashed curve and given by Eq.~\eqref{Delta_W}. Since $\Delta W^{\mathcal{P}\Pi}>0$ throughout, the POVM-based engine always possesses an advantage over the PVM-based engine in terms of optimal work extraction. We further note that this advantage is independent of the cold-bath temperature.
The figure also shows the auxiliary reset cost corresponding to the optimal POVM measurement,
$[W_{a}^{V_0}]_{m}$
represented by the orange solid curve and given by Eq.~\eqref{min aux cost}, as a function of the cold-bath temperature $T_c$. For low values of $T_c$, the reset cost remains smaller than the advantage $\Delta W^{\mathcal{P}\Pi}$ offered by the POVM-based engine. However, beyond a critical temperature $T^{\text{ad}}_c$, illustrated by the vertical dashed brown line, we obtain
$\Delta W^{\mathcal{P}\Pi}<[W_{a}^{V_0}]_{m}$.
Consequently, for operating temperatures above $T^{\text{ad}}_c$, the net work output of the POVM-based Otto engine becomes smaller than that of the PVM-based engine. In this regime, the PVM-based engine becomes more advantageous once the auxiliary reset cost is taken into account.}

In the following section, we analyze the performance of measurement-based quantum Otto engines and systematically compare them in the non-adiabatic limit.
\section{Analysis in the non-adiabatic region}
\label{sec - Non-adia_PVM_POVM}
\subsection{Non-adiabatic PVM-fueled quantum Otto engine}
\label{subsec - PV non-adia}
{In this subsection, we consider non-adiabatic work strokes in PVM-based Otto engine. The stroke-II and IV are unitary strokes, where the Hamiltonian of the qubit working substance changes with time.
Thus, the unitary operation $U$ acting on the state in such a stroke is given by
\begin{align}
    U = \mathcal{T} \left[\exp\left(-\frac{i}{\hbar} \int_{t_1}^{t_2}{H(t')dt'}\right)\right] \label{U time-ordered form},
\end{align}
where $H(t')$ is the time-dependent Hamiltonian, with $U$ being acted on the system from time $t_1$ up to time $t_2$, and  $\mathcal{T}$ denotes the time-ordering operator. For a generic time-dependent Hamiltonian $H(t')$ that interpolates between $H^{(1)}$ and $H^{(2)}$, i.e., $H(t_1)=H^{(2)}$ and $H(t_2)=H^{(1)}$, the dynamics can be characterized by the doubly-stochastic matrix $P_{ij} = |\bra{\psi^{(1)}_i} U \ket{\psi^{(2)}_j}|^2$, where $\{ \ket{\psi^{(1)}_{i}}\}$ and $\{ \ket{\psi^{(2)}_{j}}\}$ are the eigenbasis of $H^{(1)}$ and $H^{(2)}$, respectively. Let us denote $P \coloneqq |\bra{\chi^{+}} U \ket{0}|^2=|\bra{\chi^{-}} U \ket{1}|^2 = P$, and using the doubly-stochastic property the other three elements can be evaluated $|\bra{\chi^{+}} U \ket{1}|^2 = |\bra{\chi^{-}} U \ket{0}|^2 = 1-P$.

{The corresponding unitary evolution operator $U$ can be written as
\begin{align}
    U |0\rangle &= \sqrt{P} |\chi^+\rangle + e^{\iota \alpha} \sqrt{1-P} |\chi^-\rangle \coloneqq |M\rangle\\
    \text{and}~U |1\rangle &= \sqrt{1-P} |\chi^+\rangle - e^{\iota \alpha} \sqrt{P} |\chi^-\rangle \coloneqq |N\rangle,
    \label{U-def}
\end{align}
where the parameters $P$ and $\alpha$ depend on the driving protocol $H(t)$ and the total evolution time. Here, $P = |\bra{\chi^+} U \ket{0} |^2 \in [\bra{\chi^+} {0\rangle} |^2, \bra{\chi^+} {\chi^+ \rangle} |^2]$ which turns out to be $[\cos^2(\lambda/2),1]$, and $\alpha \in [0,2\pi)$. $P=1$ corresponds the adiabatic limit, whereas $P \in [\cos^2(\lambda/2),1)$ characterizes the non-adiabatic regime (see Appendix~\ref{App - Non-adiabaticity}). The parameter $P$ can be interpreted as the transition probability.

$U$ can be represented as $U = |M\rangle \langle 0| + e^{\iota \kappa} |N\rangle \langle 1|$, where $\kappa \in [0,2\pi)$.
The matrix representation of $U$ in the eigenbasis of $\sigma_z$ is given by
\begin{align*}
U = \frac{1}{\sqrt{2}}
    \begin{pmatrix}
        c_{\lambda} \sqrt{P} - e^{\iota \alpha} s_{\lambda} \sqrt{R} && e^{\iota \kappa} (c_{\lambda} \sqrt{R} + e^{\iota \alpha} s_{\lambda} \sqrt{P})\\
        s_{\lambda} \sqrt{P} + e^{\iota \alpha} c_{\lambda}\sqrt{R} && e^{\iota \kappa} (s_{\lambda} \sqrt{R} - e^{\iota \alpha} c_{\lambda} \sqrt{P})
    \end{pmatrix},
\end{align*}
where $R \coloneqq 1-P$.

Here, the additional phase $\kappa$ corresponds to the relative phase of the second column of the unitary matrix. Introducing the diagonal unitary $D = diag(1,e^{\iota \kappa})$, the above unitary can be written as $U = U_0 D$, where $U_0$ is obtained by setting $\kappa=0$. Since the initial Gibbs state $\rho_0$ is diagonal in the computational basis, it satisfies $[D,\rho_0]=0$.
Consequently, $U \rho_0 U^{\dagger} = U_0 \rho_0 U_0^{\dagger}$, implying that the state after the first unitary stroke is independent of $\kappa$. Although the state after the second unitary stroke $\rho_3 = U^{\dagger} \rho_2 U$ may acquire a $\kappa$-dependent phase in its off-diagonal elements, its diagonal elements remain unchanged. Since the Hamiltonian $H^{(1)}$ is diagonal in the computational basis, the corresponding energy $E_3 = \Tr(H^{(1)}\rho_3)$ is independent of $\kappa$. Therefore, all thermodynamic quantities derived in this work, including the work output and efficiency, are independent of $\kappa$, and we set $\kappa=0$ without loss of generality.

Let us now calculate the work output per cycle for such a non-adiabatic PVM-based Otto engine.
The thermalizing stroke-I is the same as in the conventional Otto engine. Therefore, the post thermalization-state $\tilde{\rho}_0^{\Pi} = \exp(-\beta_cH^{(1)})/\mathcal{Z}$ (see Eq.~\eqref{rho_0}) and its corresponding energy $\mathcal{E}_0^{\Pi} = -(\hbar/2)\omega_z \tau_z$ (see Eq.~\eqref{E0}). 

In the stroke-II, which is also the same as in the conventional Otto engine, the evolution of the system is given by the unitary $U$, such that $\tilde{\rho}_1^{\Pi} = U\tilde\rho_0^{\Pi}U^\dagger$, 
and its corresponding energy is given by
\begin{align}
    \mathcal{E}_1^{\Pi} = \frac{\hbar}{2} \omega \tau_z (1-2P) \label{E1_PV_nonadia},
\end{align}
{which is identical to the conventional Otto cycle one, see Eq.~\eqref{E1_2b_nonadia}}.
Hence, the work in this stroke turns out to be
\begin{align}
    \mathcal{W}_1^{\Pi} &= \mathcal{E}^{\Pi}_1 - \mathcal{E}^{\Pi}_0\nonumber\\
    &= \frac{\hbar}{2} \tau_z \left( \omega (1-2P) + \omega_z\right) \label{W1_PV_nonadia},
\end{align}

In the next stroke, the system undergoes a PV measurement. Consequently, we have, $\tilde \rho_2^{\Pi} = \bm{p}_{\psi_{\theta\phi}^+} |\psi^+_{\theta\phi}\rangle \langle \psi^+_{\theta\phi}| + \bm{p}_{\psi^-_{\theta\phi}} |\psi^-_{\theta\phi}\rangle \langle \psi^-_{\theta\phi}|$. Consequently, the energy of $\tilde\rho_2^{\Pi}$ is
\begin{align}
    \mathcal{E}^{\Pi}_2 = \Tr(H^{(2)} \tilde{\rho}_{2}^{\Pi})
    = -\frac{\hbar}{2} \omega \tau_z (2B-1)(2A-1), \label{E2_PV_nonadia}
\end{align}
where $B \coloneqq |\langle \psi^+_{\theta\phi} | \chi^+ \rangle |^{2}$ and $A \coloneqq | \langle \psi^+_{\theta\phi} | U | 0 \rangle |^{2}$ (see Appendix~\ref{App - E2_PV_nonadia} for detailed derivation).
Hence, the amount of injected heat in the system, by the measurement stroke is given by
\begin{align}
    \mathcal{Q}^{\Pi}_h &= \mathcal{E}^{\Pi}_2 - \mathcal{E}^{\Pi}_1\nonumber\\
    &= -\frac{\hbar }{2} \omega \tau_z\left[ (2B-1)(2A-1) + (1-2P) \right]
\end{align}}
{The measurement stroke is followed by another non-adiabatic unitary stroke-IV. The unitary $\tilde U$ governing this stroke, emerges from a driven Hamiltonian, that initially is $H^{(2)}$ and at the end becomes $H^{(1)}$. The unitary $\tilde U$ is of the form similar to that of Eq.~\eqref{U time-ordered form}.
The  unitary-evolved state is $\tilde{\rho}^{\Pi}_3 = \tilde{U} \rho_2 \tilde{U}^{\dagger}$ and the corresponding energy is 
\begin{align}
    \mathcal{E}^{\Pi}_3 = -\frac{\hbar }{2} \omega_z\tau_z (2Q-1) (2A-1), \label{E3_PV_nonadia}
\end{align}
where $Q \coloneqq |\langle 0 | \tilde{U} | \psi^{+}_{\theta\phi} \rangle |^{2}$. See Appendix~\ref{App - E3_PV_nonadia} for detailed derivation.

From Eq.~\eqref{E3_PV_nonadia} and Eq.~\eqref{E2_PV_nonadia}, we obtain the work associated with stroke-IV as
\begin{align}
    \mathcal{W}_2^{\Pi} &= \mathcal{E}^{\Pi}_3 - \mathcal{E}^{\Pi}_2 \nonumber\\
    &= \frac{\hbar}{2} \tau_z(2A-1) \left[ \omega (2B-1) - \omega_z (2Q-1) \right] \label{W2_PV_nonadia}.
\end{align}
From Eq.~\eqref{W1_PV_nonadia} and Eq.~\eqref{W2_PV_nonadia},  we obtain the total work as
\begin{align}
    \mathcal{W}^{\Pi} =& -(\mathcal{W}^{\Pi}_1 + \mathcal{W}^{\Pi}_2) \nonumber\\
    =& -\frac{\hbar }{2} \tau_z[ (\omega (1-2P) + \omega_z)\nonumber\\
    & \ \ + (2A-1) (\omega (2B-1) - \omega_z (2Q-1)) ] \label{W PV nonadia}.
\end{align}
The qubit is reset to the Gibbs state of $\tilde\rho_0^{\Pi}$ in the thermalizing (to the cold-bath) stroke-I, thereby starting the cycle again.
The amount of exchanged heat with the cold-bath is given by
\begin{align}
    \mathcal{Q}^{\Pi}_c &= \mathcal{E}^{\Pi}_0 - \mathcal{E}^{\Pi}_3 \nonumber\\
    &= -\frac{\hbar \tau_z}{2} \omega_z (1 + (2Q-1)(2A-1)) \label{Q_c PV nonadia}
\end{align}
Again, we find that $\mathcal{Q}^{\Pi}_h + \mathcal{Q}^{\Pi}_c = \mathcal{W}^{\Pi}$, thus satisfying the conservation of energy.

We consider the two unitary strokes to be symmetric. Thus, for the Otto cycle, the unitary $\tilde{U}$ of the stroke-IV retraces the exact path of $U$ of stroke-II reversibly. 
Consequently, we obtain $\tilde{U} = U^{\dagger}$, where $\tilde{U} = \mathcal{T} \left[\exp\left({-\frac{i}{\hbar} \int_{t_1}^{t_2}{\tilde{H}(t')dt'}}\right)\right]$ with $\tilde{H}(t') = H(t_2 + t_1 - t')$~\cite{Otto_finite_time_12}. 
With this consideration, we have $Q = |\langle 0 | U^{\dagger} | \psi^+_{\theta\phi} \rangle |^{2} = |\langle \psi^+_{\theta\phi} | U | 0 \rangle |^{2} = A$. }

{}Now, let us optimize $W$ for a given set of $P$ and $\alpha$. Expressing $|\psi \rangle$ in the Bloch sphere representation as $|\psi \rangle = \cos{({\theta}/{2})} |\chi^+\rangle + e^{\iota \phi} \sin{({\theta}/{2})} |\chi^-\rangle$, we obtain $2A-1 = (2P-1) \cos{\theta} + 2 \sqrt{P(1-P)} \sin{\theta} \cos{(\alpha - \phi)}$ and $2B-1 = \cos{\theta}$.

Now, from Eq.~\eqref{W PV nonadia}, the total work can be rewritten as
\begin{align}
    \mathcal{W}^{\Pi} = \frac{\hbar \tau_z}{2} [a \omega - \omega_z + \omega_z \mu^2 - \omega \mu \cos{\theta}]\label{W_opt PV nonadia},
\end{align}
where $a \coloneqq 2P-1$, $b \coloneqq 2 \sqrt{P(1-P)}$, and $\mu \coloneqq a \cos{\theta} + b \sin{\theta} \cos{(\alpha - \phi)}$.

Here, we analytically optimize over $\theta$ and $\phi$ for a given set of $\alpha$ and $P$. The explicit optimization and analysis are shown in Appendix~\ref{App - PV nonadia optimization}. The optimal work for a given set of $\alpha$ and $P$ is given by
\begin{align}
    \mathcal{W}^{\Pi}_{M} =& \frac{\hbar \tau_z}{4} \big(\sqrt{(\omega - \omega_z)^2 + 4 \omega \omega_z (1-P)}
    - \omega_z \nonumber\\
    &+ \omega (2P-1)\big), \label{absW_opt PV nonadia}
\end{align}
where the maximization is performed over the set of all PV measurements. With $P=1$, Eq.~\eqref{absW_opt PV nonadia} reduces to Eq.~\eqref{abs_W_max_PV}, which corresponds to the adiabatic limit.
\begin{figure*}[t]
    \centering
    \includegraphics[width=\textwidth]{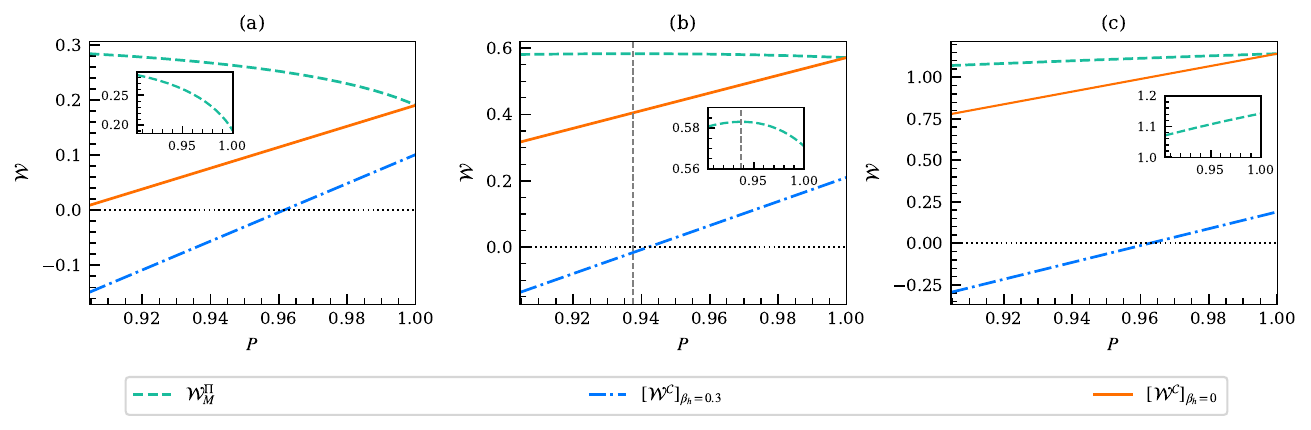}
    \caption{\textit{\textbf{Extractable work: comparison between the PVM-based and conventional two-bath Otto engines.}} 
The non-adiabatic work output, $\mathcal{W}$, is plotted as a function of the non-adiabaticity parameter $P$, where $P=1$ corresponds to the adiabatic limit. The dashed turquoise curve denotes the maximum extractable work from the PVM-based Otto engine, $\mathcal{W}^{\Pi}_{M}$, obtained after optimization over all measurement settings. For comparison, the work extracted from a conventional two-bath Otto engine is also shown for two hot-bath inverse temperatures: $\beta_h=0.3$ (dot-dashed blue) and $\beta_h=0$ (solid orange), represented by $[\mathcal{W}^{\mathcal{C}}]_{\beta_h=0.3}$ and $[\mathcal{W}^{\mathcal{C}}]_{\beta_h=0}$, respectively. 
For all the panels $\lambda = \pi/5$ is considered.
Panels (a), (b), and (c) correspond to two different compression ratios, $\gamma=\omega/\omega_z$: (a) $\gamma=1.25$ with $(\omega,\omega_z)=(2.5,2)$, (b) $\gamma=1.75$ with $(\omega,\omega_z)=(3.5,2)$, and (c) $\gamma=2.5$ with $(\omega,\omega_z)=(5,2)$.
The dotted black horizontal line corresponds to $\mathcal{W}=0$. Above this line, the cycle operates as an engine and delivers positive work output, whereas below it, the cycle absorbs work and therefore ceases to function as an engine.
The insets show $\mathcal{W}^{\Pi}_{M}$ separately in order to clearly illustrate its monotonic dependence in panels~(a) and (c), and non-monotonic dependence on $P$ in panel~(b). In panel~(b) the black dashed vertical line illustrates the value of $P$ corresponding to the maximum value of $\mathcal{W}^{\Pi}_{M}$. Note that, the work outputs are plotted along the vertical axis in the units of $\hbar{\Omega}_0$, and the non-adiabaticity paramter $P$, which is a dimensionless quantity, is plotted along the horizontal axis.
    }
    \label{fig: 2b vs PV nonadia}
\end{figure*}

\textbf{\textit{Behavior of the Otto cycle based on the compression ratio.}} Here, we focus on the maximization of $\mathcal{W}^{\Pi}_{M}$ given by Eq.~\eqref{absW_opt PV nonadia} over the set of the non-adiabaticity parameter, i.e., $P \in [\cos^{2}(\lambda/2),1]$. We denote the left boundary point $P=\cos^{2}(\lambda/2)$ and the right boundary point $P=1$ with $L$ and $R$, respectively. Here, $P \in [\cos^{2}(\lambda/2),1)$ corresponds to the non-adiabatic regime and $P=1$ implies the adiabatic limit. We observe that the maximization behaves differently in the three distinct regimes of the compression ratio $\gamma \coloneqq {\omega}/{\omega_z}$. See Appendix~\ref{App - PV nonadia optimization} for details.

For $\gamma \leq 2 \cos \lambda$, the optimal work $\mathcal{W}^{\Pi}_M$ becomes a monotonically decreasing function of the non-adiabaticity parameter $P$ and the maximum value is attained at the left boundary point $L$, i.e., $P = \cos^2(\lambda/2)$. The maximum value is given by Eq.~\eqref{PV nonadia absW P=L}, which we rewrite as
\begin{align*}
        [\mathcal{W}^{\Pi}_{M}]_{L} & = \frac{\hbar \tau_z}{4} \big( \Lambda - \xi \big),
    \end{align*}
where $\xi \coloneqq \omega_z - \omega \cos \lambda$, $\epsilon \coloneqq \omega \sin \lambda$, and $\Lambda \coloneqq \sqrt{\xi^2 + \epsilon^2}$.
For a setup with $\lambda \geq \pi/3$, this regime does not exist at all.

For $2 \cos \lambda < \gamma < 2$, the optimally extracted work $\mathcal{W}^{\Pi}_M$ shows non-monotonic dependence over $P$ and the maximum value is obtained at a particular value of $P=P_0$, where,
\begin{align*}
P_0 &\coloneqq \frac{1}{2} + \frac{\gamma}{4},
\end{align*}
with $\gamma \coloneqq \omega/\omega_z$. Interestingly, this particular value of $P$ lies in the strictly non-adiabatic region, since $P_0 < 1$ for $\gamma<2$. The maximum value of the optimal work in this regime is
\begin{align}
    [\mathcal{W}^{\Pi}_{M}]_{P=P_0} = \frac{\hbar \omega^2 \tau_z}{8 \omega_z}\label{W PV nonadia P=1/2}.
\end{align}
Note that for a setup with $\lambda = 0$, this regime does not exist.

For the two regimes discussed above, the non-adiabatic implementation is found to be beneficial because the maximum work extraction occurs for the $P$ values, which lie in the $P \in [\cos^2 (\lambda/2),1)$ regime. On the other hand, when $\gamma \geq 2$, the optimal work $\mathcal{W}^{\Pi}_M$ exhibits monotonically increasing behavior w.r.t. $P$ and becomes maximum at the right boundary point $R$ i.e., $P = 1$, which strictly represents the adiabatic limit.
The maximum value of the optimal work in this regime is given by Eq.~\eqref{PV nonadia absW P=R}, i.e.,
\begin{align}
    [\mathcal{W}^{\Pi}_{M}]_{P=1} = \frac{\hbar \tau_z}{2} (\omega - \omega_z) \label{W PV nonadia P=1},
\end{align}
which coincides with Eq.~\eqref{abs_W_max_PV}.

\textbf{\textit{PVM-based Otto cycle operates as an engine for the entire range of non-adiabaticity parameter $P$.}}
In contrast to the conventional Otto cycle, which fails to operate as an engine for certain values of the non-adiabaticity parameter $P \in [\cos^2(\lambda/2),1]$, we find from Eq.~\eqref{absW_opt PV nonadia} that
\begin{align}
    \mathcal{W}^{\Pi}_{M}>0 \label{W PV nonadia positivity}
\end{align}
throughout this entire range. Hence, the PVM-based Otto cycle functions as an engine for all values of $P$. See Appendix~\ref{App - PV nonadia optimization} for detailed derivation.

This behavior is illustrated in Fig.~\ref{fig: 2b vs PV nonadia}, where we show the variation of the optimal extracted work with the transition probability $P$. The extracted work from the conventional Otto engine with $\beta_h = 0.3$, denoted by $[\mathcal{W}^{\mathcal{C}}]{\beta_h = 0.3}$ (obtained from Eq.~\eqref{absW nonadia 2b}) is depicted by the blue dash-dotted line. Similarly, $[\mathcal{W}^{\mathcal{C}}]{\beta_h = 0}$ denotes extracted work from the conventional Otto engine with infinite temperature ($\beta_h = 0$) hot bath, illustrated by the orange solid line, given Eq.~\eqref{absW nonadia 2b, when beta_h = 0}. The turquoise dashed line represents the maximum extractable work $\mathcal{W}^{\Pi}_{M}$ from the PVM-based engine, following from Eq.~\eqref{absW_opt PV nonadia}. The blue dash-dotted and orange solid curves denote the extractable work of the conventional Otto cycle for $\beta_h = 0.3$ and $\beta_h = 0$, respectively, whereas the turquoise dashed curve corresponds to the PVM-based Otto cycle. 
Panel~(a) corresponds to the regime $\gamma \leq 2 \cos \lambda$, obtained for $\omega = 2.5$ and $\omega_z = 2$, giving $\gamma = 1.75$. Panel~(b) corresponds to $2 \cos \lambda < \gamma < 2$, where $\omega = 3.5$ and $\omega_z = 2$, yielding $\gamma = 1.75$. Panel~(c) corresponds to $\gamma \geq 2$, where $\omega = 5$ and $\omega_z = 2$, yielding $\gamma = 2.5$. For all three panels, $\lambda = \pi/5$ is considered. Furthermore, the inset provides a magnified view of $W^{\Pi}$ within the relevant parameter regime in all three panels.
Fig.\ref{fig: 2b vs PV nonadia} clearly demonstrates that, for the PVM-based Otto cycle, the extracted work remains positive over the entire range of $P$, as shown by the turquoise dashed curves. In contrast, the conventional Otto cycle with two thermal baths fails to maintain positive work extraction throughout the whole domain of $P$.

\begin{figure*}[t]
    \centering
    \includegraphics[
        width=18cm,
    ]{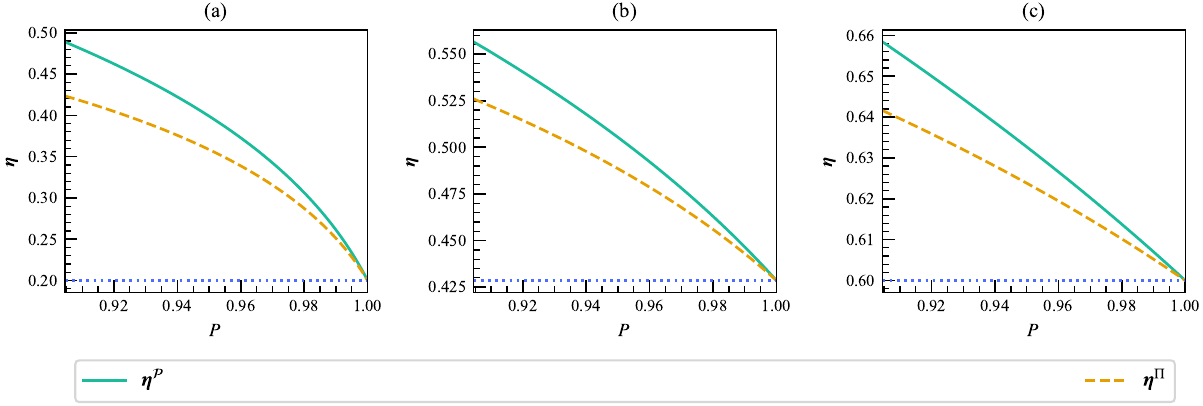}
    \caption{\textit{\textbf{Efficiency at optimal work extraction: comparison between the PVM- and POVM-based Otto engines.}} 
    The non-adiabatic efficiency, $\boldsymbol{\eta}$, is plotted as a function of the non-adiabaticity parameter $P$, where $P=1$ corresponds to the adiabatic limit. The solid turquoise curve depicts the efficiency of the POVM-based Otto engine at optimal work extraction, $\boldsymbol\eta^{\mathcal{P}}$, at the optimal work extraction. The dashed yellow curve illustrates the efficiency of the PVM-based Otto engine at its optimal work extraction, $\boldsymbol{\eta}^{\Pi}$, given by Eq.~\eqref{relabel: eta pi na}. 
    Panels (a), (b), and (c) correspond to two different compression ratios, $\gamma=\omega/\omega_z$: (a) $\gamma=1.25$ with $(\omega,\omega_z)=(2.5,2)$, (b) $\gamma=1.75$ with $(\omega,\omega_z)=(3.5,2)$, and (c) $\gamma=2.5$ with $(\omega,\omega_z)=(5,2)$. The plots clearly illustrate the monotonic decrement of $\boldsymbol{\eta}$ on $P$ in all the panels. Furthermore, the plots showcase that the POVM-based engine outperforms the PVM-based engine, also in terms of efficiency when optimal work is extracted, particularly in the non-adiabatic regime. The dotted blue horizontal line depicts the efficiency given by Eq.~\eqref{effi-rob}, corresponding to the adiabatic limit, i.e., $P=1$. Note that, the quantities, the efficiency $\boldsymbol{\eta}$ and the non-adiabaticity paramter $P$, are plotted along the vertical and horizontal axes, respectively, where both are dimensionless quantities.}
    \label{fig: eta nonadia comparison}
\end{figure*}
\textbf{\textit{Efficiency at optimal work extraction.}} Despite the efficiency remains the same for all work output and hence also for the optimal work output at the adiabatic limit (shown in Sec.~\ref{Subsec - adiabatic: efficiency can not be enhanced}), here in the non-adiabatic regime we focus on the efficiency corresponding to the maximum work extraction in particular The efficiency
$\boldsymbol{\eta}^{\Pi} \coloneqq \mathcal{W}^{\Pi}_{M}/\mathcal{Q}^{\Pi}_{h}$ for any given value of $P$ is given by Eq.~\eqref{eta PV nonadia} (for details, see Appendix~\ref{App - Efficiency at optimal work extraction}), i.e.,
\begin{align}
    \boldsymbol{\eta}^{\Pi} = \frac{1}{\big (1+\frac{\omega_z}{D}\big )} \label{relabel: eta pi na},
\end{align}
where $D \coloneqq \sqrt{(\omega - \omega_z)^2 + 4 \omega \omega_z (1-P)}$. Eq.~\eqref{relabel: eta pi na} implies that $\boldsymbol{\eta}^{\Pi}$ is a monotonically decreasing function of $P$. In Fig.~\ref{fig: eta nonadia comparison}, $\boldsymbol{\eta}^{\Pi}$ given by Eq.~\eqref{relabel: eta pi na} is depicted with the dashed yellow curve as a function of the non-adiabaticity parameter $P$, where $P=1$ corresponds to the adiabatic limit. The dashed yellow curve clearly exhibits the monotonic decrement of $\boldsymbol{\eta}^{\Pi}$ with $P$.
Panels~(a), (b), and (c) correspond to the regimes $\gamma \leq 2 \cos\lambda$, $2 \cos \lambda < \gamma < 2$, and $\gamma \geq 2$, respectively. At the adiabatic limit, i.e., $P=1$, this curve coincides with the value given by Eq.~\eqref{effi-rob}, shown by the blue dotted horizontal line, thereby confirming the consistency of the analysis.

Here, we analyze the efficiency of the PVM-based Otto cycle when operated at optimal work extraction. For $\gamma \geq 2$, the optimal work extraction occurs in the adiabatic limit, corresponding to $P=1$. In this regime, the efficiency is given by
\begin{align}
    [\boldsymbol{\eta}^{\Pi}]_{P=1} =\eta^{\Pi} = 1-\frac{\omega_z}{\omega}
    \label{eq:PV_effi_at_maxW_gamma_geq_2}.
\end{align}
For $2 \cos \lambda < \gamma < 2$, the optimal work is extracted at $P=P_0$, consequently the efficiency is given by
\begin{align}
    [\boldsymbol{\eta}^{\Pi}]_{P=P_0} = \frac{1}{2}\label{eq:PV_effi_at_maxW_gamma_less_than_2}.
\end{align}
For $\gamma \leq 2 \cos \lambda$, the optimal work is extracted at the left boundary point, i.e., $P=\cos^2(\lambda/2)$, and the corresponding efficiency is obtained
\begin{align}
    [\boldsymbol{\eta}^{\Pi}]_{P=\cos^2 (\frac{\lambda}{2})} = \frac{1}{\big (1+\frac{\omega_z}{\Lambda}\big )}\label{eq:PV_effi_at_maxW_gamma_less_than_2}.
\end{align}
Furthermore, note that the simultaneous positivity of $\boldsymbol{\eta}^{\Pi}$ (Fig.~\ref{fig: eta nonadia comparison}) and $\mathcal{W}^{\Pi}_M$ (Eq.~\eqref{W PV nonadia positivity} and Fig.~\ref{fig: 2b vs PV nonadia}) confirms the positivity of $\mathcal{Q}^{\Pi}$, implying that the PV measurement is injecting heat to the working substance.

Next we analyze the POVM-fueled quantum Otto engine in the non-adiabatic limit.
\begin{figure*}[t]
    \centering
    \includegraphics[
        width=18cm,
    ]{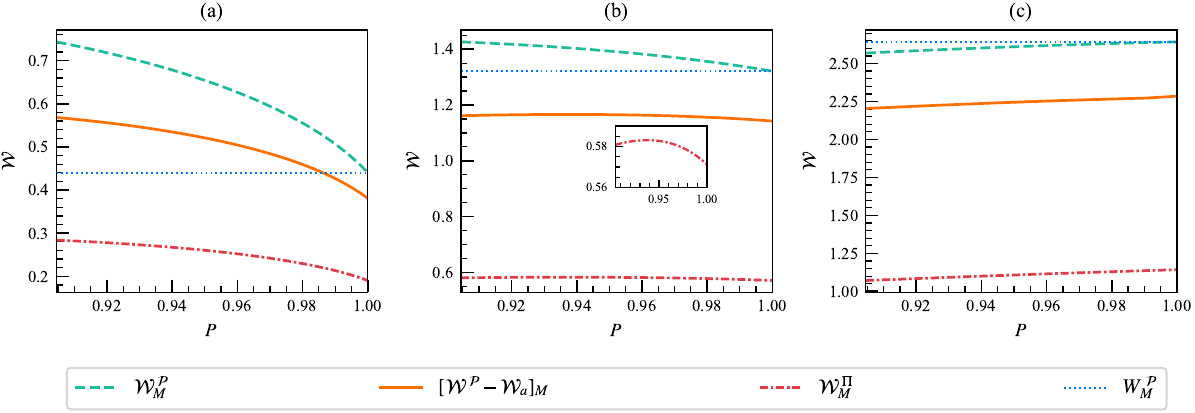}
\caption{
\textit{\textbf{Extractable work: comparison between the PVM-based and POVM-based Otto engines.}} 
The non-adiabatic work output, $\mathcal{W}$, is plotted as a function of the non-adiabaticity parameter $P$, where $P=1$ corresponds to the adiabatic limit. The turquoise dashed curve denotes the maximum extractable work from the POVM-based Otto engine, $\mathcal{W}^{\mathcal{P}}_{M}$, obtained after optimization over all measurement settings. At the adiabatic limit, i.e., $P=1$, this curve touches with the analytically derived value, i.e., $W^{\mathcal{P}}_{M}$, represented by the blue-colored dotted horizontal line.
The numerical optimization over the net work $[\mathcal{W}-\mathcal{W}_{a}]_{M}$ is illustrated by the orange solid curve. 
For comparison, the optimal work extracted from a PVM-based Otto engine is also shown (red tiny dot-dash), represented by $\mathcal{W}^{\Pi}_{M}$. For all the panels $\lambda = \pi/5$ is considered.
Panels (a), (b), and (c) correspond to two different compression ratios, $\gamma=\omega/\omega_z$: (a) $\gamma=1.25$ with $(\omega,\omega_z)=(2.5,2)$, (b) $\gamma=1.75$ with $(\omega,\omega_z)=(3.5,2)$, and (c) $\gamma=2.5$ with $(\omega,\omega_z)=(5,2)$. The inset corresponding to panel~(b) magnifies $\mathcal{W}^{\Pi}_{M}$ to exhibit its non-monotonic dependence over $P$. Note that, the work outputs are plotted along the vertical axis in the units of $\hbar{\Omega}_0$, and the non-adiabaticity paramter $P$, which is a dimensionless quantity, is plotted along the horizontal axis.}

    \label{fig: POVM nonadia}
\end{figure*}

\subsection{Non-adiabatic POVM-fueled quantum Otto engine}
\label{POVM nonadia}
{We now consider the POVM-based engine with non-adiabatic work strokes. 
Here, we numerically maximize the extractable work over all two-outcome POVM measurements to obtain $\mathcal{W}^{\mathcal{P}}_{M}$. Specifically, for a given value of the transition probability $P$, we optimize the extractable work with respect to the system--auxiliary joint unitary operator $V$ appearing in Eq.~\eqref{rhopsa}. 

{The global unitary operator $V$ acts on a four-dimensional Hilbert space and therefore possesses $4^2=16$ independent real parameters, since $V \in U(4)$. However, every unitary matrix connected to a special unitary matrix by a global factor, it is sufficient to optimize over the $4^2 - 1 = 15$ independent real parameters of the $SU(4)$ group. We parametrize it according to the Kraus-Cirac representation~\cite{Kraus-Cirac}, as it provides finite bounds on the parameters.
During the numerical optimization, we fix the measurement basis as$\{\Pi_i\}=\left\{|\chi^+\rangle\langle\chi^+|,|\chi^-\rangle\langle\chi^-|\right\}$,
since optimization over the $15$ parameters of the unitary operator $V$ alone is sufficient to span the entire set of two-outcome POVMs. For any given value of $P$ the numerical optimization is performed over a set of total $16$ parameters consisting of $\alpha$ and the $15$ parameters associated with the global unitary operator V, using the ``dual annealing'' global optimization algorithm from SciPy, followed by a local optimization routine.}

The numerically optimized maximum extractable work $\mathcal{W}_{M}^{\mathcal{P}}$ (turquoise dashed) as a function of the non-adiabaticity parameter $P$ is shown in Fig.~\ref{fig: POVM nonadia}, where $P=1$ corresponds to the adiabatic limit. 
Panels~(a), (b), and (c) correspond to the regimes $\gamma \leq 2\cos \lambda $, $2\cos \lambda < \gamma < 2$, and $\gamma \geq 2$, respectively.  At the adiabatic limit ($P=1$), this curve coincides with the analytical value given by Eq.~\eqref{abs_W_max_POVM}, shown by the blue dotted horizontal line, thereby confirming the consistency of the numerical optimization.

\textbf{\textit{Efficiency at optimal work extraction.}} Although the efficiency remains the same for all work output including the optimal work output at the adiabatic limit (shown in Sec.~\ref{Subsec - adiabatic: efficiency can not be enhanced}), here in the non-adiabatic regime we particularly analyze the efficiency for the maximum work extraction.
The numerically computed efficiency $\boldsymbol{\eta}^{\mathcal{P}} \coloneqq \mathcal{W}^{\mathcal{P}}_{M}/\mathcal{Q}^{\mathcal{P}}_{h}$ (turquoise solid) for the optimal maximum work extraction as a function of the non-adiabaticity parameter $P$ is shown in Fig.~\ref{fig: eta nonadia comparison}, where $P=1$ corresponds to the adiabatic limit. 
Panels~(a), (b), and (c) correspond to the regimes $\gamma \leq 2\cos\lambda$, $2\cos\lambda < \gamma<2$, and $\gamma \geq 2$, respectively. At the adiabatic limit ($P=1$), this curve coincides with the analytical value given by Eq.~\eqref{effi-rob}, shown by the blue dotted horizontal line, thereby confirming the consistency of the numerical optimization. Fig.~\ref{fig: eta nonadia comparison} clearly shows that the efficiency at optimal work extraction has a monotonically decreasing behavior with the non-adiabaticity parameter $P$, in both classes of the measurement-driven engines. Moreover, the POVM-based engine can outperform the PVM-based engine in terms of efficiency at optimal work extraction also, particularly in the non-adiabatic regime. Furthermore, owing to the simultaneous positivity of $\boldsymbol{\eta}^{\mathcal{P}}$ (Fig.~\ref{fig: eta nonadia comparison}) and $\mathcal{W}^{\mathcal{P}}_M$ (Fig.~\ref{fig: POVM nonadia}) we note the positivity of $\mathcal{Q}^{\mathcal{P}}$, which implies that the POVM is injecting heat to the working substance.

In the following section, we analyze how accounting for the auxiliary resetting cost modifies the work output per cycle of a POVM-based engine in the non-adiabatic limit.

\subsection{Incorporation of the auxiliary resetting cost}
\label{subsec - aux cost - POVM - non-adia}
The auxiliary after the measurement stroke is in some mixed state,
$\rho_a'$. The auxiliary is reset to the pure state of $\ket{\chi^+}$ while being in contact with the cold-bath with temperature $T_c$. So, the cost of recycling the auxiliary is $W_a = k_B T_c S(\rho_a')\ln 2$. 
It is essential to remove this amount of work from the total work output of the POVM-based Otto engine. In the next section, we discuss when this net work output from a POVM is larger than the corresponding PVM and two-bath cases.

While performing the numerical optimization, we adopted two approaches. One way is to perform the optimization on $V$ itself to maximize solely the extractable work $\mathcal{W}^{\mathcal{P}}$. From the maximum value $\mathcal{W}^{\mathcal{P}}_M$, obtained from the optimization, the corresponding resetting cost can be subtracted. The turquoise dashed line in Fig.~\ref{fig: POVM nonadia} illustrates $\mathcal{W}^{\mathcal{P}}_M$ before subtracting the resetting cost. The other way is to maximize over itself the net work $\mathcal{W}^{\mathcal{P}} - \mathcal{W}_a$, obtained after subtracting the resetting cost $\mathcal{W}_a$ from the extractable work $\mathcal{W}^{\mathcal{P}}$. The maximization in the latter approach yields the quantity $[\mathcal{W}^{\mathcal{P}} - \mathcal{W}_a]_M$, depicted by the orange solid line in Fig.~\ref{fig: POVM nonadia}. Note that, in the adiabatic limit, we adopt the former approach for the optimization, analytically performed in Sec.~\ref{POVM adia} followed by Sec.~\ref{subsec - POVM vs PV, adia}.

In the following section, we present a comparative analysis of conventional, PVM-based, and POVM-based quantum Otto engines in terms of their work output in the non-adiabatic limit. For the POVM-based engine, we consider two scenarios: one in which the auxiliary resetting cost is neglected and another in which it is explicitly accounted for.

\section{Optimal Work Extraction: Comparative Analysis in the non-Adiabatic Limit}
\label{sec - comparison_non-adia}
\subsection{Two-bath vs. PVM-based Otto engines}
\label{subsec - PV vs trad}
{
\label{comp: PV - trad, non-adia}
{ In the regime of non-adiabatic work strokes, the work output of the two-bath conventional Otto engine~\cite{Otto_finite_time_12} (see Appendix~\ref{App - nonadia 2b} for details), after maximizing over the temperature of the hot bath, is given by $$ \mathcal{W}^{\mathcal{C}}_{M} = \frac{\hbar}{2} \tau_z\left[(\omega (2P-1) - \omega_z)\right],$$ as shown in Eq.~\eqref{absW nonadia 2b, when beta_h = 0}.
From Eq.~\eqref{absW_opt PV nonadia}, we know that $$\mathcal{W}_{M}^{\Pi} = \frac{\hbar}{4}\tau_z \left[ \sqrt{\Delta \omega^2 + 4\omega \omega_z (1-P)} - \omega_z +\omega (2P-1)\right],$$
where $\Delta\omega \coloneqq \omega - \omega_z$. 
Subtracting $\mathcal{W}_{M}^{\mathcal{C}}$ from $\mathcal{W}_{M}^{\Pi}$, we obtain
\begin{align}
    \Delta\mathcal{W}^{\Pi\mathcal{C}} &\coloneqq \mathcal{W}^{\Pi}_{M} - \mathcal{W}^{\mathcal{C}}_{M} \nonumber \\
    &= \frac{\hbar }{4}\tau_z \left[D - \Delta\omega+2 \omega (1-P) \right] \label{diff: PV - trad},
\end{align}
where $D=\sqrt{\Delta\omega^2 + 4\omega \omega_z (1-P)}$.

Clearly as $D\geq\Delta\omega$ and $0\leq P \leq 1$, we have $\Delta\mathcal{W}^{\Pi\mathcal{C}} \geq  0$, with equality holding for $P = 1$.
Thus, in the non-adiabatic work stroke regime, the PVM-based engine provides more work output than the conventional two-bath Otto engine.


{
In Fig.~\ref{fig: 2b vs PV nonadia} we present the change in $\mathcal{W}^{\mathcal{C}}_{M}$ (turquoise-dashed line) and $\mathcal{W}^{\Pi}_{M}$ (orange-solid line) with change in non-adiabatic parameter $P$. We confirm that $\Delta \mathcal{W}^{\Pi\mathcal{C}} \geq 0$, in other words, $\mathcal{W}^{\Pi}_{M} \geq \mathcal{W}^{\mathcal{C}}_{M}$, with the equality attained at $P = 1$.  We also observe that for compression ration $\gamma = \omega/\omega_z \leq 2 \cos \lambda$, the value of $P$ corresponding to maximum work output is $\left[P^{\gamma}_{\mathcal{W}_{M}}\right]_{\gamma\leq 2 \cos \lambda} = \cos^2 (\lambda/2)$. For $2 \cos \lambda <\gamma<2$, we have $\left[P^{\gamma}_{\mathcal{W}_{M}}\right]_{2 \cos \lambda < \gamma< 2} = (2+\gamma)/{4}$. For compression ration $\gamma = \omega/\omega_z \geq 2$, the value of $P$ corresponding to maximum work output is $\left[P^{\gamma}_{\mathcal{W}_{M}}\right]_{\gamma\geq 2} = 1$. We present this in Fig.~\ref{fig: 2b vs PV nonadia}~(a), (b), and (c), where we have set $\gamma = 1.25$, $\gamma = 1.75$, and $\gamma = 2.5$, respectively. In Fig.~\ref{fig: 2b vs PV nonadia}~(b) and also in the corresponding inset, $\left[P^{\gamma}_{\mathcal{W}_{M}}\right]_{2 \cos \lambda < \gamma< 2} = (2+\gamma)/{4}$ is depicted by the black dashed vertical line.
}
\begin{table*}[t]
\centering
\setlength{\tabcolsep}{1pt}
\begin{tabular}{c c c c}
\hline \hline
\multirow{2}{*}{} 
& {{Conventional two-bath~[Ref.~\cite{Otto_finite_time_12}]}} 
& {{PVM-based }} 
& {POVM-based } \\

\hline
\makecell{Optimal work \\ (adiabatic)}& \makecell{$W_{M}^{\mathcal{C}} = \frac{\hbar}{2} \Delta\omega\Delta\tau$~[Eq.\eqref{W_2b}]} &{\Large$\leq$} \makecell{  $W_{M}^{\Pi} = \frac{\hbar}{2}\Delta\omega\tau_z$~[Eq.~\eqref{abs_W_max_PV}]}  &{\Large$<$} \makecell{    \ \ $W_{M}^{\mathcal{P}}=\frac{\hbar}{2}\Delta\omega(1 + \tau_z)$~[Eq.~\eqref{abs_W_max_POVM}] \ \ }\\
\text{} & \text{} & \text{} & ~\text{} \\
    \makecell{Efficiency \\(adiabatic)} & $\eta^{\mathcal{C}} = \eta_0$~[Eq.~\eqref{effi-rob}] & $\eta^{\Pi} = \eta_0$~[Eq.~\eqref{effi-rob}] & $\eta^{\mathcal{P}} = \eta_0$~[Eq.~\eqref{effi-rob}]\\
\text{} & \text{} & \text{} & ~\text{} \\
   \makecell{Optimal work \\ (non-adiabatic)} & \makecell{$\mathcal{W}_{M}^{\mathcal{C}} = \frac{\hbar}{2} \big(\tau_z\omega_{\chi z}^a + \tau_x \omega_{z \chi}^a\big)$  \\ {[Eq.~\eqref{absW nonadia 2b}]}} &{\Large$<$} 
\makecell{ $\mathcal{W}_{M}^{\Pi} = \frac{\hbar}{4}{\tau_z} \big(D - \omega_{\chi z}^a\big) $ \\{[Eq.~\eqref{absW_opt PV nonadia}]}}  &{\Large$<$} \makecell{ $\mathcal{W}^{\mathcal{P}}_M$  (Numerically studied)~[Fig.~\ref{fig: POVM nonadia}]} \\
    \text{} & \text{} & \text{} & ~\text{} \\
\makecell{Efficiency at\\optimal work\\extraction \\ (non-adiabatic)} & $\boldsymbol{\eta}^{\mathcal{C}} = 1 + \frac{\omega_z (\tau_z - \tau  a)}{\omega (\tau - \tau_z a)}$~[Eq.~\eqref{eta_2b_nonadia}] &\hspace{3mm} \makecell{$\boldsymbol{\eta}^{\Pi} = \frac{1}{1+\frac{\omega_z}{D}}$~[Eq.~\eqref{relabel: eta pi na}]} &{\Large$\leq$} \makecell{$\boldsymbol{\eta}^{\mathcal{P}}$ (Numerically studied)~[Fig.~\ref{fig: eta nonadia comparison}]} \\
\hline \hline
\end{tabular}
\caption{\textit{\textbf{Comparison between conventional two-bath, PVM-based, and POVM-based Otto engines.}} 
The table summarizes the optimal work extraction and efficiencies for the three classes of quantum Otto engines in both adiabatic and non-adiabatic regimes. The maximum extractable works satisfy the hierarchy
$W_{M}^{\mathcal{C}} \leq W_{M}^{\Pi} < W_{M}^{\mathcal{P}}$, where the equality holds only with the infinite-temperature hot bath in the conventional Otto cycle, i.e., $\beta_h = 0$, despite all three engines attain the same efficiency, i.e., $\eta_0$, in the adiabatic limit.
In the non-adiabatic regime, the corresponding optimized works also exhibit the ordering
$\mathcal{W}_{M}^{\mathcal{C}} < \mathcal{W}_{M}^{\Pi} < \mathcal{W}_{M}^{\mathcal{P}}$,
where the POVM-based case has been analyzed numerically. 
For the conventional two-bath case the maximization is over the temperature of the hot-bath, while for the other two it is on the measurement settings, where we keep the temperature of the cold bath the same in all the engines. Despite the all three engines for all work outputs exhibit the same efficiency for all work outputs in the adiabatic limit, the same does not hold in general in the non-adiabatic regimes. In the non-adiabatic regime, we focus on the efficiency for the maximal work extraction. For the conventional two-bath and PVM-based Otto engines the analytical expressions are enlisted, while the POVM-based engine is analyzed numerically, showcasing that $\boldsymbol{\eta}^{\Pi} \leq \boldsymbol{\eta}^{\mathcal{P}}$, where the equality holds at the adiabatic limit. Note that the results shown here for the POVM-based engine in the non-adiabatic regime, do not consider the work cost associated with resetting the auxiliary and such an analysis is undertaken in Sec.~\ref{subsec - POVM vs PV, adia} and Sec.~\ref{subsec - aux cost - POVM - non-adia}. Here, $\eta_0 \coloneqq 1-(\omega_z/\omega)$, $D \coloneqq \sqrt{(\omega - \omega_{z})^{2} + 4 (1-P) \omega \omega_z}$,  $a \coloneqq 2P-1$, $\Delta \omega \coloneqq\omega - \omega_z$, $\Delta\tau \coloneqq\tau_z - \tau$, $\omega_{\chi z}^a\coloneqq(\omega a - \omega_z)$, and $\omega_{z \chi}^a\coloneqq(\omega_z a - \omega)$.}
\label{table-comp}
\end{table*}

{\subsection{PVM-based vs. POVM-based Otto engines}
\label{POVM vs PV, nonadia}
In Fig.~\ref{fig: POVM nonadia}, we compare the maximum work extractable from the PVM-based engine, $\mathcal{W}^{\Pi}_{M}$ (red small dot-dashed), with the net work extractable from the POVM-based engine, for a fixed cold-bath inverse temperature $\beta_c = 1$. 
In Fig.~\ref{fig: POVM nonadia} the turquoise dashed line illustrates the maximal work from the POVM-based engine, i.e., $\mathcal{W}^{\mathcal{P}}_M$ before subtracting the resetting cost. Fig.~\ref{fig: POVM nonadia} clearly exhibits that $\mathcal{W}^{\mathcal{P}}_M$ attains significantly higher values than that of the PVM-based engine, i.e., $\mathcal{W}^{\Pi}_M$, depicted by the red dash-dot line. Moreover, on the other hand, when the net work $\mathcal{W}^{\mathcal{P}} - \mathcal{W}_a$ itself is is maximized, yielding $[\mathcal{W}^{\mathcal{P}} - \mathcal{W}_a]_M$, represented by the orange solid line, clearly attains larger value than $\mathcal{W}^{\Pi}_M$, depicted by the red dash-dot line. Therefore, in the parameter regime considered here, the POVM-based engine provides a clear advantage over the PVM-based engine, even after accounting for the auxiliary reset cost.

In the subsequent section, we provide a tabular summary comparing the performance of the three types of quantum Otto engines—conventional, PVM-based, and POVM-based—in both the adiabatic limit and non-adiabatic regime.

\section{Comparative Summary and Key Findings}
\label{sec - key results}
Table~\ref{table-comp} provides a comprehensive summary of our findings, highlighting the key similarities and differences among conventional two-bath, PVM-based, and POVM-based quantum Otto engines. It also offers a comparative assessment of their performance to identify the engine that delivers the highest work output in both the adiabatic and non-adiabatic limits. In particular, we find the following features:
\begin{itemize}
    \item In the adiabatic limit, the maximum extractable works satisfy the hierarchy
$W_{M}^{\mathcal{C}}$ [Eq.~\eqref{W_2b}] $\leq W_{M}^{\Pi}$ [Eq.~\eqref{abs_W_max_PV}]$< W_{M}^{\mathcal{P}}$ [Eq.~\eqref{abs_W_max_POVM}], where the equality holds only with the infinite-temperature hot bath in the conventional Otto cycle, i.e., $\beta_h = 0$.
\item In the adiabatic limit, all three engines attain the same efficiency, i.e., $\eta_0 \coloneqq 1-(\omega_z/\omega)$ [Eq.~\eqref{effi-rob}] for all work outputs,.
\item In the non-adiabatic regime, the corresponding optimized works also exhibit the ordering
$\mathcal{W}_{M}^{\mathcal{C}}$ [Eq.~\eqref{absW nonadia 2b}] $< \mathcal{W}_{M}^{\Pi}$ [Eq.~\eqref{absW_opt PV nonadia}] $ < \mathcal{W}_{M}^{\mathcal{P}}$ [Fig.~\ref{fig: POVM nonadia}],
where the POVM-based case has been analyzed numerically. The maximization is over the temperature of the hot-bath for the conventional two-bath engine, while for the other two engines it is performed on the measurement settings, while the temperature of the cold bath the is considered equal in all the engines.
\item Despite the all three engines exhibit the same efficiency for all work outputs in the adiabatic limit, the same does not hold in general in the non-adiabatic regimes. In the non-adiabatic regime, we focus on the efficiency for the maximal work extraction. For the conventional two-bath and PVM-based Otto engines, the quantities are given by Eq.~\eqref{eta_2b_nonadia} and Eq.~\eqref{relabel: eta pi na}, respectively. On the other hand, the POVM-based engine is analyzed numerically in Fig.~\ref{fig: eta nonadia comparison}, which showcases that $\boldsymbol{\eta}^{\Pi} \leq \boldsymbol{\eta}^{\mathcal{P}}$, where the equality holds at the adiabatic limit.
\end{itemize}

Beyond Table~\ref{table-comp}, moreover we find various realizable advantageous aspects, which are as follows:
\begin{itemize}
    \item The PVM-based Otto cycle can operate as an engine (Sec.~\ref{subsec - PV non-adia}), i.e., $\mathcal{W}^{\Pi}_M>0$ and $Q_h^{\Pi}>0$ hold simultaneously, for the entire domain of the non-adiabaticity parameter $P$. This is not always valid, in general, for the conventional two-bath engine, as shown in Fig.~\ref{fig: 2b vs PV nonadia}.
    \item Despite the results presented in Table~\ref{table-comp}, for the POVM-based engine in the non-adiabatic regime do not consider the work cost associated with resetting the auxiliary, such an analysis is undertaken in Sec.~\ref{subsec - POVM vs PV, adia} and Sec.~\ref{subsec - aux cost - POVM - non-adia}.\\
    In Sec.~\ref{subsec - POVM vs PV, adia}, we show that the positivity of $W^{\mathcal{P}\Pi} \coloneqq W^{\mathcal{P}}_M - W^{\Pi}_M$ can still hold (Fig.~\ref{fig: surpass}) even when we take the auxiliary resetting cost into account, and corresponding that we identify a sufficient condition $T_c < \hbar(\omega-\omega_z)/(2k_B \ln 2)$ based on the spectral gaps and the temperature of the cold bath.\\
    Sec.~\ref{subsec - aux cost - POVM - non-adia} exhibits that $[\mathcal{W}^{\mathcal{P}}-\mathcal{W}_a]_M > \mathcal{W}^{\Pi}_M$ can be obtained (Fig.~\ref{fig: POVM nonadia}), when the maximization is performed itself on the net work $\mathcal{W}^{\mathcal{P}}-\mathcal{W}_a$ obtaine after subtracting the auxiliary resetting cost.
\end{itemize}
}

\section{Conclusion}
\label{sec - Conclusion}

In this work, we investigated a measurement-driven quantum Otto engine in which the conventional hot thermalization stroke was replaced by quantum measurements. Motivated by the growing interest in measurement-based quantum thermal machines and their potential advantages in controllability, operational simplicity, and enhanced performance, we systematically analyzed the optimal work extraction achievable from such engines using a qubit working substance. The study was carried out in both the infinite-time (adiabatic) and finite-time (non-adiabatic) regimes, considering implementations based on both PVMs and generalized two-outcome POVMs.

We show that measurement-based Otto engines retain the same efficiency as conventional two-bath Otto engines in the adiabatic limit, while nevertheless enabling enhanced work extraction. By optimizing over all allowed measurement protocols, we identified the maximum extractable work for both PVM- and two-outcome POVM-based engines and compared with that of conventional hot-bath-driven quantum Otto engines. {For both adiabatic and non-adiabatic implementation} we showed that measurement-based engines can yield higher work output than the conventional two-bath Otto cycles within suitable parameter regimes, particularly in experimentally relevant finite-temperature settings. 
Furthermore, we found that non-adiabatic implementations of the PVM- as well as POVM-fueled engines can outperform than their adiabatic counterparts. We also identified the corresponding operational conditions, particularly in the case of PVM-fueled Otto engines. Moreover, measurement-fueled Otto cycles are able to operate as engines even in parameter regimes, where conventional Otto cycles fail to function as engines.

A central result of the present study was that, in the context of optimal work extraction, POVM-based engines can outperform their PVM-based counterparts in both the adiabatic limit and non-adiabatic regime. Even after incorporating the thermodynamic cost associated with resetting the auxiliary system required for POVM implementation, the resulting net work output can still exceed that achievable with PVM-based engines. In this regard, we identified parameter regimes, particularly for the adiabatic limit. Furthermore, despite the efficiency retaining the same value irrespective of the amount of work output in all three classes of quantum Otto engines in the adiabatic limit, the POVM-based engines can exhibit higher efficiency at optimal work extraction than their PVM-based counterparts in the non-adiabatic regime.


Overall, our work provided a comprehensive comparative analysis of PVM-based, POVM-based, and conventional two-bath quantum Otto engines, clarifying the regimes in which measurement-driven engines exhibit enhanced thermodynamic performance.  More broadly, the present study highlighted the potential of quantum measurements as useful resources for energy conversion at the quantum scale and may motivate further theoretical and experimental investigations toward the development of efficient finite-time quantum thermal devices and advanced measurement-assisted quantum heat engines.

\acknowledgements
S. M. and D. S. acknowledge support from the Infosys scholarship for senior students at Harish-Chandra Research Institute. U. S. acknowledges financial support from the Anusandhan National Research Foundation (ANRF), Government of India, under the Grant No. ANRF/ARG/2025/004617/PS.



\section*{Appendices}

\label{sec-Appendix}
\appendix
\section{Brief discussion on the conventional two-bath Otto engine}
\label{App - 2b conventional Otto adia}
Let us discuss here the working of the conventional two-bath Otto cycle. During the stroke-I of the cycle, the thermalization process associated with the cold bath occurs. The system is connected to a cold bath of the inverse temperature $\beta_c$ during this stroke. This stroke yields the thermal state (Gibbs state), given by

\begin{align}
    \rho_0^{\mathcal{C}} &= \frac{e^{-\beta_c H^{(1)}}}{\Tr(e^{-\beta_c H^{(1)}})}\nonumber\\
    &= \frac {e^{-v_z} | 0 \rangle \langle 0 | + e^{v_z} | 1 \rangle \langle 1 |}{2 \cosh{v_z}} \label{rho_0},
\end{align}
where $v_z \coloneqq (\hbar \beta_c \omega_z)/2$.
The energy after the thermalizing stroke is evaluated as
\begin{align}
    E_0^{\mathcal{C}} = \Tr(H^{(1)} \rho_0^{\mathcal{C}}) = - \frac{\hbar}{2} \omega_z \tau_z \label{E_0},
\end{align}
with $\tau_z \coloneqq \tanh{v_z}$

The stroke-II consists of an evolution of the Hamiltonian to $H^{(2)}$ starting from $H^{(1)}$. The system is kept detached from the cold bath during this stroke. The state undergoes a unitary transformation. 
{In the adiabatic limit, evolution from \(H^{(1)}\) to \(H^{(2)}\) preserves the populations in the instantaneous energy eigenbasis. Thus, the final populations in \(\{|\chi^+\rangle,|\chi^-\rangle\}\) equal those of the corresponding initial eigenstates of \(H^{(1)}\).}
The final state and the corresponding energy after the completion of this stroke are given by, respectively,
\begin{align}
    \rho_1^{\mathcal{C}} = p_1 | \chi^+ \rangle \langle \chi^+ | + p_2 | \chi^- \rangle \langle \chi^- | \label{rho_1},
\end{align}
where $p_1 \coloneqq {e^{-v_z}}/{(2 \cosh{v_z})}$ and $p_2 \coloneqq {e^{v_z}}/{(2 \cosh{v_z})}$,
and
\begin{align}
    E_1^{\mathcal{C}} = \Tr(\rho_1 H^{(2)}) = -\frac{\hbar}{2} \omega \tau_z \label{E_1}.
\end{align}
Here, we consider that the two unitary strokes are adiabatic. Therefore, each of the two unitary operations during these two unitary strokes effectively transforms between the eigenbasis of $\sigma_z$ and $(\cos \lambda) \sigma_z + (\sin \lambda) \sigma_x$. The work done during the first unitary stroke is given by
\begin{align}
    W_1^{\mathcal{C}} = E_1^{\mathcal{C}} - E_0^{\mathcal{C}} = \frac{\hbar}{2} \tau_z (\omega_z - \omega) \label{W_1}.
\end{align}

During the stroke-III, the system is attached to a hot bath of inverse temperature $\beta_h$, where $\beta_h < \beta_c$. This stroke yields the thermal state
\begin{align*}
    \rho_2^{\mathcal{C}} &= \frac{e^{-\beta_h H^{(2)}}}{\Tr(e^{-\beta_h H^{(2)}})}\\
    &= \frac{e^{-v} | \chi^+ \rangle \langle \chi^+ | + e^{v} | \chi^- \rangle \langle \chi^- |}{2 \cosh{v}},
\end{align*}
where $v \coloneqq (\hbar \beta_h \omega)/ 2$.
The energy after the thermalizing stroke is evaluated as
\begin{align}
    E_2^{\mathcal{C}} = \Tr(H^{(2)} \rho_2) = - \frac{\hbar}{2} \omega \tau \label{E2_2b},
\end{align}
where $\tau \coloneqq \tanh{v_x}$.
The change of energy due to this stroke is considered as heat because the Hamiltonian remains fixed during this stroke. The amount of transferred heat is given by
\begin{align}
    Q_h^{\mathcal{C}} = E_2^{\mathcal{C}} - E_1^{\mathcal{C}} = \frac{\hbar}{2} \omega (\tau_z - \tau)\label{QH_2b}.
\end{align}

The stroke-IV drives the Hamiltonian of the system to $H^{(1)}$ from $H^{(2)}$. Consequently, the state undergoes a unitary evolution. 
{In the adiabatic limit, the populations of the two-level system are preserved, and the populations of $\{|0\rangle,|1\rangle\}$ are equal to those of the \(\{|\chi^+\rangle,|\chi^-\rangle\}\)  in the initial state.} The final state and the corresponding energy after the completion of this stroke are given by, respectively,
\begin{align*}
    \rho_3^{\mathcal{C}} = \frac{e^{-v} | \chi^+ \rangle \langle \chi^+ | + e^{v} | \chi^- \rangle \langle \chi^- |}{2 \cosh{v_x}},
\end{align*}
and $E_3^{\mathcal{C}} = \Tr(\rho_3^{\mathcal{C}} H^{(1)}) = -(\hbar \omega_z \tau)/2$.
The work done due to this stroke is evaluated as
\begin{align}
    W_2^{\mathcal{C}} = E_3^{\mathcal{C}} - E_2^{\mathcal{C}} = \frac{\hbar}{2} (\omega - \omega_z) \tau. \label{W2_2b}
\end{align}
From Eq.~\eqref{W1} and Eq.~\eqref{W2_2b}, according to our adopted convention discussed in Sec.~\ref{subsec - convention_work}, the total work $W$ can be computed as
\begin{align}
    W^{\mathcal{C}} &= -(W_1^{\mathcal{C}} + W_2^{\mathcal{C}}) \nonumber \\
    &= \frac{\hbar}{2} (\omega - \omega_z) (\tau_z - \tau). \label{W_2b}
\end{align}
The amount of exchanged heat with the cold bath is given by
\begin{align}
    Q_c^{\mathcal{C}} &= E_0^{\mathcal{C}} - E_3^{\mathcal{C}} \nonumber\\
    &= \frac{\hbar}{2} \omega_z (\tau - \tau_z) \label{QC_2b}.
\end{align}
From Eq.~\eqref{W_2b}, Eq.~\eqref{QH_2b} and Eq.~\eqref{QC_2b}, we obtain $Q_h^{\mathcal{C}} + Q_c^{\mathcal{C}} = W^{\mathcal{C}}$, which verifies the consistency with conservation of energy.

The efficiency of the engine is given by
\begin{align}
    \eta^{\mathcal{C}} = \frac{W^{\mathcal{C}}}{Q_h^{\mathcal{C}}}= 1 - \frac{\omega_z}{\omega} \label{efficiency traditional nonadia}.
\end{align}
The requirement for the device to operate as an engine is that the work done is performed by the engine, i.e., $W^{\mathcal{C}}>0$ according to our convention. The physical requirement of $\eta^{\mathcal{C}} \leq 1$ gives us the criterion, i.e., $\omega_z \leq \omega$. Now, $\omega_z = \omega$ is not practically acceptable for operating as an engine, because it makes the net work $W^{\mathcal{C}}=0$ according to Eq.~\eqref{W_2b}. Therefore, the equality does not hold, and we obtain $\omega > \omega_z$, which also leads that $\eta^{\mathcal{C}} < 1$.

When $\omega > \omega_z$ is satisfied, the required positivity of the RHS of Eq.~\eqref{W_2b} gives us $\tau_z - \tau>0$ which yields
\begin{align}
    \frac{\omega_z}{\omega} > \frac{\beta_h}{\beta_c} \label{Carnot}.
\end{align}
Eq.~\eqref{Carnot} confirms that the Otto engine can not overcome the Carnot's efficiency, i.e., $1-(\beta_h/\beta_c)$, verifying that the framework is physically consistent.
\section{Considering mixed auxiliary states}
\label{App - Mixed state aux}
Let us consider that the auxiliary system is initially in an arbitrary mixed state $\rho_a = q |\tilde{\psi}\rangle \langle \tilde{\psi} | +  (1-q) |\tilde{\psi}^{\perp}\rangle \langle \tilde{\psi}^{\perp} |$ in its spectral decomposed form, where $q$ is the mixing parameter which satisfies $0 \leq q \leq 1$. Here, the eigenvalues of $\rho_a$ are $q$ and $1-q$. Therefore, the four eigenvalues of $\rho_{1}^{\mathcal{P}} \otimes \rho_a$ are ${q e^{-v_z}}/{(2 \cosh{v_z})}$, ${(1-q) e^{-v_z}}/{(2 \cosh{v_z})}$, ${q e^{v_z}}/{(2 \cosh{v_z})}$, and ${(1-q) e^{v_z}}/{(2 \cosh{v_z})}$. Without the loss of generality, we can consider $q \geq 1-q$. Furthermore, we restrict ourselves to the non-inverted population regime (excluding the effective negative temperature scenario), which implies the ordering $(1-q) e^{v_z} > q e^{-v_z}$. Therefore, the eigenvalues are sorted as follows:
\begin{align*}
    \frac{q e^{v_z}}{2 \cosh{v_z}} > \frac{(1-q) e^{v_z}}{2 \cosh{v_z}} > \frac{q e^{-v_z}}{2 \cosh{v_z}} > \frac{(1-q) e^{-v_z}}{2 \cosh{v_z}}.
\end{align*}

Here, we obtain the maximum value of $E_2$ as the following:
\begin{align}
    (E_{2}^{\mathcal{P}})_{M} &=  \Big(\frac{\hbar}{2} \omega \Big) \frac{q e^{v_z}}{2 \cosh{v_z}} +  \Big(\frac{\hbar}{2} \omega \Big) \frac{(1-q) e^{v_z}}{2 \cosh{v_z}} \nonumber\\ &+  \Big(-\frac{\hbar}{2} \omega \Big) \frac{q e^{-v_z}}{2 \cosh{v_z}} +  \Big(-\frac{\hbar}{2} \omega \Big) \frac{(1-q) e^{-v_z}}{2 \cosh{v_z}} \nonumber\\
    &= \frac{\hbar}{2} \omega \tanh{v_z}.
\end{align}
Thus we obtain the maximum value of $W$ as
\begin{align}
    W^{\mathcal{P}}_{M} = \frac{\hbar}{2} (\omega - \omega_z) \tanh{(v_z)}, \label{abs_W_max_POVM_mixed}
\end{align}
which is identical with the RHS of Eq.~\eqref{abs_W_max_PV}, i.e., the maximum extractable work from the PVM-based Otto engine. We conclude that taking the initial state of the auxiliary system as an arbitrary mixed state does not give any advantage over the PVM-based Otto engine.
\section{Construction of the swap operator}
\label{App - Swapping_stage}
Expressing the the identity operator as ${I} = |\chi^+\rangle \langle \chi^+| + |\chi^-\rangle \langle \chi^-|$ and $H^{(2)}$ as given by Eq.~\eqref{H2}, we can write $H^{(2)} \otimes {I}$ as
\begin{align*}
    H^{(2)} \otimes {I} = \frac{\hbar}{2} \omega (|\chi^+ \chi^+ \rangle \langle \chi^+ \chi^+ | + |\chi^+ \chi^- \rangle \langle \chi^+ \chi^- | \nonumber\\
    - |\chi^- \chi^+ \rangle \langle \chi^- \chi^+ | - |\chi^- \chi^- \rangle \langle \chi^- \chi^- |),
\end{align*}
which clearly shows that the eigenstates $\{ |\chi^+\chi^+\rangle, |\chi^+\chi^-\rangle \}$ are associated with the two degenerate higher energy eigenvalues $\hbar \omega / 2$ and $\{ |\chi^-\chi^+\rangle, |\chi^-\chi^-\rangle \}$ are associated with the two degenerate lower energy eigenvalues $- \hbar \omega / 2$.

Furthermore, with the help of Eq.~\eqref{rho_1} and plugging $\rho_a = |\chi^+ \rangle \langle \chi^+  |$, the bipartite state $\rho_{sa}$ is obtained as ${(e^{-v_z}} |\chi^+ \chi^+\rangle \langle \chi^+ \chi^+ | + {e^{v_z}} |\chi^- \chi^+\rangle \langle \chi^- \chi^+|)/(2 \cosh{v_z})$, showing that $\operatorname{supp}
(\rho_{sa}) = \{ |\chi^+ \chi^+ \rangle, |\chi^- \chi^+ \rangle \}$, being associated with only two non-zero eigenvalues.
\section{Non-adiabaticity}
\label{App - Non-adiabaticity}
Let us discuss how the single parameter $P$ incorporates the signature of non-adiabaticity as well as the adiabatic limit. When the duration of the unitary stroke is sufficiently large in comparison to the minimum spectral gap of the Hamiltonian, the adiabatic limit is reached. Although our analysis regarding the engine is generic, there are several possible paths for driving the time-dependent Hamiltonian from $H^{(1)}$ to $H^{(2)}$, provided that the gap never closes at any point of time during the evolution. In the adiabatic limit, characterized by $[P]_{t_f \rightarrow \infty}=1$, the Hamiltonian-driving is so slow that the elements of the density matrix of the state is preserved, while the eigenbasis of $H^{(1)}$ gets completely mapped to the eigenbasis of $H^{(2)}$.
\begin{figure*}[t]
    \centering
    \includegraphics[
        width=18cm,
    ]{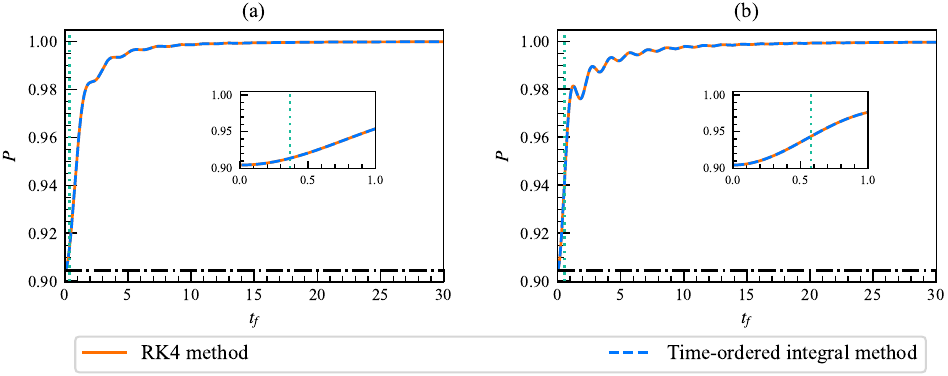}
\caption{\textbf{Dependence of the non-adiabaticity parameter $P$ on the duration of evolution $t_f$.}~Panels~(a) and (b) correspond to $H_{1}$ and $H_{2}$, given by Eq.~\eqref{Time dependent Hamiltonian: ramp} and Eq.~\eqref{Time dependent Hamiltonian: sinusoidal}, respectively. In both of the panels, the curves show the numerically computed values of $P$, where the dashed blue and solid orange lines depict the result generated by the time-ordered integral and 4th-order Runge-Kutta methods in Schrödinger picture and interaction picture, respectively. The turquoise dotted vertical line illustrates the time-scale $t_f^c$ given by the inverse of the minimum spectral gap during the evolution. The plot clearly exhibits that the adiabatic limit, i.e., $P=1$ is achieved when $t_f \gg t_f^c$, whereas $P\in [\cos^2 (\lambda/2),1)$ characterizes the non-adiabatic regime. The horizontal black dash-dot line illustrates the constant value $\cos^2(\lambda/2)$. The numerical analysis is performed with the values $(\omega,\omega_z)=(5,2)$ and $\lambda = \pi/5$. Note that, the quantities, the non-adiabaticity parameter $P$ and the duration of the drive $t_f$, are plotted along the vertical and horizontal axes, respectively, where the $t_f$ is expressed in the units of $\Omega_0^{-1}$ and $P$ is dimensionless.}

    \label{fig: P vs tau}
\end{figure*}

As an example, we take the time-dependent Hamiltonian
\begin{align}
    H_1(t) = (1-\tau') H^{(1)} + \tau' H^{(2)} \label{Time dependent Hamiltonian: ramp},
\end{align}
with $\tau' \coloneqq t/t_f$, where the boundary values render $H_1(t=0)=H^{(1)}$ and $H_1(t=t_f)=H^{(2)}$, with $H^{(1)}$ and $H^{(2)}$ given by Eq.~\eqref{H1} and Eq.~\eqref{H2 parameterized}, respectively. The energy spectrum of $H_1(t)$ is given by the two energy levels $\pm \sqrt{g}/(2t_f)$, resulting the spectral gap $\Delta E = \sqrt{g}/t_f$, where $g \coloneqq \hbar^2 (t^2 \omega^2+\omega_z^2 (t-t_f )^2 + 2 t (t_f - t) \omega \omega_z \cos \lambda)$.
The minimum spectral gap,
$G \coloneqq \min_{0 \leq \tau' \leq 1} \Delta E(\tau')$,
is given by
\begin{equation}
G =
\begin{cases}
\hbar \omega_z,
& u \leq 0, \\[1ex]
\dfrac{\hbar \omega \omega_z \sin\lambda}
{\sqrt{u^2 + \omega^2 \sin^2\lambda}},
& u>0,
\end{cases}
\end{equation}
where $u \coloneqq \omega_z - \omega \cos\lambda $.
The minimum is attained at $t=0$ in the first case and at
$t = \omega_z t_f u/(u^2+\omega^2\sin^2\lambda)$ in the second.
We see that $G \neq 0$, which ensures that the gap never closes during the evolution. Following from~\cite{Adiabatic_approximation}, we evaluate that the adiabatic approximation is valid when $t_f \gg t_f^c$ with the critical value $t_f^c =(\hbar^3 \omega_z \omega \sin \lambda)/(2 G^3)$.

Similarly, we can take another example of such time-dependent Hamiltonian corresponding to a different driving path, as follows:
\begin{align}
    H_2(t) = \Big[\cos\Big(\frac{\pi \tau'}{2}\Big) H^{(1)} + \sin\Big(\frac{\pi \tau'}{2}\Big) H^{(2)} \Big] \label{Time dependent Hamiltonian: sinusoidal},
\end{align}
with $\tau' \coloneqq t/t_f$, where the boundary values also satisfy $H_2(t=0)=H^{(1)}$ and $H_2(t=t_f)=H^{(2)}$, with $H^{(1)}$ and $H^{(2)}$ given by Eq.~\eqref{H1} and Eq.~\eqref{H2 parameterized}, respectively.
The energy spectrum of $H_2(t)$ is given by the two energy levels $\pm \sqrt{\tilde{g}}/(2\sqrt{2})$, where $\tilde{g} \coloneqq \hbar^2 (\omega^2 + \omega_z^2 - (\omega^2 - \omega_z^2) \cos \delta + 2 \omega \omega_z \cos \lambda \sin \delta)$ with $\delta \coloneqq \pi t/t_f$. Therefore, the energy-gap is given by $\Delta E \coloneqq \sqrt{\tilde{g}/2}$, with the minimum value $G \coloneqq \min_{0 \leq \tau' \leq 1} \Delta E = \hbar \omega_z$, which exists at $t=0$. Hence, we can conclude that the gap never closes throughout the evolution. With~\cite{Adiabatic_approximation}, we evaluate that the adiabatic approximation is valid when $t_f \gg t_f^c$ with the critical value $t_f^c =(\pi \omega \sin \lambda)/(4 \omega_z^2)$.

We numerically solve Eq.~\eqref{U time-ordered form} by performing the time-ordered integral and compute $P$ (dashed blue line), for each value of $t_f$. Furthermore, we solve the equivalent differential equation governing the evolution in the interaction picture, by using the 4th-order Runge-Kutta (RK4) method, and yielding identical result (orange solid line) up to the numerical precision of the order of $10^{-6}$. In both of the panels of Fig.~\ref{fig: P vs tau}, we plot $P$ as a function of $t_f$. Panels~(a) and (b) correspond to the driving governed by the time-dependent Hamiltonians $H^{(1)}$ and $H^{(2)}$, given by Eq.~\eqref{Time dependent Hamiltonian: ramp} and Eq.~\eqref{Time dependent Hamiltonian: sinusoidal}, respectively. In each of the panels, for the corresponding $H(t)$, the turquoise dotted vertical lines depict the critical value $t_f^c$, evaluated following~\cite{Adiabatic_approximation}. The plot shows that the adiabatic limit, i.e., $P=1$ is reached at sufficiently large value of $t_f$, i.e., $t_f \gg t_f^c$. The non-adiabatic regime is characterized by $P \in [\cos^2 (\lambda/2),1)$. For both of the panels, the numerical analysis is performed with the values $(\omega,\omega_z)=(5,2)$ and $\lambda = \pi/5$.
\section{Conventional Otto engine in non-adiabatic regime}
\label{App - nonadia 2b}
The behavior of the quantum Otto cycle, including the non-adiabatic regime, is discussed in~\cite{Otto_finite_time_12}.
The energies after the cold bath thermalization (stroke-I) and the following first unitary stroke (stroke-II) are given by
\begin{align}
    \mathcal{E}_0^{\mathcal{C}} &= -\frac{\hbar}{2}\omega_z \tau_z\label{E0_2b_nonadia},\\
    \text{and}~\mathcal{E}_1^{\mathcal{C}} &= \frac{\hbar \tau_z}{2} \omega (1-2P)\label{E1_2b_nonadia},
\end{align}
respectively, where $P \coloneqq |\langle \chi^+ | U | 0 \rangle |^2$. Subtracting, the work $\mathcal{W}_1^{\mathcal{C}} = \mathcal{E}_1^{\mathcal{C}} - \mathcal{E}_0^{\mathcal{C}}$ associated with the first unitary stroke (stroke-II) is obtained
\begin{align}
    \mathcal{W}_1^{\mathcal{C}} = \frac{\hbar \tau_z}{2} (\omega (1-2P) + \omega_z) \label{W1_2b_nonadia}.
\end{align}
Similarly, the energies after the hot bath thermalization (stroke-III) and the following second unitary stroke (stroke-IV) are given by 
\begin{align}
    \mathcal{E}^{\mathcal{C}}_2 = -\frac{\hbar}{2}\omega \tau \label{E2_2b_nonadia},\\
    \text{and}~\mathcal{E}^{\mathcal{C}}_3 = \frac{\hbar \tau}{2} \omega_z (1-2 \tilde{P})\label{E3_2b_nonadia},
\end{align}
respectively, where $\tilde{P} \coloneqq |\langle 0 | \tilde{U} | \chi^+ \rangle |^2$. Subtracting, the work $\mathcal{W}^{\mathcal{C}}_2 = \mathcal{E}^{\mathcal{C}}_3 - \mathcal{E}^{\mathcal{C}}_2$ associated with the second unitary stroke (stroke-IV) is obtained
\begin{align}
    \mathcal{W}^{\mathcal{C}}_2 = \frac{\hbar \tau}{2} (\omega_z (1-2\tilde{P}) + \omega)\label{W2_2b_nonadia},
\end{align}
Therefore, from Eq.~\eqref{W1_2b_nonadia} and Eq.~\eqref{W2_2b_nonadia}, the total work $\mathcal{W}^{\mathcal{C}} = -(\mathcal{W}^{\mathcal{C}}_1 + \mathcal{W}^{\mathcal{C}}_2)$ is obtained
\begin{align}
    \mathcal{W}^{\mathcal{C}} = \frac{\hbar}{2} [\tau_z(\omega (2P-1) - \omega_z) + \tau (\omega_z (2\tilde{P}-1) - \omega)]\label{W_2b_nonadia_asym}
\end{align}
The exchanged heat associated with the stroke-I and stroke-III are given by
\begin{align}
    \mathcal{Q}^{\mathcal{C}}_c &= \mathcal{E}^{\mathcal{C}}_0 - \mathcal{E}^{\mathcal{C}}_3 = -\frac{\hbar \omega_z}{2}\Big( \tau_z + \tau  (1-2 \tilde{P})\Big)\label{Qc_2b_nonadia}~\text{and}\\
    \mathcal{Q}^{\mathcal{C}}_h &= \mathcal{E}^{\mathcal{C}}_2 - \mathcal{E}^{\mathcal{C}}_1 = -\frac{\hbar \omega}{2}\Big(\tau+ \tau_z (1-2P)\Big)\label{Qh_2b_nonadia},
\end{align}
respectively.
From Eq.~\eqref{W_2b_nonadia_asym}, Eq.~\eqref{Qc_2b_nonadia}, and Eq.~\eqref{Qh_2b_nonadia}, it can be checked that $\mathcal{Q}^{\mathcal{C}}_c + \mathcal{Q}^{\mathcal{C}}_h = \mathcal{W}^{\mathcal{C}}$ is satisfied, showing that the conservation of energy is not violated.

Therefore, from Eq.~\eqref{W_2b_nonadia_asym} and Eq.~\eqref{Qh_2b_nonadia}, the efficiency $\boldsymbol{\eta}^{\mathcal{C}} = \mathcal{W}^{\mathcal{C}}/\mathcal{Q}^{\mathcal{C}}_h = 1+(\mathcal{Q}^{\mathcal{C}}_c / \mathcal{Q}^{\mathcal{C}}_h)$ of the engine can be evaluated
\begin{align}
    \boldsymbol{\eta}^{\mathcal{C}} = 1 + \frac{\omega_z \big(\tau_z + \tau  (1-2 \tilde{P})\big)}{\omega \big(\tau+ \tau_z (1-2P)\big)} \label{eta_2b_nonadia_asym}
\end{align}

Considering $\tilde{U} = U^{\dagger}$, we obtain $\tilde{P} = |\langle 0 | U^{\dagger} | \chi^+ \rangle |^2 = \langle 0 | U^{\dagger} | \chi^+ \rangle \langle 0 | U^{\dagger} | \chi^+ \rangle ^{*} = \langle \chi^+ | U | 0 \rangle ^{*} \langle \chi^+ | U | 0 \rangle = |\langle \chi^+ | U | 0 \rangle|^2 = P$, with which Eq.~\eqref{W_2b_nonadia_asym} and Eq.~\eqref{eta_2b_nonadia_asym} reduce to
\begin{align}
    \mathcal{W}^{\mathcal{C}} &= \frac{\hbar}{2} [\tau_z(\omega (2P-1) - \omega_z) + \tau (\omega_z (2P-1) - \omega)]\label{absW nonadia 2b},\\
    \text{and}~\boldsymbol{\eta}^{\mathcal{C}} &= 1 + \frac{\omega_z \big(\tau_z + \tau  (1-2P)\big)}{\omega \big(\tau + \tau_z (1-2P)\big)} \label{eta_2b_nonadia},
\end{align}
respectively.
$W^{\mathcal{C}}$ is not necessarily positive for the entire range of $P$, i.e., $P \in [\cos^2(\lambda/2),1]$. The regimes where $\mathcal{W}^{\mathcal{C}}>0$ holds for a particular setup, the Otto cycle operates as an engine.

Now, let us investigate when $\mathcal{W}^{\mathcal{C}}$ is maximum. Differentiating the both sides of Eq.~\eqref{absW nonadia 2b} w.r.t. $P$, we obtain $\frac{\partial\mathcal{W}^{\mathcal{C}}}{\partial P} = \hbar (\tau_z \omega + \tau \omega_z)$, which is a strictly positive quantity for the entire domain of $P$, i.e., $P \in [\cos^2(\lambda/2),1]$. As $\frac{\partial\mathcal{W}^{\mathcal{C}}}{\partial P}>0$, the maximum value of $\mathcal{W}^{\mathcal{C}}$ is attained at $P=1$, i.e., the adiabatic limit. Therefore, The efficiency at optimal work extraction is found $\boldsymbol{\eta}^{\mathcal{C}} = 1 - (\omega_z / \omega)$.

Furthermore, since $\omega > \omega_z$ and $2P-1  \in [\cos \lambda,1]$ hold, $\omega > \omega_z (2P-1)$ also holds good. Therefore, for a given set of $\omega_z$, $\omega$, $\beta_h$, and $P$, the extractable work $\mathcal{W}^{\mathcal{C}}$, given by Eq.~\eqref{absW nonadia 2b}, is a monotonically decreasing function of $\tau$, consequently implying that $\mathcal{W}^{\mathcal{C}}$ monotonically decreases with $\beta_h$ also. When $\beta_h = 0$, the extracted work is given by
\begin{align}
    [\mathcal{W}^{\mathcal{C}}]_{\beta_h=0} = \frac{\hbar}{2} [\tau_z(\omega (2P-1) - \omega_z)]\label{absW nonadia 2b, when beta_h = 0}.
\end{align}
Subtracting Eq.~\eqref{absW nonadia 2b, when beta_h = 0} from Eq.~\eqref{absW nonadia 2b}, we obtain
\begin{align}
    \mathcal{W}^{\mathcal{C}}-[\mathcal{W}^{\mathcal{C}}]_{\beta_h=0} = \tau (\omega_z (2P-1) - \omega) \label{W_2b inside comparison},
\end{align}
which is negative, leading to $[\mathcal{W}^{\mathcal{C}}]_{\beta_h=0}>\mathcal{W}^{\mathcal{C}}$. Therefore, when the Otto cycle, characterized by $\omega_z$ and $\omega$, performs as an engine for a given set of $\beta_h$ and $P$, then the extractable work $\mathcal{W}^{\mathcal{C}}$ can be increased by lowering $\beta_h$, and finally it becomes maximum when $\beta_h = 0$, i.e., the hot bath is taken to infinite temperature.
In both the panels of Fig.~\ref{fig: 2b vs PV nonadia}, Eq.~\eqref{absW nonadia 2b} and Eq.~\eqref{absW nonadia 2b, when beta_h = 0} are illustrated by the blue dashed and orange solid lines, respectively.
\section{Derivation of Eq.~\eqref{E2_PV_nonadia}}
\label{App - E2_PV_nonadia}
Let us evaluate $\mathcal{E}^{\Pi}_{2} = \Tr(H^{(2)} \tilde{\rho}^{\Pi}_{2})$ for the PVM-based Otto cycle. Putting $\tilde{\rho}^{\Pi}_{2} = \bm{p}_{\psi_{\theta\phi}^{+}} |\psi_{\theta\phi}^{+}\rangle \langle \psi_{\theta\phi}^{+}| + \bm{p}_{\psi_{\theta\phi}^{-}} |\psi_{\theta\phi}^{-}\rangle \langle \psi_{\theta\phi}^{-}|$, we obtain
\begin{align}
    \mathcal{E}^{\Pi}_{2} = \bm{p}_{\psi_{\theta\phi}^{+}} \langle \psi_{\theta\phi}^{+} | H^{(2)} | \psi_{\theta\phi}^{+}\rangle +  \bm{p}_{\psi_{\theta\phi}^{-}} \langle \psi_{\theta\phi}^{-} | H^{(2)} | \psi_{\theta\phi}^{-}\rangle\label{E2 intermediate_1}.
\end{align}
The spectral decomposition of $H^{(1)}$ can be written as $H^{(1)} = \sum_{k=1}^{2}{E_k^{(1)} |\psi_k^{(1)}\rangle \langle \psi_k^{(1)}|}$, where $k=1$ corresponds to $E_{1}^{(1)} = \hbar\omega_z/2$ and $|\psi_{k}^{(1)}\rangle = |0\rangle$, and $k=2$ corresponds to $E_2^{(1)} = -\hbar\omega_z/2$ and $|\psi_{2}^{(1)}\rangle = |1\rangle$. Similarly, the spectral decomposition of $H^{(2)}$ can be written as $H^{(2)} = \sum_{k=1}^{2}{E_k^{(2)} |\psi_k^{(2)}\rangle \langle \psi_k^{(2)}|}$, where $k=1$ corresponds to $E_{1}^{(2)} = \hbar \omega/2$ and $|\psi_{1}^{(2)}\rangle = |\chi^+\rangle$, and $k=2$ corresponds to $E_2^{(2)} = -\hbar \omega/2$ and $|\psi_{2}^{(1)}\rangle = |\chi^-\rangle$. 

With the spectral decomposition of $H^{(2)}$, we obtain 
\begin{align*}
\langle \psi_{\theta\phi}^{+} | H^{(2)} | \psi_{\theta\phi}^{+} \rangle &= \sum_{k}{E_{k}^{(2)} |\langle \psi_{\theta\phi}^{+} | \psi_k^{(2)}\rangle|^2}~\text{and}\\
\langle \psi_{\theta\phi}^{-} | H^{(2)} | \psi_{\theta\phi}^{-} \rangle &= \sum_{k}{E_{k}^{(2)} |\langle \psi_{\theta\phi}^{-} | \psi_k^{(2)}\rangle|^2}.
\end{align*}
Let us consider $B \coloneqq |\langle \psi_{\theta\phi}^{+} | \psi_1^{(2)}\rangle|^2$, and then we obtain $|\langle \psi_{\theta\phi}^{-} | \psi_1^{(2)}\rangle|^2 = |\langle \psi_{\theta\phi}^{+} | \psi_2^{(2)}\rangle|^2 = 1-B$ and $|\langle \psi_{\theta\phi}^{-} | \psi_2^{(2)}\rangle|^2 = B$, which give $| \langle \psi_{\theta\phi}^{+}| H^{(2)} | \psi_{\theta\phi}^{+}\rangle| = \hbar \omega (2B-1)/2$ and $| \langle \psi_{\theta\phi}^{-}| H^{(2)} | \psi_{\theta\phi}^{-}\rangle| = \hbar \omega (1-2B)/2$. Now, with $\bm{p}_{\psi_{\theta\phi}^{-}} = 1-\bm{p}_{\psi_{\theta\phi}^{+}}$, Eq.~\eqref{E2 intermediate_1} becomes
\begin{align}
    \mathcal{E}^{\Pi}_{2} = \frac{\hbar\omega}{2} (2B-1) (2 \bm{p}_{\psi_{\theta\phi}^{+}}-1)\label{E2 intermediate_2}.
\end{align}
Now, $p_{\psi^{+}_{\theta \phi}} = \langle \psi_{\theta\phi}^{+} | \tilde{\rho}^{\Pi}_1 | \psi_{\theta\phi}^{+} \rangle$. With $\tilde{\rho}^{\Pi}_1 = U \tilde{\rho}^{\Pi}_0 U^{\dagger}$ and $\tilde{\rho}^{\Pi}_0 = \sum_{j}{p_j |\psi_j^{(1)}\rangle \langle \psi_j^{(1)}|}$, we obtain $\bm{p}_{\psi_{\theta\phi}^{+}} = \sum_{j}{p_j |\langle \psi_{\theta\phi}^{+} | U | \psi_j^{(1)}\rangle |^2}$. Let, $A \coloneqq |\langle \psi_{\theta\phi}^{+} | U | \psi_1^{(1)}\rangle|^2$ and we obtain $|\langle \psi_{\theta\phi}^{-} | U | \psi_1^{(1)}\rangle|^2 = 1-A$. Now, $\bm{p}_{\psi_{\theta\phi}^{+}}$ can be rewritten as $2\bm{p}_{\psi_{\theta\phi}^{+}}-1 = p_1 A + p_2 (1-A)$. Putting $p_1$ and $p_2$ given in Sec.~\ref{tradional Otto 2b adia}, we obtain $2\bm{p}_{\psi_{\theta\phi}^{+}}-1 = \tau_z (1-2A)$, which reduces Eq.~\eqref{E2 intermediate_2} to
\begin{align*}
    \mathcal{E}^{\Pi}_{2} = -\frac{\hbar}{2} \omega \tau_z(2B-1)(2A-1),
\end{align*}
which is Eq.~\eqref{E2_PV_nonadia}. Hence, Eq.~\eqref{E2_PV_nonadia} is derived.

\section{Derivation of Eq.~\eqref{E3_PV_nonadia}}
\label{App - E3_PV_nonadia}
Let us evaluate $\mathcal{E}^{\Pi}_{3} = \Tr(H^{(1)} \rho_3)$ for the PVM-based Otto cycle. Plugging the spectral decomposition of $H^{(1)}$ and $\rho_3 = \tilde{U} \tilde{\rho}^{\Pi}_2 \tilde{U}^{\dagger}$, we obtain
\begin{align}
    \mathcal{E}^{\Pi}_{3} = \sum_{i,j,k}{E_{k}^{(1)} p_j |\langle \psi_i | U | \psi_{j}^{(1)} \rangle |^2 Q_{ki}} \label{E3_intermediate},
\end{align}
where $Q_{ki} \coloneqq |\langle \psi_{k}^{(1)}| \tilde{U} | \psi_i \rangle|^2$. Let us denote $Q \coloneqq |\langle \psi_{1}^{(1)}| \tilde{U} |\psi_{\theta\phi}^{+} \rangle |^2 $. Accordingly, we obtain $|\langle \psi_{1}^{(1)}| \tilde{U} |\psi_{\theta\phi}^{-} \rangle |^2 = |\langle \psi_{2}^{(1)}| \tilde{U} |\psi_{\theta\phi}^{+} \rangle |^2 = 1-Q$ and $|\langle \psi_{2}^{(1)}| \tilde{U} |\psi_{\theta\phi}^{-} \rangle |^2 = Q$. Now, Eq.~\eqref{E3_intermediate} becomes
\begin{align}
    \mathcal{E}_{3}^{\Pi} =& h_{\psi_{\theta\phi}^{+}}^{(1)} \sum_{j}{\big(p_j |\langle \psi_{\theta\phi}^{+} | U | \psi_j^{(1)}\rangle|^2 \big)} \nonumber\\
    &+ h_{\psi_{\theta\phi}^{-}}^{(1)} \sum_{j}{\big(p_j |\langle \psi_{\theta\phi}^{-} | U | \psi_j^{(1)}\rangle|^2 \big)}\label{E3 intermediate_2},
\end{align}
where $h_{\psi}^{(1)} \coloneqq \sum_{k}{E_k^{(1)} |\langle \psi_k^{(1)}| \tilde{U} | \psi_{\theta\phi}^{+}\rangle |^2} = {\hbar\omega_z} (2Q-1)/2$ and $h^{(1)}_{\psi_{\theta\phi}^{-}} \coloneqq \sum_{k}{E_k^{(1)} |\langle \psi_k^{(1)}| \tilde{U} | \psi_{\theta\phi}^{-}\rangle |^2} = {\hbar\omega_z} (1-2Q)/2$. Hence, Eq.~\eqref{E3 intermediate_2} reduces to
\begin{align}
    \mathcal{E}_{3}^{\Pi} = \frac{\hbar\omega_z}{2} (2Q-1) (2\bm{p}_{\psi_{\theta\phi}^{+}} - 1) \label{E3 intermediate_3},
\end{align}
with $\bm{p}_{\psi_{\theta\phi}^{+}} = \sum_{j}{p_j |\langle \psi_{\theta\phi}^{+}|U|\psi_j^{(1)}\rangle|^2}$. In Appendix~\ref{App - E2_PV_nonadia}, we have already derived $2p_{\psi^+_{\theta \phi}}-1 = \tau_z (1-2A)$, which reduces Eq.~\eqref{E3 intermediate_3} to
\begin{align*}
    \mathcal{E}_{3}^{\Pi} = -\frac{\hbar}{2} \tau_z \omega_z(2Q-1) (2A-1),
\end{align*}
which is Eq.~\eqref{E3_PV_nonadia}. Hence, Eq.~\eqref{E3_PV_nonadia} is derived.

\section{Analytical optimization of work extraction in the non-adiabatic PVM-based Otto engine}
\label{App - PV nonadia optimization}
To find the optimal value of the work, given by Eq.~\eqref{W_opt PV nonadia}, we analyze the extremum points as well as the boundary values.
Putting $\theta=0$, Eq.~\eqref{W_opt PV nonadia} gives
\begin{align}
    [\mathcal{W}^{\Pi}]_{\theta = 0} = -2 \omega_z \tau_z P (1-P) \leq 0,
\end{align}
for $P \in [\cos^2{(\lambda/2)},1]$, where the equality holds for $P=1$.
The two boundary values $\theta = 0 ,\pi$ correspond to the same basis of measurement, $\{ |\chi^+\rangle, |\chi^-\rangle \}$, leading to $\mathcal{W}^{\Pi} \leq 0$, concluding that the Otto cycle does not run as an engine at the two boundary values of $\theta$.

Now, let us find the intermediate extremum point. From the discussion in Sec.~\ref{subsec - PV non-adia}, we re-write Eq.~\eqref{W_opt PV nonadia}
\begin{align}
    \mathcal{W}^{\Pi} = \frac{\hbar \tau_z}{2} [a \omega - \omega_z + \mu (\omega_z \mu - \omega \cos{\theta})] \label{W_PV in terms of mu}
\end{align}
where $a \coloneqq 2P-1$, $b \coloneqq 2 \sqrt{P(1-P)}$, and
\begin{align}
    \mu = a \cos{\theta} + b \sin{\theta} \cos{( \phi - \alpha)} \label{mu}.
\end{align}
Intuitively, $\mathcal{W}^{\Pi}$ given by Eq.~\eqref{W_PV in terms of mu} is maximum when $\mu$ is maximum. Owing to the fact that $\sin \theta \geq 0$ for $\theta \in [0,\pi]$, we see that $\mu$ is maximum when 
\begin{align}
    \cos(\phi - \alpha) = 1\label{s_related},
\end{align}
for a given value of $\theta$.
However, we do the derivation formally with the differentiation-based method with the help of the Hessian matrix.

Now, partially differentiating $\mu$ w.r.t. $\phi$ and $\theta$, respectively, we obtain
\begin{align}
    \frac{\partial \mu}{\partial {\phi}} &= -b \sin{\theta} \sin{(\phi - \alpha)}~\text{and} \label{partial diff mu phi}\\
    \frac{\partial \mu}{\partial {\theta}} &= -a \sin{\theta} + b \cos{\theta} \cos{( \phi - \alpha)} \label{partial diff mu theta}.
\end{align}
Partially differentiating $W$ given by Eq.~\eqref{W_opt PV nonadia} w.r.t. $\phi$ and $\theta$, respectively, we obtain
\begin{align}
    \frac{\partial \mathcal{W}^{\Pi}}{\partial {\phi}} &= \frac{\hbar \tau_z}{2} \Big(2 \omega_z \mu - \omega \cos{\theta}\Big) \frac{\partial \mu}{\partial {\phi}} \label{partial diff W phi}~\text{and}\\
    \frac{\partial \mathcal{W}^{\Pi}}{\partial {\theta}} &= \frac{\hbar \tau_z}{2} \Big( (2 \omega_z \mu - \omega \cos{\theta}) \frac{\partial \mu}{\partial {\theta}} + \omega \mu \sin{\theta} \Big) \label{partial diff W theta},
\end{align}
where $\frac{\partial \mu}{\partial {\phi}}$ and $\frac{\partial \mu}{\partial {\theta}}$ are to be plugged from Eq.~\eqref{partial diff mu phi} and Eq.~\eqref{partial diff mu theta}, respectively.\\
Equating $\frac{\partial \mathcal{W}^{\Pi}}{\partial \phi}$ given by Eq.~\eqref{partial diff W phi} with zero, we obtain the two following cases -
\textbf{case~1:} $ \omega \cos{\theta} = 2 \omega_z \mu$ and \textbf{case~2:} $\frac{\partial \mu}{\partial {\phi}} = 0$.\\
\textbf{Case~1:} If case~1 holds, then Eq.~\eqref{partial diff W theta} reduces to $\sin{\theta} = 0$, leading to $\theta \in \{0, \pi \}$, which in turn corresponds to the boundary cases of $\theta$ in the domain of $\theta$, i.e., $\theta \in [0,\pi]$. We have already discussed this scenario; accordingly, this case can not lead to the optimal work extraction.\\
\textbf{Case~2:} Case~2 gives $\sin{\theta} \sin{(\phi - \alpha)} = 0$, which yields two subcases - \textbf{2(a):} $\sin{\theta} = 0$, which essentially leads to Case~1. \textbf{2(b):} $\sin{( \phi - \alpha)} = 0$, which leads to $\phi - \alpha = 0, \pm \pi$.\\
Therefore, case~2 gives us a valid condition for optimality, as the following: $\phi - \alpha = 0, \pm \pi$, which gives $\cos{(\phi - \alpha)} = s \in \{ 1, -1\}$.

With this optimality condition, Eq.~\eqref{mu}, Eq.~\eqref{partial diff mu theta}, and Eq.~\eqref{W_opt PV nonadia} reduce to, respectively,
\begin{align}
    \mu &= a \cos{\theta} + sb \sin{\theta} \label{mu reduced},\\
    \frac{\partial \mu}{\partial {\theta}} &= -a \sin{\theta} + s b \cos{\theta} \label{partial mu theta reduced},~\text{and}\\
    \mathcal{W}^{\Pi} = &\frac{\hbar \tau_z}{4} \Big[  \omega a -\omega_z - \Big ( a \omega 
    - \omega_z (a^2 - b^2) \Big) \cos{x} \nonumber\\ & - sb (\omega - 2a \omega_z) \sin{x} \Big] \label{W reduced},
\end{align}
where $s \in \{ 1, -1\}$. As Eq.~\eqref{mu reduced} and Eq.~\eqref{partial mu theta reduced} hold, Eq.~\eqref{partial diff W theta} reduces to $\frac{\partial \mathcal{W}^{\Pi}}{\partial {\theta}} = {\hbar \tau_z} ( -A \sin{x} + sB \cos{x} )/2$, where $A \coloneqq a \omega - \omega_z (a^2 - b^2)$, $B \coloneqq b (\omega - 2a \omega_z)$, and $x \coloneqq 2\theta$. Equating $\frac{\partial \mathcal{W}^{\Pi}}{\partial {\theta}}$ with zero, we obtain
\begin{align}
    A \sin{x} = sB \cos{x} \label{Optimality eqn}.
\end{align}
Squaring both sides of Eq.~\eqref{Optimality eqn} and using the identity $\sin^{2}{x} + \cos^{2}{x}=1$, we obtain
\begin{align}
    \cos{x} &= \frac{s'A}{\sqrt{A^2 + B^2}} \label{cos}\\
    \text{and}~\sin{x} &= \frac{sB}{A} \cos{x} \nonumber\\
    &= \frac{ss'B}{\sqrt{A^2 + B^2}}, \label{sin}
\end{align}
where $s'$ corresponds to the two solutions of the quadratic equation $s' \in \{1, -1\}$. Plugging the RHSs of Eq.~\eqref{sin} and Eq.~\eqref{cos} into Eq.~\eqref{W reduced}, we obtain
\begin{align}
    \mathcal{W}^{\Pi} =& \frac{\hbar \tau_z}{4} \big( a \omega -\omega_z - s' \sqrt{A^2 + B^2} \big)\nonumber\\
    =& \frac{\hbar \tau_z}{4} \Big( (2P-1) \omega -\omega_z \nonumber\\
    & - s' \sqrt{(\omega - \omega_z)^2 + 4 \omega \omega_z (1-P)} \Big) \label{W with s'},
\end{align}
with $P \in [\cos^2(\lambda/2),1]$ and $s' \in \{ 1,-1\}$.
    
Let us compute the Hessian matrix. Partially differentiating the both sides of Eq.~\eqref{partial diff W phi} w.r.t. $\phi_x$, we obtain
\begin{align}
    \frac{\partial^2 \mathcal{W}^{\Pi}}{\partial \phi^2} = \frac{\hbar \tau_z}{2} \Big( \big(2 \omega_z \mu - \omega \cos{\theta}\big) \frac{\partial^2 \mu}{\partial {\phi^{2}}} +2 \omega_z \big(\frac{\partial \mu}{\partial \phi} \big)^2 \Big) \label{partial double diff W phi}.
\end{align}
Now, $\frac{\partial \mu}{\partial \phi}=0$ at the optimality condition. Also, partially differentiating both sides of Eq.~\eqref{partial diff mu phi} w.r.t. $\phi$, we obtain $\frac{\partial^2 \mu}{\partial \phi^{2}} = -b \sin{\theta} \cos{(\phi-\alpha)}$, which reduces to $-sb \sin{\theta}$ at the optimality condition. So, the RHS of Eq.~\eqref{partial double diff W phi} becomes $-\tau_z sb ( \omega_z \mu \sin{\theta} - (\omega/4)\sin{x})$, with $2\theta = x$. Now, we plug $\sin{x}$ from Eq.~\eqref{sin}, simplify with $s^2=1$, and obtain
\begin{align}
    \frac{\partial^2 \mathcal{W}^{\Pi}}{\partial \phi^{2}} = -\hbar \tau_z b \Big( s \omega_z \mu \sin{\theta} - \frac{\omega}{4}\frac{s'B}{\sqrt{A^2 + B^2}}\Big) \label{diag phi}.
\end{align}
We also perform partial differentiation of the both sides of Eq.~\eqref{partial diff W theta} w.r.t. $\theta$, then plug $\cos{x}$ and $\sin{x}$ from Eq.~\eqref{cos} and Eq.~\eqref{sin} respectively, and finally obtain
\begin{align}
    \frac{\partial^2 \mathcal{W}^{\Pi}}{\partial \theta^{2}} = \hbar s' \tau_z \sqrt{A^2 + B^2} \label{diag theta}
\end{align}
Now, let us compute the off-diagonal elements of the Hessian matrix. Partially differentiating both sides of Eq.~\eqref{partial diff W phi} w.r.t. $\theta$, we obtain
\begin{align}
    \frac{\partial^2 \mathcal{W}^{\Pi}}{\partial \theta \partial \phi}  = \frac{\hbar \tau_z}{2} \Big[& (2 \omega_z \mu - \omega \cos{\theta}) \frac{\partial^2 \mu}{\partial \theta \partial \phi}  \nonumber\\
    &+ (2 \omega_z \frac{\partial \mu}{\partial \theta} + \omega \sin{\theta})\frac{\partial \mu}{\partial \phi} \Big] \label{off-diag}.
\end{align}
Partially differentiating both sides of Eq.~\eqref{partial diff mu phi} w.r.t. $\theta$, we get $\frac{\partial^2 \mu}{\partial \theta \partial \phi} = -b \cos{\theta} \sin{(\phi - \alpha)}$, which vanishes at the optimality condition. As $\frac{\partial^2 \mu}{\partial \theta \partial \phi} = 0$ and $\frac{\partial \mu}{\partial \phi}=0$ hold at the optimality condition, the RHS of Eq.~\eqref{off-diag} becomes zero. As, $W$ is a regular function, the two off-diagonal elements are the same due to the property of the partial differentiation.
Eq.~\eqref{diag phi} and Eq.~\eqref{diag theta}, with $s'=-1$, gives the Hessian matrix:
    \begin{align*}
            &\begin{pmatrix}
            \frac{\partial^2 \mathcal{W}^{\Pi}}{\partial \phi^{2}} & \frac{\partial^2 \mathcal{W}^{\Pi}}{\partial \theta \partial \phi}\\
            \frac{\partial^2 \mathcal{W}^{\Pi}}{\partial \phi \partial \theta} & \frac{\partial^2 \mathcal{W}^{\Pi}}{\partial \theta^{2}}
        \end{pmatrix}_{\text{optimal}}\\
        =&
            \begin{pmatrix}
            -\hbar \tau_z b \Big(s \omega_z \mu \sin{\theta} - \frac{\omega}{4}\frac{s' B}{\sqrt{A^2 + B^2}}\Big) & 0\\
            0 & \hbar s' \tau_z \sqrt{A^2 + B^2}
        \end{pmatrix},
    \end{align*}
of which the two diagonal elements are the two eigenvalues because the off-diagonal elements are zero. The required negativity of the eigenvalue $\hbar s' \tau_z \sqrt{A^2 + B^2}$ enforces $s'=-1$. Since $s'=-1$ and $\theta \in [0,\pi]$ implies $\sin{\theta}>0$, therefore, $-\hbar \tau_z b \Big(s \omega_z \mu \sin{\theta} - \frac{\omega}{4}\frac{s' B}{\sqrt{A^2 + B^2}}\Big)<0$ when $s=1$, being consistent with Eq.~\eqref{s_related}. Thus $s=1$ and $s'=-1$ together maximize $\mathcal{W}^{\Pi}$. With $s=1$ and $s'=-1$, the optimal $\theta$ can be obtained from either of Eq.~\eqref{cos} and Eq.~\eqref{sin}. Hence, with $s'=-1$, the optimal extractable work from the PVM-based Otto cycle is given by
\begin{align}
    \mathcal{W}^{\Pi}_{M} = \frac{\hbar \tau_z}{4} \Big( \sqrt{(\omega - \omega_z)^2 + 4 \omega \omega_z (1-P)} \nonumber\\
    - \omega_z \nonumber + \omega (2P-1) \Big),
\end{align}
which is Eq.~\eqref{absW_opt PV nonadia}, depicted by the turquoise dashed lines in Fig.~\ref{fig: 2b vs PV nonadia} as well as in the inset. Thus, we have obtained the optimal work for a given $P$ value.

Now, we analyze the value of $P$, for which $\mathcal{W}^{\Pi}_{M}$ is maximum. At first, we analyze the two boundary values of $P$. At the left boundary point $L \coloneqq \cos^2(\lambda/2)$ and at the right boundary point $R \coloneqq 1$, we obtain, respectively,
\begin{align}
        [\mathcal{W}^{\Pi}_{M}]_{L} & = \frac{\hbar \tau_z}{4} \big( \sqrt{\xi^2 + \epsilon^2} - \xi \big)\label{PV nonadia absW P=L}~\text{and}\\
        [\mathcal{W}^{\Pi}_{M}]_{R} & = \frac{\hbar \tau_z}{2} (\omega - \omega_z)\label{PV nonadia absW P=R},
    \end{align}
where $\xi \coloneqq \omega_z - \omega \cos \lambda$ and $\epsilon \coloneqq \omega \sin \lambda$.
Since, $\xi^2 + \epsilon^2 \geq \xi^2$, therefore, $\sqrt{\xi^2 + \epsilon^2} - \xi \geq 0$, which renders the RHS of Eq.~\eqref{PV nonadia absW P=L} positive-semidefinite. The equality holds particularly for the setup with $\lambda = 0$, which leads to $\epsilon = 0$. Similarly, since $\omega_z < \omega$, the RHS of Eq.~\eqref{PV nonadia absW P=R} is also positive. Now, we analyze whether any intermediate extremum point exists. We evaluate
\begin{align}
    \frac{\partial \mathcal{W}^{\Pi}}{\partial P} &= \frac{\hbar \tau_z \omega_z}{2} \Big(1-\frac{\omega_z}{D} \Big), \label{del W del P}
\end{align}
where $D \coloneqq \sqrt{(\omega - \omega_z)^2 + 4 \omega \omega_z (1-P)}$. Eq.~\eqref{del W del P} can also be rewritten as
\begin{align}
    \frac{\partial \mathcal{W}^{\Pi}}{\partial P} &= \frac{\hbar \tau_z \omega_z}{2} \Big(1-\frac{1}{\zeta} \Big), \label{del W del P, in terms of gamma}
\end{align}
with $\zeta \coloneqq \sqrt{(\gamma - 1)^2 + 4 \gamma (1-P)}$ and $\gamma \coloneqq {\omega}/{\omega_z}$ is the compression ratio.
Equating $\frac{\partial \mathcal{W}^{\Pi}}{\partial P} = 0$, we obtain that there is one extremum point at the particular value of $P$, i.e., $P_0 \coloneqq (1/2) + (\gamma/4)$. We also find that
\begin{align*}
    \Big[\frac{\partial^2 \mathcal{W}^{\Pi}}{\partial P^2}\Big]_{P = P_0} = - \frac{\hbar \tau_z \omega^2}{\omega_z} <0,
\end{align*}
which ensures that this point gives the maximum of $\mathcal{W}^{\Pi}$. Since the domain of $P$ is given by $P \in [\cos^2(\lambda/2),1]$, this intermediate extremum value exists when $2 \cos \lambda < \gamma < 2$.

Hence, we can classify the operational manifold in three regimes.

{\textbf{Case~1:}~{$2 \cos \lambda < \gamma < 2$.}}~ There is only one intermediate extremum point, which is found to be a maximum. Hence, this point is the global maxima. In this regime, non-adiabaticity is found to be beneficial. Furthermore, for a setup with $\lambda = 0$, i.e., no rotation of basis in the driven Hamiltonian, there is no existence of the regime characterized by $2 \cos \lambda < \gamma < 2$. 

{\textbf{Case~2:}~$\gamma \geq 2$.}~
For $\gamma = 2$, we find that $P_0 = 1$, hence coincides with the boundary point $R$.
For $\gamma > 2$, we find that $P_0 > 1$, which lies outside the domain of $P$, implying that the intermediate extremum point does not exist in that regime, and the maximum of $W$ is at any of the boundary values of $P$. At this regime, $\gamma > 2$ enforces $(\gamma-1)^2 > 1$, and owing to the fact that $1-P$ is positive semidefinite, finally it is implied that $\zeta > 1$, which renders $\frac{\partial \mathcal{W}^{\Pi}}{\partial P}>0$, leading to $[\mathcal{W}^{\Pi}_M]_{R}>[\mathcal{W}^{\Pi}_M]_{L}$ for all $\gamma \geq 2$. So, the global maximum lies at $R$, i.e., $P=1$, which is the adiabatic limit. Here, the maximum value of the work leads to Eq.~\eqref{PV nonadia absW P=R}, which is identical with Eq.~\eqref{abs_W_max_PV}.

{\textbf{Case~3:}~$\gamma \leq 2 \cos \lambda$.}~For $\gamma = 2 \cos \lambda$, we obtain that $P_0 = \cos^2 (\lambda/2)$ which turns out to be the left side boundary point $L$. For $\gamma \leq 2 \cos \lambda$, we find that $P_0 < \cos^2 (\lambda/2)$, which lies outside the domain of $P$, implying that the intermediate extremum point does not exist in that regime, and the maximum of $W$ is at any of the boundary values of $P$. Now, the lower bound of the domain of $P$ implies $1-P \leq (1- \cos \lambda)/2$. Then multiplying both sides with $4 \gamma$ (which is positive) and adding $(\gamma-1)^2$ into both sides, we obtain $\zeta^2 \leq 1 + \gamma (\gamma - 2 \cos \lambda)$. At this regime, $\gamma < 2 \cos \lambda$ enforces $\gamma (\gamma - 2 \cos \lambda) < 0$, owing to the fact that $\gamma$ is a positive quantity. Adding $1$ into both sides gives $1 + \gamma (\gamma - 2 \cos \lambda) < 1$, which implies $\zeta^2 < 1$, finally rendering $\frac{\partial \mathcal{W}^{\Pi}}{\partial P}<0$, which leads to $[\mathcal{W}^{\Pi}_M]_{R}<[\mathcal{W}^{\Pi}_M]_{L}$ for all $\gamma \leq 2 \cos \lambda$. So, the global maxima lies at the left side boundary point $L$. Here, the maximum value of the work leads to Eq.~\eqref{PV nonadia absW P=L}. In this regime also, non-adiabatic implementation becomes advantageous. Furthermore, for a setup with $\lambda \geq \pi/3$, we find that $2 \cos \lambda \leq 1$, therefore the regime characterized by $\gamma < 2 \cos \lambda$ does not exist at all, owing to the fact that $\gamma>1$ is for the operation of the cycle to effectively act as an engine.

Moreover, we see that $\mathcal{W}^{\Pi}>0$ holds for the two boundary points. There is no intermediate extremum point for $\gamma \geq 2$ and $\gamma \leq 2 \cos \lambda$. For $2 \cos \lambda < \gamma < 2$ there is only one extremum point which is found to be a maximum. Therefore, we can conclude that $\mathcal{W}^{\Pi}>0$ holds good for all values of $P \in [\cos^2(\lambda/2),1]$, implying that the PVM-based Otto cycle runs as an engine for the entire domain of $P$.

\section{Efficiency at the optimal work extraction in the non-adiabatic PVM-based Otto engine}
\label{App - Efficiency at optimal work extraction}
For any given value of the non-adiabaticity parameter $P$, the optimally extractable work $\mathcal{W}^{\Pi}$ is given by Eq.~\eqref{absW_opt PV nonadia}. The heat injected by the measurement stroke is evaluated
\begin{align}
    \mathcal{Q}^{\Pi}_{h} = \frac{\hbar \omega \tau_z (a (D-\omega_z)+\omega)}{4 D}\label{Q_H PV nonadia},
\end{align}
with $a \coloneqq 2P-1$ and $D \coloneqq \sqrt{(\omega - \omega_z)^2 + 4 \omega \omega_z (1-P)}$.
From Eq.~\eqref{absW_opt PV nonadia} and Eq.~\eqref{Q_H PV nonadia}, we obtain the efficiency $\boldsymbol{\eta}^{\Pi} = \mathcal{W}^{\Pi}_{M}/\mathcal{Q}^{\Pi}_{h}$ for any given value of $P$,
\begin{align}
\boldsymbol{\eta}^{\Pi} = \frac{1}{1+(\omega_z/D)} \label{eta PV nonadia},
\end{align}
which is a positive quantity for the entire range of the non-adiabaticity parameter $P$. In Fig.~\ref{fig: eta nonadia comparison}, the yellow dashed line illustrates $\boldsymbol{\eta}^{\Pi}$, according to Eq.~\eqref{eta PV nonadia}.
\bibliography{citations.bib}

@article{Engine_1,
  title={The quantum open system as a model of the heat engine},
  author={Robert Alicki},
  journal={Journal of Physics A},
  year={1979},
  volume={12},
  url={https://api.semanticscholar.org/CorpusID:120159113}
}

@article{Engine_2,
    author = {Kosloff, Ronnie},
    title = {A quantum mechanical open system as a model of a heat engine},
    journal = {The Journal of Chemical Physics},
    volume = {80},
    number = {4},
    pages = {1625-1631},
    year = {1984},
    month = {02},
    issn = {0021-9606},
    doi = {10.1063/1.446862},
    url = {https://doi.org/10.1063/1.446862}
}

@article{Engine_3,
  title = {Quantum thermodynamic cycles and quantum heat engines},
  author = {Quan, H. T. and Liu, Yu-xi and Sun, C. P. and Nori, Franco},
  journal = {Phys. Rev. E},
  volume = {76},
  issue = {3},
  pages = {031105},
  numpages = {18},
  year = {2007},
  month = {Sep},
  publisher = {American Physical Society},
  doi = {10.1103/PhysRevE.76.031105},
  url = {https://link.aps.org/doi/10.1103/PhysRevE.76.031105}
}

@article{Engine_4,
  title = {Work extremum principle: Structure and function of quantum heat engines},
  author = {Allahverdyan, Armen E. and Johal, Ramandeep S. and Mahler, Guenter},
  journal = {Phys. Rev. E},
  volume = {77},
  issue = {4},
  pages = {041118},
  numpages = {17},
  year = {2008},
  month = {Apr},
  publisher = {American Physical Society},
  doi = {10.1103/PhysRevE.77.041118},
  url = {https://link.aps.org/doi/10.1103/PhysRevE.77.041118}
}

@article{Engine_5,
author = {Marlan O. Scully  and Kimberly R. Chapin  and Konstantin E. Dorfman  and Moochan Barnabas Kim  and Anatoly Svidzinsky },
title = {Quantum heat engine power can be increased by noise-induced coherence},
journal = {Proceedings of the National Academy of Sciences},
volume = {108},
number = {37},
pages = {15097-15100},
year = {2011},
doi = {10.1073/pnas.1110234108},
URL = {https://www.pnas.org/doi/abs/10.1073/pnas.1110234108},
}

@article{Engine_6,
  title = {Isolated Quantum Heat Engine},
  author = {Fialko, O. and Hallwood, D. W.},
  journal = {Phys. Rev. Lett.},
  volume = {108},
  issue = {8},
  pages = {085303},
  numpages = {4},
  year = {2012},
  month = {Feb},
  publisher = {American Physical Society},
  doi = {10.1103/PhysRevLett.108.085303},
  url = {https://link.aps.org/doi/10.1103/PhysRevLett.108.085303}
}

@article{Engine_7,
   author = "Kosloff, Ronnie and Levy, Amikam",
   title = "Quantum Heat Engines and Refrigerators: Continuous Devices", 
   journal = "Annu. Rev. of Phys. Chem.",
   year = "2014",
   volume = "65",
   number = "Volume 65, 2014",
   pages = "365-393",
   doi = "https://doi.org/10.1146/annurev-physchem-040513-103724",
   url = "https://www.annualreviews.org/content/journals/10.1146/annurev-physchem-040513-103724",
   publisher = "Annual Reviews",
   issn = "1545-1593",
   type = "Journal Article",
  }

@article{Engine_8,
doi = {10.1088/1367-2630/18/8/083012},
url = {https://doi.org/10.1088/1367-2630/18/8/083012},
year = {2016},
month = {aug},
publisher = {IOP Publishing},
volume = {18},
number = {8},
pages = {083012},
author = {Niedenzu, Wolfgang and Gelbwaser-Klimovsky, David and Kofman, Abraham G and Kurizki, Gershon},
title = {On the operation of machines powered by quantum non-thermal baths},
journal = {New Journal of Physics}
}

@article{Engine_9,
  title = {Coherence-Induced Reversibility and Collective Operation of Quantum Heat Machines via Coherence Recycling},
  author = {Uzdin, Raam},
  journal = {Phys. Rev. Appl.},
  volume = {6},
  issue = {2},
  pages = {024004},
  numpages = {16},
  year = {2016},
  month = {Aug},
  publisher = {American Physical Society},
  doi = {10.1103/PhysRevApplied.6.024004},
  url = {https://link.aps.org/doi/10.1103/PhysRevApplied.6.024004}
}

@article{Engine_10,
doi = {10.1209/0295-5075/120/10002},
url = {https://doi.org/10.1209/0295-5075/120/10002},
year = {2017},
month = {dec},
publisher = {EDP Sciences, IOP Publishing and Società Italiana di Fisica},
volume = {120},
number = {1},
pages = {10002},
author = {Friedenberger, Alexander and Lutz, Eric},
title = {When is a quantum heat engine quantum?},
journal = {Europhysics Letters}
}

@Article{Engine_11,
author={Niedenzu, Wolfgang
and Mukherjee, Victor
and Ghosh, Arnab
and Kofman, Abraham G.
and Kurizki, Gershon},
title={Quantum engine efficiency bound beyond the second law of thermodynamics},
journal={Nature Communications},
year={2018},
month={Jan},
day={11},
volume={9},
number={1},
pages={165},
issn={2041-1723},
doi={10.1038/s41467-017-01991-6},
url={https://doi.org/10.1038/s41467-017-01991-6}
}

@article{Carnot_1,
    author = {Bender, Carl M. and Brody, Dorje C. and Meister, Bernhard K.},
    title = {Entropy and temperature of a quantum Carnot engine},
    journal = {Proceedings of the Royal Society A: Mathematical, Physical and Engineering Sciences},
    volume = {458},
    number = {2022},
    pages = {1519-1526},
    year = {2002},
    month = {06},
    issn = {1364-5021},
    doi = {10.1098/rspa.2001.0928},
    url = {https://doi.org/10.1098/rspa.2001.0928},
}

@article{Carnot_2,
  title = {Minimal universal quantum heat machine},
  author = {Gelbwaser-Klimovsky, D. and Alicki, R. and Kurizki, G.},
  journal = {Phys. Rev. E},
  volume = {87},
  issue = {1},
  pages = {012140},
  numpages = {9},
  year = {2013},
  month = {Jan},
  publisher = {American Physical Society},
  doi = {10.1103/PhysRevE.87.012140},
  url = {https://link.aps.org/doi/10.1103/PhysRevE.87.012140}
}

@article{Carnot_3,
  title = {Optimal Cycles for Low-Dissipation Heat Engines},
  author = {Abiuso, Paolo and Perarnau-Llobet, Mart\'{\i}},
  journal = {Phys. Rev. Lett.},
  volume = {124},
  issue = {11},
  pages = {110606},
  numpages = {7},
  year = {2020},
  month = {Mar},
  publisher = {American Physical Society},
  doi = {10.1103/PhysRevLett.124.110606},
  url = {https://link.aps.org/doi/10.1103/PhysRevLett.124.110606}
}

@Article{Otto_1,
author={Henrich, M. J.
and Rempp, F.
and Mahler, G.},
title={Quantum thermodynamic Otto machines: A  spin-system approach},
journal={EPJ ST},
year={2007},
month={Dec},
day={01},
volume={151},
number={1},
pages={157-165},
issn={1951-6401},
doi={10.1140/epjst/e2007-00371-8},
url={https://doi.org/10.1140/epjst/e2007-00371-8}
}

@article{Otto_2,
  title = {Quantum thermodynamic cycles and quantum heat engines},
  author = {Quan, H. T. and Liu, Yu-xi and Sun, C. P. and Nori, Franco},
  journal = {Phys. Rev. E},
  volume = {76},
  issue = {3},
  pages = {031105},
  numpages = {18},
  year = {2007},
  month = {Sep},
  publisher = {American Physical Society},
  doi = {10.1103/PhysRevE.76.031105},
  url = {https://link.aps.org/doi/10.1103/PhysRevE.76.031105}
}

@article{Otto_3,
  title = {Single-Ion Heat Engine at Maximum Power},
  author = {Abah, O. and Ro\ss{}nagel, J. and Jacob, G. and Deffner, S. and Schmidt-Kaler, F. and Singer, K. and Lutz, E.},
  journal = {Phys. Rev. Lett.},
  volume = {109},
  issue = {20},
  pages = {203006},
  numpages = {6},
  year = {2012},
  month = {Nov},
  publisher = {American Physical Society},
  doi = {10.1103/PhysRevLett.109.203006},
  url = {https://link.aps.org/doi/10.1103/PhysRevLett.109.203006}
}

@article{Otto_4,
  title={Universal features in the efficiency at maximal work of hot quantum Otto engines},
  author={Raam Uzdin and Ronnie Kosloff},
  journal={EPL},
  year={2014},
  volume={108},
  url={https://api.semanticscholar.org/CorpusID:119110107}
}

@article{Otto_5,
  title = {Otto engine beyond its standard quantum limit},
  author = {Leggio, Bruno and Antezza, Mauro},
  journal = {Phys. Rev. E},
  volume = {93},
  issue = {2},
  pages = {022122},
  numpages = {7},
  year = {2016},
  month = {Feb},
  publisher = {American Physical Society},
  doi = {10.1103/PhysRevE.93.022122},
  url = {https://link.aps.org/doi/10.1103/PhysRevE.93.022122}
}

@Article{Otto_6,
AUTHOR = {Kosloff, Ronnie and Rezek, Yair},
TITLE = {The Quantum Harmonic Otto Cycle},
JOURNAL = {Entropy},
VOLUME = {19},
YEAR = {2017},
NUMBER = {4},
ARTICLE-NUMBER = {136},
URL = {https://www.mdpi.com/1099-4300/19/4/136},
ISSN = {1099-4300},
}

@article{Otto_7,
  title = {Quantum Otto engine with exchange coupling in the presence of level degeneracy},
  author = {Mehta, Venu and Johal, Ramandeep S.},
  journal = {Phys. Rev. E},
  volume = {96},
  issue = {3},
  pages = {032110},
  numpages = {7},
  year = {2017},
  month = {Sep},
  publisher = {American Physical Society},
  doi = {10.1103/PhysRevE.96.032110},
  url = {https://link.aps.org/doi/10.1103/PhysRevE.96.032110}
}

@article{Otto_8,
  title = {Role of quantum correlations in light-matter quantum heat engines},
  author = {Barrios, G. Alvarado and Albarr\'an-Arriagada, F. and C\'ardenas-L\'opez, F. A. and Romero, G. and Retamal, J. C.},
  journal = {Phys. Rev. A},
  volume = {96},
  issue = {5},
  pages = {052119},
  numpages = {9},
  year = {2017},
  month = {Nov},
  publisher = {American Physical Society},
  doi = {10.1103/PhysRevA.96.052119},
  url = {https://link.aps.org/doi/10.1103/PhysRevA.96.052119}
}

@article{Otto_10,
  title = {Quantum Otto cycle in a superconducting cavity in the nonadiabatic regime},
  author = {Del Grosso, Nicol\'as F. and Lombardo, Fernando C. and Mazzitelli, Francisco D. and Villar, Paula I.},
  journal = {Phys. Rev. A},
  volume = {105},
  issue = {2},
  pages = {022202},
  numpages = {10},
  year = {2022},
  month = {Feb},
  publisher = {American Physical Society},
  doi = {10.1103/PhysRevA.105.022202},
  url = {https://link.aps.org/doi/10.1103/PhysRevA.105.022202}
}

@article{Otto_finite_time_1,
  title={A quantum mechanical open system as a model of a heat engine},
  author={Kosloff, R.},
  journal={J. Chem. Phys.},
  volume={80},
  pages={1625},
  year={1984},
  publisher={American Institute of Physics},
  url={https://api.semanticscholar.org/CorpusID:120159113}
}

@article{Otto_finite_time_2,
  title={On the classical limit of quantum thermodynamics in finite time},
  author={Geva, Eitan and Kosloff, Ronnie},
  journal={J.Chem.Phys.},
  volume={97},
  number={6},
  pages={4398},
  year={1992},
  publisher={American Institute of Physics},
url={https://doi.org/10.1063/1.463909}
}

@article{Otto_finite_time_3,
  title={Three-level quantum amplifier as a heat engine: A study in finite-time thermodynamics},
  author={Geva, E.and Kosloff, R.},
  journal={Phys. Rev. E.},
  volume={49},
  number={5},
  pages={3903},
  year={1994},
  publisher={APS},
  url = {https://link.aps.org/doi/10.1103/PhysRevE.49.3903}
}

@inproceedings{Otto_finite_time_4,
  title={Performance analysis of an irreversible Otto cycle using finite time thermodynamics},
  author={Mehta, H. B. and Bharti, O.S},
  booktitle={Proceedings of the World Congress on Engineering},
  volume={2},
  pages={1},
  year={2009},
  organization={WCE London, UK},
  url={https://api.semanticscholar.org/CorpusID:2534467}
}

@article{Otto_finite_time_5,
  title={Efficiency at maximum power of a quantum Otto cycle within finite-time or irreversible thermodynamics},
  author={Wu, F. and He, J. and Ma, Y. and Wang, J.},
  journal={Phys. Rev. E},
  volume={90},
  pages={062134},
  year={2014},
  publisher={APS},
  url = {https://link.aps.org/doi/10.1103/PhysRevE.90.062134}
}

@article{Otto_finite_time_6,
  title={Quantum Otto cycle with inner friction: finite-time and disorder effects},
  author={Alecce, A. and Galve, F. and Gullo, N. Lo. and Dell’Anna, L. and Plastina, F. and Zambrini, R.},
  journal={New.J.Phys},
  volume={17},
  pages={075007},
  year={2015},
  publisher={IOP Publishing},
url={
https://doi.org/10.1088/1367-2630/17/7/075007
}
}

@article{Otto_finite_time_7,
  title={Boosting the performance of quantum Otto heat engines},
  author={Chen, J.F. and Sun, C.P. and Dong, H.},
  journal={Phys. Rev. E},
  volume={100},
  pages={032144},
  year={2019},
  publisher={APS},
  url = {https://link.aps.org/doi/10.1103/PhysRevE.100.032144}
}

@article{Otto_finite_time_8,
  title={Shortcut-to-adiabaticity Otto engine: A twist to finite-time thermodynamics},
  author={Abah, O. and Paternostro, M.},
  journal={Phys. Rev. E},
  volume={99},
  pages={022110},
  year={2019},
  publisher={APS},
url={
https://doi.org/10.1103/PhysRevE.99.022110
}
}

@article{Otto_finite_time_9,
  title={Collective performance of a finite-time quantum Otto cycle},
  author={Kloc,M. and Cejnar,P. and Schaller, G.},
  journal={Phys. Rev. E},
  volume={100},
  pages={042126},
  year={2019},
  publisher={APS},
url={
https://doi.org/10.48550/arXiv.2205.13290
}
}

@article{Otto_finite_time_10,
  title={Finite-time performance of a quantum heat engine with a squeezed thermal bath},
  author={Wang, J. and He, J. and Ma, Y.},
  journal={Phys. Rev. E},
  volume={100},
  pages={052126},
  year={2019},
  publisher={APS},
url={
https://doi.org/10.48550/arXiv.2205.13290
}
}

@article{Otto_finite_time_11,
  title={Finite-time quantum Otto engine: Surpassing the quasistatic efficiency due to friction},
  author={Lee,S. and Ha,M.and Park, J.M. and Jeong, H.},
  journal={Phys. Rev. E},
  volume={101},
  pages={022127},
  year={2020},
  publisher={APS},
 url = {https://link.aps.org/doi/10.1103/PhysRevE.101.022127}
}

@article{Otto_finite_time_12,
  title = {Nonadiabatic single-qubit quantum Otto engine},
  author = {Solfanelli, Andrea and Falsetti, Marco and Campisi, Michele},
  journal = {Phys. Rev. B},
  volume = {101},
  issue = {5},
  pages = {054513},
  numpages = {9},
  year = {2020},
  month = {Feb},
  publisher = {American Physical Society},
  doi = {10.1103/PhysRevB.101.054513},
  url = {https://link.aps.org/doi/10.1103/PhysRevB.101.054513}
}

@article{Otto_finite_time_13,
  title={Quantum-enhanced finite-time Otto cycle},
  author={Das, A. and Mukherjee, V.},
  journal={Phys.Rev.R},
  volume={2},
  pages={033083},
  year={2020},
  publisher={APS},
url = {https://link.aps.org/doi/10.1103/PhysRevResearch.2.033083}
}

@article{Otto_finite_time_14,
  title={Bounds on fluctuations for finite-time quantum Otto cycle},
  author={Saryal, S. and Agarwalla, B.},
  journal={Phys. Rev. E.},
  volume={103},
  pages={L060103},
  year={2021},
  publisher={APS},
  url = {https://link.aps.org/doi/10.1103/PhysRevE.103.L060103}
}

@article{Otto_finite_time_15,
  title={Non-Markovian effect on quantum Otto engine: Role of system-reservoir interaction},
  author={Shirai, Y. and Hashimoto, K. and Tezuka, R. and Uchiyama, C. and Hatano, N.},
  journal={Phys.Rev.R},
  volume={3},
  pages={023078},
  year={2021},
  publisher={APS},
 url = {https://link.aps.org/doi/10.1103/PhysRevResearch.3.023078}
}

@article{Otto_finite_time_16,
      title={Temperature- and interaction-tweaked efficiency boost of finite-time robust quantum Otto engines}, 
      author={Debarupa Saha and Ahana Ghoshal and Ujjwal Sen},
      year={2023},
      journal={arXiv:2309.11483},
      primaryClass={quant-ph},
      url={https://arxiv.org/abs/2309.11483}, 
}

@article{MBO_1,
  title = {Measurement-based formulation of quantum heat engines},
  author = {Hayashi, Masahito and Tajima, Hiroyasu},
  journal = {Phys. Rev. A},
  volume = {95},
  issue = {3},
  pages = {032132},
  numpages = {21},
  year = {2017},
  month = {Mar},
  publisher = {American Physical Society},
  doi = {10.1103/PhysRevA.95.032132},
  url = {https://link.aps.org/doi/10.1103/PhysRevA.95.032132}
}

@article{MBO_2,
  title = {Single-temperature quantum engine without feedback control},
  author = {Yi, Juyeon and Talkner, Peter and Kim, Yong Woon},
  journal = {Phys. Rev. E},
  volume = {96},
  issue = {2},
  pages = {022108},
  numpages = {5},
  year = {2017},
  month = {Aug},
  publisher = {American Physical Society},
  doi = {10.1103/PhysRevE.96.022108},
  url = {https://link.aps.org/doi/10.1103/PhysRevE.96.022108}
}

@Article{PV_1,
AUTHOR = {Das, Arpan and Ghosh, Sibasish},
TITLE = {Measurement Based Quantum Heat Engine with Coupled Working Medium},
JOURNAL = {Entropy},
VOLUME = {21},
YEAR = {2019},
NUMBER = {11},
ARTICLE-NUMBER = {1131},
URL = {https://www.mdpi.com/1099-4300/21/11/1131},
ISSN = {1099-4300}
}

@article{PV_2,
  title = {Critical-point behavior of a measurement-based quantum heat engine},
  author = {Chand, Suman and Biswas, Asoka},
  journal = {Phys. Rev. E},
  volume = {98},
  issue = {5},
  pages = {052147},
  numpages = {8},
  year = {2018},
  month = {Nov},
  publisher = {American Physical Society},
  doi = {10.1103/PhysRevE.98.052147},
  url = {https://link.aps.org/doi/10.1103/PhysRevE.98.052147}
}

@article{PV_3,
  title = {Measurement-based quantum heat engine in a multilevel system},
  author = {Anka, Maron F. and de Oliveira, Thiago R. and Jonathan, Daniel},
  journal = {Phys. Rev. E},
  volume = {104},
  issue = {5},
  pages = {054128},
  numpages = {11},
  year = {2021},
  month = {Nov},
  publisher = {American Physical Society},
  doi = {10.1103/PhysRevE.104.054128},
  url = {https://link.aps.org/doi/10.1103/PhysRevE.104.054128}
}

@article{PV_4_&_finite_time_2,
  title = {Suppressing coherence effects in quantum-measurement-based engines},
  author = {Lin, Zhiyuan and Su, Shanhe and Chen, Jingyi and Chen, Jincan and Santos, Jonas F. G.},
  journal = {Phys. Rev. A},
  volume = {104},
  issue = {6},
  pages = {062210},
  numpages = {8},
  year = {2021},
  month = {Dec},
  publisher = {American Physical Society},
  doi = {10.1103/PhysRevA.104.062210},
  url = {https://link.aps.org/doi/10.1103/PhysRevA.104.062210}
}

@article{PV_5_&_finite_time_3,
  title = {Measurement-based quantum Otto engine with a two-spin system coupled by anisotropic interaction: Enhanced efficiency at finite times},
  author = {Purkait, Chayan and Biswas, Asoka},
  journal = {Phys. Rev. E},
  volume = {107},
  issue = {5},
  pages = {054110},
  numpages = {11},
  year = {2023},
  month = {May},
  publisher = {American Physical Society},
  doi = {10.1103/PhysRevE.107.054110},
  url = {https://link.aps.org/doi/10.1103/PhysRevE.107.054110}
}

@article{Ent_measurement,
  title = {Enhancing the efficiency of quantum measurement-based engines with entangling measurements},
  author = {Mayo, Franco and Roncaglia, Augusto J.},
  journal = {Phys. Rev. E},
  volume = {113},
  issue = {1},
  pages = {014118},
  numpages = {6},
  year = {2026},
  month = {Jan},
  publisher = {American Physical Society},
  doi = {10.1103/bx1f-wjn8},
  url = {https://link.aps.org/doi/10.1103/bx1f-wjn8}
}

@article{PV_6_&_finite_time_4,
title = {Quasi-probability distribution of work in a measurement-based quantum Otto engine},
journal = {Physica A: Statistical Mechanics and its Applications},
volume = {673},
pages = {130650},
year = {2025},
issn = {0378-4371},
doi = {https://doi.org/10.1016/j.physa.2025.130650},
url = {https://www.sciencedirect.com/science/article/pii/S0378437125003024},
author = {Chayan Purkait and Shubhrangshu Dasgupta and Asoka Biswas}
}

@article{aux_asst_1_&_finite_time_5,
  title = {Ancilla measurement-based quantum Otto engine using double-pair spin architecture},
  author = {Rathnakaran, S. R. and Biswas, Asoka},
  journal = {Phys. Rev. E},
  volume = {111},
  issue = {6},
  pages = {064116},
  numpages = {12},
  year = {2025},
  month = {Jun},
  publisher = {American Physical Society},
  doi = {10.1103/w4j5-gftl},
  url = {https://link.aps.org/doi/10.1103/w4j5-gftl}
}

@article{Three_&_five_stroke_engines_with_POVMs,
doi = {10.1088/1751-8121/abca74},
url = {https://doi.org/10.1088/1751-8121/abca74},
year = {2020},
month = {dec},
publisher = {IOP Publishing},
volume = {54},
number = {1},
pages = {015304},
author = {Behzadi, Naghi},
title = {Quantum engine based on general measurements},
journal = {Journal of Physics A: Mathematical and Theoretical}
}

@article{Swith_based_engine,
  title = {Correlations in a quantum switch-based heat engine with measurements: a proof-of-principle demonstration},
  volume = {11},
  ISSN = {2058-9565},
  url = {http://dx.doi.org/10.1088/2058-9565/ae34e0},
  DOI = {10.1088/2058-9565/ae34e0},
  number = {1},
  journal = {Quantum Science and Technology},
  publisher = {IOP Publishing},
  author = {Lisboa,  Vinicius F and Dieguez,  Pedro R and Simonov,  Kyrylo and Serra,  Roberto M},
  year = {2026},
  month = Feb,
  pages = {015058}
}

@article{transistor_1,
  title = {Quantum Thermal Transistor},
  author = {Joulain, Karl and Drevillon, J\'er\'emie and Ezzahri, Youn\`es and Ordonez-Miranda, Jose},
  journal = {Phys. Rev. Lett.},
  volume = {116},
  issue = {20},
  pages = {200601},
  numpages = {5},
  year = {2016},
  month = {May},
  publisher = {American Physical Society},
  doi = {10.1103/PhysRevLett.116.200601},
  url = {https://link.aps.org/doi/10.1103/PhysRevLett.116.200601}
}

@article{transistor_2,
  title = {Dynamically induced heat rectification in quantum systems},
  author = {Riera-Campeny, Andreu and Mehboudi, Mohammad and Pons, Marisa and Sanpera, Anna},
  journal = {Phys. Rev. E},
  volume = {99},
  issue = {3},
  pages = {032126},
  numpages = {10},
  year = {2019},
  month = {Mar},
  publisher = {American Physical Society},
  doi = {10.1103/PhysRevE.99.032126},
  url = {https://link.aps.org/doi/10.1103/PhysRevE.99.032126}
}

@article{transistor_3,
  title = {Thermal Transistor Effect in Quantum Systems},
  author = {Mandarino, Antonio and Joulain, Karl and G\'omez, Melisa Dom\'{\i}nguez and Bellomo, Bruno},
  journal = {Phys. Rev. Appl.},
  volume = {16},
  issue = {3},
  pages = {034026},
  numpages = {15},
  year = {2021},
  month = {Sep},
  publisher = {American Physical Society},
  doi = {10.1103/PhysRevApplied.16.034026},
  url = {https://link.aps.org/doi/10.1103/PhysRevApplied.16.034026}
}

@article{transistor_4,
  title = {Quantum thermal transistor in superconducting circuits},
  author = {Majland, Marco and Christensen, Kasper Sangild and Zinner, Nikolaj Thomas},
  journal = {Phys. Rev. B},
  volume = {101},
  issue = {18},
  pages = {184510},
  numpages = {10},
  year = {2020},
  month = {May},
  publisher = {American Physical Society},
  doi = {10.1103/PhysRevB.101.184510},
  url = {https://link.aps.org/doi/10.1103/PhysRevB.101.184510}
}

@article{transistor_5,
  title = {Quantum thermal transistors: Operation characteristics in steady state versus transient regimes},
  author = {Ghosh, Riddhi and Ghoshal, Ahana and Sen, Ujjwal},
  journal = {Phys. Rev. A},
  volume = {103},
  issue = {5},
  pages = {052613},
  numpages = {14},
  year = {2021},
  month = {May},
  publisher = {American Physical Society},
  doi = {10.1103/PhysRevA.103.052613},
  url = {https://link.aps.org/doi/10.1103/PhysRevA.103.052613}
}

@article{transistor_6,
doi = {10.1088/2058-9565/ae94a2},
url = {https://doi.org/10.1088/2058-9565/ae94a2},
year = {2026},
month = {aug},
publisher = {IOP Publishing},
volume = {11},
number = {4},
pages = {045010},
author = {Bhargava, Deepika and Chaki, Paranjoy and Bhattacharyya, Aparajita and Sen, Ujjwal},
title = {Quantum transistors for heat flux in and out of working substance parts: harmonic vs transmon and Kerr environs},
journal = {Quantum Science and Technology}
}

@article{refrigerator_1,
doi = {10.1209/0295-5075/85/30008},
url = {https://doi.org/10.1209/0295-5075/85/30008},
year = {2009},
month = {feb},
publisher = {},
volume = {85},
number = {3},
pages = {30008},
author = {Rezek, Y. and Salamon, P. and Hoffmann, K. H. and Kosloff, R.},
title = {The quantum refrigerator: The quest for absolute zero},
journal = {Europhysics Letters}
}

@article{refrigerator_2,
      title={Optimal Performance of Quantum Refrigerators}, 
      author={Tova Feldmann and Ronnie Kosloff},
      year={2009},
      journal={arXiv:0906.0986},
      primaryClass={quant-ph},
      url={https://arxiv.org/abs/0906.0986}, 
}

@article{transformer_1,
doi = {10.1088/1367-2630/ae2e34},
url = {https://doi.org/10.1088/1367-2630/ae2e34},
year = {2025},
month = {dec},
publisher = {IOP Publishing},
volume = {28},
number = {1},
pages = {014501},
author = {Maity, Arghya and Chaki, Paranjoy and Ghoshal, Ahana and Sen, Ujjwal},
title = {Quantum heat transformers},
journal = {New Journal of Physics}
}

@article{battery_1,
  title = {Entanglement boost for extractable work from ensembles of quantum batteries},
  author = {Alicki, Robert and Fannes, Mark},
  journal = {Phys. Rev. E},
  volume = {87},
  issue = {4},
  pages = {042123},
  numpages = {4},
  year = {2013},
  month = {Apr},
  publisher = {American Physical Society},
  doi = {10.1103/PhysRevE.87.042123},
  url = {https://link.aps.org/doi/10.1103/PhysRevE.87.042123}
}

@Article{battery_2,
author={Bhattacharjee, Sourav
and Dutta, Amit},
title={Quantum thermal machines and batteries},
journal={The European Physical Journal B},
year={2021},
month={Dec},
day={08},
volume={94},
number={12},
pages={239},
issn={1434-6036},
doi={10.1140/epjb/s10051-021-00235-3},
url={https://doi.org/10.1140/epjb/s10051-021-00235-3}
}

@article{battery_3,
  title = {Colloquium: Quantum batteries},
  author = {Campaioli, Francesco and Gherardini, Stefano and Quach, James Q. and Polini, Marco and Andolina, Gian Marcello},
  journal = {Rev. Mod. Phys.},
  volume = {96},
  issue = {3},
  pages = {031001},
  numpages = {30},
  year = {2024},
  month = {Jul},
  publisher = {American Physical Society},
  doi = {10.1103/RevModPhys.96.031001},
  url = {https://link.aps.org/doi/10.1103/RevModPhys.96.031001}
}

@article{Experimental_1,
  title = {Maxwell's Demon Assisted Thermodynamic Cycle in Superconducting Quantum Circuits},
  author = {Quan, H. T. and Wang, Y. D. and Liu, Yu-xi and Sun, C. P. and Nori, Franco},
  journal = {Phys. Rev. Lett.},
  volume = {97},
  issue = {18},
  pages = {180402},
  numpages = {4},
  year = {2006},
  month = {Oct},
  publisher = {American Physical Society},
  doi = {10.1103/PhysRevLett.97.180402},
  url = {https://link.aps.org/doi/10.1103/PhysRevLett.97.180402}
}

@article{Experimental_2,
doi = {10.1088/1367-2630/ab2684},
url = {https://doi.org/10.1088/1367-2630/ab2684},
year = {2019},
month = {jun},
publisher = {IOP Publishing},
volume = {21},
number = {6},
pages = {063019},
author = {Barontini, Giovanni and Paternostro, Mauro},
title = {Ultra-cold single-atom quantum heat engines},
journal = {New Journal of Physics}
}

@article{Experimental_3,
  title = {Experimental Characterization of a Spin Quantum Heat Engine},
  author = {Peterson, John P. S. and Batalh\~ao, Tiago B. and Herrera, Marcela and Souza, Alexandre M. and Sarthour, Roberto S. and Oliveira, Ivan S. and Serra, Roberto M.},
  journal = {Phys. Rev. Lett.},
  volume = {123},
  issue = {24},
  pages = {240601},
  numpages = {7},
  year = {2019},
  month = {Dec},
  publisher = {American Physical Society},
  doi = {10.1103/PhysRevLett.123.240601},
  url = {https://link.aps.org/doi/10.1103/PhysRevLett.123.240601}
}

@article{Experimental_4,
  title = {Optical simulation of a quantum thermal machine},
  author = {Passos, M. H. M. and Santos, Alan C. and Sarandy, Marcelo S. and Huguenin, J. A. O.},
  journal = {Phys. Rev. A},
  volume = {100},
  issue = {2},
  pages = {022113},
  numpages = {9},
  year = {2019},
  month = {Aug},
  publisher = {American Physical Society},
  doi = {10.1103/PhysRevA.100.022113},
  url = {https://link.aps.org/doi/10.1103/PhysRevA.100.022113}
}

@article{Experimental_5,
      title={Realization of a coupled-mode heat engine with cavity-mediated nanoresonators}, 
      author={Jiteng Sheng and Cheng Yang and Haibin Wu},
      year={2021},
      journal={arXiv:2110.13022},
      primaryClass={quant-ph},
      url={https://arxiv.org/abs/2110.13022}, 
}

@article{Experimental_6,
      title={Exploring the role of criticality in the quantum Otto cycle fueled by the anisotropic quantum Rabi-Stark model}, 
      author={He-Guang Xu and Jiasen Jin and Norton G. de Almeida and G. D. de Moraes Neto},
      year={2024},
      journal={arXiv:2407.09027},
      primaryClass={quant-ph},
      url={https://arxiv.org/abs/2407.09027}, 
}

@article{Experimental_7,
      title={Experimental investigation of a quantum Otto heat engine with shortcuts to adiabaticity implemented using counter-adiabatic driving}, 
      author={Krishna Shende and Matreyee Kandpal and Arvind and Kavita Dorai},
      year={2024},
      journal={arXiv:2412.20194},
      primaryClass={quant-ph},
      url={https://arxiv.org/abs/2412.20194}, 
}

@article{Experimental_8,
      title={Experimental realization of a quantum heat engine based on dissipation-engineered superconducting circuits}, 
      author={Tuomas Uusnäkki and Timm Mörstedt and Wallace Teixeira and Miika Rasola and Mikko Möttönen},
      year={2025},
      journal={arXiv:2502.20143},
      primaryClass={quant-ph},
      url={https://arxiv.org/abs/2502.20143}, 
}

@article{Experimental_9,
      title={Autonomous quantum heat engine}, 
      author={Tuomas Uusnäkki and Miika Rasola and Vasilii Vadimov and Priyank Singh and Ahmad Darwish and Mikko Möttönen},
      year={2026},
      journal={arXiv:2603.15355},
      primaryClass={quant-ph},
      url={https://arxiv.org/abs/2603.15355}, 
}

@article{transformer2,
  title = {Quantum Thermal Analogs of Electric Circuits: A Universal Approach},
  author = {Tiwari, Devvrat and Bhattacharya, Samyadeb and Banerjee, Subhashish},
  journal = {Phys. Rev. Lett.},
  volume = {135},
  issue = {2},
  pages = {020404},
  numpages = {6},
  year = {2025},
  month = {Jul},
  publisher = {American Physical Society},
  doi = {10.1103/5x8m-bhgd},
  url = {https://link.aps.org/doi/10.1103/5x8m-bhgd}
}

@article{No-Go-MBO,
      title={No-Go Theorem for Quantum Heat Engines Powered Purely by Quantum Measurements in the Steady Regime}, 
      author={Kenta Koshihara and Kazuya Yuasa},
      year={2026},
      journal={arXiv:2604.22376},
      primaryClass={quant-ph},
      url={https://arxiv.org/abs/2604.22376}, 
}

@book{Deffner-book2019,
author = {Deffner, Sebastian and Campbell, Steve},
title = {Quantum Thermodynamics},
publisher = {Morgan \& Claypool Publishers},
year = {2019},
series = {2053-2571},
isbn = {978-1-64327-658-8},
url = {https://doi.org/10.1088/2053-2571/ab21c6},
doi = {10.1088/2053-2571/ab21c6}
}

@article{Vinjanampathy01102016,
author = {Sai Vinjanampathy and Janet Anders},
title = {Quantum thermodynamics},
journal = {Contemporary Physics},
volume = {57},
number = {4},
pages = {545--579},
year = {2016},
publisher = {Taylor \& Francis},
doi = {10.1080/00107514.2016.1201896},
URL = {https://doi.org/10.1080/00107514.2016.1201896}
}

@book{Gemmer2009,
  title = {Quantum Thermodynamics: Emergence of Thermodynamic Behavior Within Composite Quantum Systems},
  ISBN = {9783540705109},
  ISSN = {1616-6361},
  url = {http://dx.doi.org/10.1007/978-3-540-70510-9},
  DOI = {10.1007/978-3-540-70510-9},
  journal = {Lecture Notes in Physics},
  publisher = {Springer Berlin Heidelberg},
  author = {Gemmer,  Jochen and Michel,  M. and Mahler,  G\"{u}nter},
  year = {2009}
}

@book{ThermoBook2018,
  title = {Thermodynamics in the Quantum Regime: Fundamental Aspects and New Directions},
  ISBN = {9783319990460},
  ISSN = {2365-6425},
  url = {http://dx.doi.org/10.1007/978-3-319-99046-0},
  DOI = {10.1007/978-3-319-99046-0},
  journal = {Fundamental Theories of Physics},
  publisher = {Springer International Publishing},
  year = {2018}
}

@article{PhysRevLett.105.130401,
  title = {How Small Can Thermal Machines Be? The Smallest Possible Refrigerator},
  author = {Linden, Noah and Popescu, Sandu and Skrzypczyk, Paul},
  journal = {Phys. Rev. Lett.},
  volume = {105},
  issue = {13},
  pages = {130401},
  numpages = {4},
  year = {2010},
  month = {Sep},
  publisher = {American Physical Society},
  doi = {10.1103/PhysRevLett.105.130401},
  url = {https://link.aps.org/doi/10.1103/PhysRevLett.105.130401}
}

@article{Maruyama2009,
  title = {Colloquium: The physics of Maxwell’s demon and information},
  volume = {81},
  ISSN = {1539-0756},
  url = {http://dx.doi.org/10.1103/RevModPhys.81.1},
  DOI = {10.1103/revmodphys.81.1},
  number = {1},
  journal = {Reviews of Modern Physics},
  publisher = {American Physical Society (APS)},
  author = {Maruyama,  Koji and Nori,  Franco and Vedral,  Vlatko},
  year = {2009},
  month = Jan,
  pages = {1–23}
}

@article{Reeb2014,
  title = {An improved Landauer principle with finite-size corrections},
  volume = {16},
  ISSN = {1367-2630},
  url = {http://dx.doi.org/10.1088/1367-2630/16/10/103011},
  DOI = {10.1088/1367-2630/16/10/103011},
  number = {10},
  journal = {New Journal of Physics},
  publisher = {IOP Publishing},
  author = {Reeb,  David and Wolf,  Michael M},
  year = {2014},
  month = Oct,
  pages = {103011}
}

@article{Chattopadhyay2025,
  title = {Landauer principle and thermodynamics of computation},
  volume = {88},
  ISSN = {1361-6633},
  url = {http://dx.doi.org/10.1088/1361-6633/add6b3},
  DOI = {10.1088/1361-6633/add6b3},
  number = {8},
  journal = {Reports on Progress in Physics},
  publisher = {IOP Publishing},
  author = {Chattopadhyay,  Pritam and Misra,  Avijit and Pandit,  Tanmoy and Paul,  Goutam},
  year = {2025},
  month = July,
  pages = {086001}
}

@article{CANGEMI20241,
title = {Quantum engines and refrigerators},
journal = {Physics Reports},
volume = {1087},
pages = {1-71},
year = {2024},
note = {Quantum engines and refrigerators},
issn = {0370-1573},
doi = {https://doi.org/10.1016/j.physrep.2024.07.001},
url = {https://www.sciencedirect.com/science/article/pii/S0370157324002710},
author = {Loris Maria Cangemi and Chitrak Bhadra and Amikam Levy}
}

@article{refrigerator_0,
  title = {Quantum thermodynamic cooling cycle},
  author = {Palao, Jos\'e P. and Kosloff, Ronnie and Gordon, Jeffrey M.},
  journal = {Phys. Rev. E},
  volume = {64},
  issue = {5},
  pages = {056130},
  numpages = {8},
  year = {2001},
  month = {Oct},
  publisher = {American Physical Society},
  doi = {10.1103/PhysRevE.64.056130},
  url = {https://link.aps.org/doi/10.1103/PhysRevE.64.056130}
}

@article{Yunger-Halpern2019,
  title = {Quantum engine based on many-body localization},
  author = {Yunger Halpern, Nicole and White, Christopher David and Gopalakrishnan, Sarang and Refael, Gil},
  journal = {Phys. Rev. B},
  volume = {99},
  issue = {2},
  pages = {024203},
  numpages = {24},
  year = {2019},
  month = {Jan},
  publisher = {American Physical Society},
  doi = {10.1103/PhysRevB.99.024203},
  url = {https://link.aps.org/doi/10.1103/PhysRevB.99.024203}
}

@article{Jakhar2026,
  title = {Ergotropic advantage in a measurement-fueled quantum heat engine},
  author = {Jakhar, Sidhant and Johal, Ramandeep S.},
  journal = {Phys. Rev. E},
  volume = {113},
  issue = {2},
  pages = {024142},
  numpages = {8},
  year = {2026},
  month = {Feb},
  publisher = {American Physical Society},
  doi = {10.1103/npcw-l6gn},
  url = {https://link.aps.org/doi/10.1103/npcw-l6gn}
}

@article{Kraus-Cirac,
  title = {Optimal creation of entanglement using a two-qubit gate},
  author = {Kraus, B. and Cirac, J. I.},
  journal = {Phys. Rev. A},
  volume = {63},
  issue = {6},
  pages = {062309},
  numpages = {8},
  year = {2001},
  month = {May},
  publisher = {American Physical Society},
  doi = {10.1103/PhysRevA.63.062309},
  url = {https://link.aps.org/doi/10.1103/PhysRevA.63.062309}
}

@article{Luders_2006,
author = {Lüders, G.},
title = {Concerning the state-change due to the measurement process},
journal = {Annalen der Physik},
volume = {518},
number = {9},
pages = {663-670},
url = {https://onlinelibrary.wiley.com/doi/abs/10.1002/andp.20065180904},
year = {2006}
}

@article{Busch_1998,
title = {Lüders theorem for unsharp quantum measurements},
journal = {Physics Letters A},
volume = {249},
number = {1},
pages = {10-12},
year = {1998},
issn = {0375-9601},
doi = {https://doi.org/10.1016/S0375-9601(98)00704-X},
url = {https://www.sciencedirect.com/science/article/pii/S037596019800704X},
author = {Paul Busch and Javed Singh}
}

@article{Adiabatic_approximation,
  title = {Adiabatic approximation in open quantum systems},
  author = {Sarandy, M. S. and Lidar, D. A.},
  journal = {Phys. Rev. A},
  volume = {71},
  issue = {1},
  pages = {012331},
  numpages = {12},
  year = {2005},
  month = {Jan},
  publisher = {American Physical Society},
  doi = {10.1103/PhysRevA.71.012331},
  url = {https://link.aps.org/doi/10.1103/PhysRevA.71.012331}
}

@book{von_Neumann_entropy_English_translated_original_paper,
    author = {},
    editor = {von Neumann, John and Beyer, Robert T.},
    title = {Mathematical Foundations of Quantum Mechanics: New Edition},
    publisher = {Princeton University Press},
    year = {2018},
    month = {02},
    isbn = {9780691178561},
    doi = {10.23943/princeton/9780691178561.001.0001},
    url = {https://doi.org/10.23943/princeton/9780691178561.001.0001},
}

@article{Paris_2012,
  author = {Paris, M. G. A.},
  title = {The modern tools of quantum mechanics},
  journal = {The European Physical Journal Special Topics},
  year = {2012},
  volume = {203},
  number = {1},
  pages = {61--86},
  doi = {10.1140/epjst/e2012-01535-1},
  url = {https://doi.org/10.1140/epjst/e2012-01535-1},
  isbn = {1951-6401}
}

@book{Nielsen_Chuang_2010, place={Cambridge}, title={Quantum Computation and Quantum Information: 10th Anniversary Edition}, publisher={Cambridge University Press}, author={Nielsen, Michael A. and Chuang, Isaac L.}, year={2010}}

@manual{preskill2018quantum_ch3,
  title        = {Quantum Information, Chapter 3: Foundations II: Measurement and Evolution},
  author       = {Preskill, John},
  organization = {California Institute of Technology},
  year         = {2018},
  note         = {Lecture Notes for Ph219/CS219},
  url          = {https://www.preskill.caltech.edu/ph219/chap3_15.pdf},
  urldate      = {2025-12-11}
}

@article{Peres1990Neumark,
  author = {Peres, Asher},
  title = {Neumark's theorem and quantum inseparability},
  journal = {Foundations of Physics},
  year = {1990},
  volume = {20},
  number = {12},
  pages = {1441--1453},
  doi = {10.1007/BF01883517},
  url = {https://doi.org/10.1007/BF01883517},
  isbn = {1572-9516}
}

@article{Naimark1976Non,
  author = {Naimark, M. A. and Loginov, A. I. and Shul'man, V. S.},
  title = {Non-self-adjoint operator algebras in Hilbert space},
  journal = {Journal of Soviet Mathematics},
  year = {1976},
  volume = {5},
  number = {2},
  pages = {250--278},
  doi = {10.1007/BF01247398},
  url = {https://doi.org/10.1007/BF01247398},
  isbn = {1573-8795}
}
\end{document}